\documentclass[a4paper,11pt]{article}
\pdfoutput=1 

\usepackage{jcappub} 

\usepackage[T1]{fontenc} 

\usepackage{graphicx}
\usepackage{amssymb}
\usepackage{ulem}
\usepackage{multirow}
\usepackage{color}
\usepackage{rotating}
\usepackage{placeins} 
\usepackage{cleveref}
\usepackage[font=footnotesize]{caption}
\usepackage{dcolumn}
\usepackage{pifont}
\usepackage{float}

\def\be{\begin{equation}}
\def\ee{\end{equation}}

\def\bi{\begin{itemize}}
\def\ei{\end{itemize}}

\def\ben{\begin{enumerate}}
\def\een{\end{enumerate}}

\def\bt{\begin{tabular}}
\def\et{\end{tabular}}

\def\bc{\begin{center}}
\def\ec{\end{center}}

\def\be{\begin{equation}}
\def\ee{\end{equation}}

\newcommand{\bes}{\begin{subequations}}
\newcommand{\ees}{\end{subequations}}
\def\bea{\begin{eqnarray}}
\def\eea{\end{eqnarray}}

\def\tcr{\textcolor{red}}

\def\tcg{\textcolor{green}}

\allowdisplaybreaks

\newcount\hh
\newcount\mm
\mm=\time
\hh=\time
\divide\hh by 60
\divide\mm by 60
\multiply\mm by 60
\mm=-\mm
\advance\mm by \time
\def\hhmm{\number\hh:\ifnum\mm<10{}0\fi\number\mm}

\title{Constraining Violations of the Weak Equivalence Principle in the Dark Sector}

\author{Sina Bahrami$^{1,2}$}
\affiliation{$1)$ Institute for Gravitation and the Cosmos, Pennsylvania State University, State College, PA 16803, USA.

$2)$ Department of Physics, Cornell University, Ithaca, NY 14853, USA.
}

\emailAdd{sina.bahrami@psu.edu}

\abstract{The effective field theory of dark energy is generalized
to incorporate dark matter, which is modelled using a complex scalar
field with a global $U(1)$ symmetry.
The dark matter model used here has similarities to models of ultralight axion. Generic interaction terms in the dark matter sector violate the weak equivalence principle. The degree to which such violations can occur is constrained using a number of recent analyses and forecasts in cosmology.}

\begin{document}
\maketitle
\flushbottom

\def\tcr{\textcolor{red}}
\def\tcb{\textcolor{blue}}
\definecolor{darkgreen}{rgb}{0.0, 0.6, 0.0}
\def\tcg{\textcolor{darkgreen}}

\section{Introduction}\label{intro}

\subsection{Background and motivation}\label{context}

Over the past few decades, evidence for a recent era of cosmic acceleration has accumulated from a
diverse set of cosmological observations and is now overwhelming.  These observations include
type Ia supernovae
\cite{supernovae1,supernovae2,supernovae3,supernovae4}, anisotropies
of the cosmic microwave background radiation (CMB)
\cite{cmb-planck,cmb-act}, and large scale structure (LSS)
surveys \cite{lss-1,lss-2,lss-3}.  The most common explanation for the
current
phase of cosmic acceleration is dark energy, with a cosmological
constant being its simplest realization\footnote{See \cite{khoury}
for a comprehensive review of the many theories of cosmic acceleration.}. Besides dark energy, which accounts for about 70 percent of the Universe's overall energy, observations have long
supported the notion that about a quarter of the energy content
of the Universe is in the form of  non-luminous, non-relativistic, and weakly interacting matter that is commonly referred to as (cold) dark matter. As for dark energy, the evidence for dark matter has various origins, with the oldest pieces of evidence coming from the observations of galaxy rotation curves \cite{grc-1,grc-2}
and the more recent though somewhat indirect ones coming from
observational cosmology, e.g.\ the CMB anisotropies
\cite{cmb-planck}. Non-dynamical dark energy, or  cosmological constant, along with cold dark matter form the so-called $\Lambda$CDM model of cosmology, which
is by far the most successful model describing the history of our
Universe from well before the epoch of matter-radiation equality.
In conjunction with the standard predictions of
inflation, the $\Lambda$CDM model provides nearly perfect theoretical
fits to the CMB anisotropy data \cite{planck-inflation}.

Despite its overwhelming success, the $\Lambda$CDM model faces some theoretical and observational challenges. A long standing theoretical challenge
to non-dynamical dark energy is the well known cosmological constant problem \cite{weinberg-cc}. With observations indicating that the
cosmological constant has a present value $\Lambda_0 \sim (10^{-3} {\rm eV})^4$ \cite{cmb-planck}, a large amount of fine tuning is necessary to cancel out loop corrections of about $(10^2 {\rm GeV})^4$ that it receives
from the Standard Model fields alone over the energy ranges at which the
Standard Model is known to be valid. On the observational end, a prominent challenge
to the $\Lambda$CDM model is the discrepancy in the deduced values of the Hubble constant $H_0$ from observations of Cepheid variables \cite{cepheid} and the CMB data collected and analysed by the Planck collaborations \cite{cmb-planck}.

Dynamical dark energy models, of which there are many \cite{khoury}, are a
useful theoretical foil for analysing dark energy observations.  They can also potentially
address the Hubble discrepancy problem \cite{pettorino}.  Generically
they do not address the fine tuning problems of the cosmological
constant \cite{weinberg-cc}, and in fact generically have additional
fine tuning problems associated with matter couplings \cite{PhysRevLett.81.3067}.


If dark energy is indeed dynamical and driven by a single
degree of freedom, then its dynamics can be described by an effective field theory (EFT) where the dark energy perturbations are the pseudo Nambu-Goldstone bosons that spontaneously break the  approximate de Sitter symmetry of the background spacetime \cite{Creminelli2009,Gubitosi:2012hu,eft-de-2}\footnote{This EFT formulation
is analogous to the effective field theory of inflation developed in
Refs.\ \cite{eft-inflation-1,eft-inflation-2}.}. A question then arises as to how the dark energy field couples to other forms of matter, including dark matter.
If the matter sectors are incorporated in the EFT of dark energy,
all generic interaction terms that are compatible with the assumed
internal symmetries of the matter sectors must be included.
Such interaction terms amount to the dark energy field mediating a new kind of force, the so-called ``fifth'' force, among particles of various kind, in violation of the weak  equivalence principle (WEP) or the universality of free fall for test particles in an external gravitational field
\footnote{The weak equivalence principle is expected to be violated in generic theories where dynamical dark energy interacts with dark matter. See \cite{cho} and \cite{damour} for some of the first studies of WEP violation for dark matter.}.

More specifically, one can classify violations of the WEP into three types:

\begin{enumerate}
\item {\it Baryonic - Baryonic (BB):} This is the type of WEP violation
  most commonly considered in the past.  An additional scalar field
  (here the dynamical dark energy field) couples in
  different ways to different sectors of the standard model.  Then the
  total gravitational force between two bodies (tensor plus fifth
  force) can depend on for example baryon number as well as total mass.
  This scenario is now tightly constrained thanks to a variety of
  satellite and Earth-based experiments \footnote{The so-called
  E\(\ddot{o}\)tv\(\ddot{o}\)s parameter, which is a dimensionless
  measure of the amount of the WEP violation for baryons, is currently
  constrained down to $\big[-1 \pm 9 (\text{syst}) \pm 9
  (\text{stat})\big]\times 10^{-15}$ by the recent MICROSCOPE
  experiment \cite{eotvos1}.}.

\item {\it Baryonic - Dark Matter (BD):} This occurs when dark matter
  particles and baryonic matter experience different accelerations.
  As an example\footnote{More general WEP violating interactions are
    discussed in the following sections.}, suppose that the action for the baryonic fields
  $\Psi_b$ and dark matter fields $\Psi_d$ can be written as
\be
S_b[e^{2 \alpha_b(\phi)} g_{ab}, \Psi_b] +
S_d[e^{2 \alpha_d(\phi)} g_{ab}, \Psi_d]
\label{bbdd}
\ee
for two different coupling functions $\alpha_b$ and $\alpha_d$ to the
spacetime metric $g_{ab}$.
The effective Newton's constant then becomes a matrix with elements
\cite{PhysRevD.78.023009}
\be
G_{ij} \propto 1 + 2 m_p^2 \alpha_i'(\phi)  \alpha_j'(\phi) ,
\ee
where $i$ and $j$ run over the dark and baryonic sectors $d$ and $b$.  Since the dark matter is observed
only through its gravitational effects, the individual constants
$G_{dd}$ and $G_{db}$ are not directly observable.  A rescaling of the
form $\rho_d \to e^\nu \rho_d$, $G_{dd} \to e^{-2 \nu} G_{dd}$,
$G_{db} \to e^{-\nu} G_{db}$, where $\rho_d$ is the dark matter
density, has no effect on the dynamics of visible
matter and is thus not observable.  Hence only the
combination $G_{db} / \sqrt{G_{dd} G_{bb}}$ is observable, and its
deviations from unity are a measure of the WEP violation \cite{kamion}.

\item {\it Dark Matter - Dark Matter (DD):} Finally, if the couplings between the
dark energy and dark matter fields have generic forms, then not
only is there a violation of the WEP between dark matter and baryons, but also within the dark matter
sector alone.  This means that the dark matter couples not just to a
combination of the metric and dark energy field, as in Eq.\ (\ref{bbdd})
above, but to the metric and
dark energy fields individually.  In this case, if there is more than
one species of dark matter particle, the acceleration experienced by a
freely falling particle can vary from one species to another.

\end{enumerate}

In the absence of direct detection of dark matter, the status of  WEP
violations of the DD type is currently unknown.  Violations of the BD
type can be constrained by a variety of astrophysical and cosmological observations.
One constraint comes from observations of tidal streams of tidally
disrupted satellite galaxies \cite{kamion}.
An analysis of dark matter clustering in
the Abell Cluster A586 also hinted at the possibility of a dark
energy-enhanced gravitational interaction for dark matter particles
over baryons \cite{abell1,abell2}.
On the other end, in observational cosmology, it has been suggested that weak lensing
and redshift space distortions can be used to search for the WEP violation between dark matter and baryons if the dark matter density perturbations obey the $\Lambda$CDM continuity equation \cite{maartens1}.  Moreover, existing and upcoming CMB and LSS observations provide a large amount of information that can be used to constrain dark energy interactions with dark matter, which in turn can be used to constrain WEP violations between the dark and baryonic matter sectors 
\cite{maccio,he,keselman,barrow,baldi,delliou,morris,yangbai,santos, miranda, Yang:2018euj}.

\subsection{Outline of the paper}\label{results}

In this paper, we generalize the EFT of dark energy by incorporating a dark matter sector.
We model dark matter as
a complex scalar field with a global $U(1)$ symmetry \footnote{The number of the EFT operators are reduced if fermions are used in place of  scalar fields.}, which has some similarities to axion dark matter models \footnote{For a recent review of axion cosmology and axion models of dark matter see \cite{marsh-axion-cosmology}.}.
Coherent excitations of this field can act like a pressureless fluid, at the level of the background
cosmology and of linearized perturbations, as we discuss in more detail below.
Since the WEP is not a symmetry, it is generically violated by EFT operators in the dark matter sector
in this and similar models \cite{ArmendarizPicon:2011ys}.

The structure of this paper is as follows. In Sec.\ \ref{sec.intro-model} we introduce the basics of the dark matter model under study. After
a brief review of the systematics of the WEP in a field theoretic Lagrangian, we show in Secs.\ \ref{sec.back.1} and \ref{sec.linear.pert} that with a suitable choice of parameters,
the basic dark matter model gives rise to the same background and linear perturbations phenomenology as cold dark matter.
We then move on to formulating the EFT of dark matter interactions
with dark energy in Sec.\ \ref{sec.eft}. After identifying the EFT
operators and expressing the action in Secs.\ \ref{sec.operators} and
\ref{sec.action}, we discuss the regime of validity of the EFT
in Sec.\ \ref{sec.scale}. We work out the EFT dynamical equations
for the background cosmology and linear perturbations in Sec.\ \ref{sec.eft.eqm}. Finally, as an example of how the WEP violation modifies the cosmological and astrophysical observables, we derive the effective Newton's constant for dark matter in Sec.\ \ref{sec.observables}. 

\subsection{Summary of the main results}
The following summarizes our main findings in the present analysis:
\begin{itemize}
\item In the space of solutions of the background equations, there are
  fairly generic solutions for which the background stress energy
  tensor of the dark matter candidate converges exactly to that of a
  pressureless fluid in the limit of complete cosmological constant
  domination, without the need to fine tune its mass. This is in
  contrast to axion models, where the equation of state parameter
  for dark matter becomes nearly zero after averaging over a period
  large compared to the period of oscillations
  \cite{marsh-axion-cosmology}.
\item As for axions, the sound speed of dark matter linear
  perturbation modes\footnote{i.e. modes with physical momenta
    $k_{\rm phys} \gtrsim 0.1 \ {\rm Mpc}^{-1}$. } is close to zero when
  averaged over a period large compared to the period of
  oscillations. For this to hold today, the dark matter mass should be
  at least a few orders of magnitude larger than the Hubble
  constant. If this is to be true
at the time when the shortest linear mode became subhorizon,
then the dark matter mass is required to be at least ten orders of magnitude larger than the Hubble constant. This is consistent with
the lower bound on the mass of ultralight axion dark matter that was found in \cite{marsh-cosmo}.
\item We have not analysed the dark matter perturbations in non-linear regime to see if they continue to behave as perturbations of a pressureless fluid.  This is an open question.
\item  There are thirteen dark energy-dark matter interaction operators in the EFT with dimension $\leq 4$ that are relevant
for analysing the background and linear perturbations. The results
of analyses performed in  \cite{yangbai} and \cite{santos} are used to constrain the coefficients of some of these operators. However, most of these constraints are weak. 
\item Some fine tuning in the EFT coefficients is necessary if the model is to successfully reproduce the $\Lambda$CDM background cosmology. The fine tuning becomes more severe as the dark matter mass becomes lighter.
\item As far as observations are concerned, we can only talk about the WEP violations of the BD type. An astrophysical signature of this effect is a modification of the Newton's constant for dark
  matter as compared to baryons.  However, we show that a similar modification can be obtained from a non-minimal coupling of dark matter to
  gravity. In order to differentiate the two effects, one can
potentially exploit the fact that the latter correction to the Newton's constant is scale dependent. 
\end{itemize}

\section{A first step: a simplified model of gravity-dark matter-dark energy interactions} \label{sec.intro-model}

Before introducing our more comprehensive effective action for the
interactions of dark matter and dark energy in Sec.\ \ref{sec.eft},
we find it instructive to study the preliminary aspects
of such interactions using a simpler model.
We model the dark energy field
using a single real scalar field $\phi$ and the dark matter field using
a complex scalar field $\Pi$. We require
the dark matter sector to respect a global $U(1)$ symmetry.

The action is given by \footnote{Our convention is to set $c=\hbar=1$ and define the reduced Planck mass $m_p^2 = (8\pi G)^{-1}$. The metric has signature $(-+++)$.}
\bea\label{act.back.1}
S[g_{ab},\phi,\Pi^{\dagger},\Pi,\psi] = S_{\text{gravity}}[g_{ab}]+S_{\text{DE}}[g_{ab}, \phi]+S_{\text{DM-DE}}[g_{ab}, \phi, \Pi^{\dagger}
,\Pi] + S_b[g_{ab},\psi],
\eea
where the gravitational action, dark energy and dark matter actions are
\begin{subequations}
\label{act.back.2}
\bea
&& S_{\text{gravity}}[g_{ab}]\equiv S_{\text{EH}}[g_{ab}] = \int d^4x \sqrt{-g} \ \frac{m_p^2}{2} R,\\
&& S_{\text{DE}}[g_{ab}, \phi] = \int d^4x \sqrt{-g} \ \bigg[
X - V_1(\phi) \bigg], \\
&& S_{\text{DM-DE}}[g_{ab}, \phi, \Pi^{\dagger}
,\Pi] = \int d^4x \sqrt{-g} \ e^{-2 \alpha}
\bigg[e^{\alpha } Y
- V_2(\phi,X, \Pi^{\dagger}\Pi) \bigg].
\label{act.dark1}
\eea
\end{subequations}
Also $S_b$ is the baryonic action of the Standard Model fields collectively denoted by $\psi$.
Here we have defined
\begin{subequations}
\bea
&& X \equiv -\frac{1}{2} g^{ab} \nabla_a \phi \nabla_{b} \phi,\\
&& Y \equiv  -g^{ab}\nabla_a \Pi^{\dagger} \nabla_{b} \Pi.
\eea
\end{subequations}
In Eq.\ \eqref{act.back.2} ``$\text{EH}$'' stands for Einstein-Hilbert, $R$ is the Ricci scalar, and $\alpha$ is a smooth but otherwise arbitrary function of $\phi$.

If the dark matter potential $V_2$ does not depend on $\phi$ and $X$,
\be
V_2(\phi,X, \Pi^{\dagger}\Pi)  = V_2(\Pi^{\dagger}\Pi),
\label{simplify}
\ee
then
the conformal transformation
\be   \label{trans}
\hat{g}_{ab} = e^{-\alpha}g_{ab}
\ee
removes all direct couplings between the dark matter and the dark energy fields.
The action is then cast in the form
\be  \label{act.back.3}
\hat{S}[\hat{g}_{ab},\phi,\Pi^{\dagger},\Pi,\psi]=\hat{S}_{\text{gravity}}[\hat{g}_{ab}, \phi]+\hat{S}_{\text{DE}}[\hat{g}_{ab}, \phi]
+\hat{S}_{\text{DM}}[\hat{g}_{ab}, \Pi^{\dagger},\Pi] + S_b[e^\alpha
{\hat g}_{ab},\psi],
\ee
where
\begin{subequations}
\label{act.back.4}
\bea
&& \hat{S}_{\text{gravity}}[\hat{g}_{ab}, \phi]= \int d^4x \sqrt{-\hat{g}}
\  \frac{e^{ \alpha}m_p ^2}{2}\bigg[\hat{R}-3 \hat{g}^{ab} \hat{\nabla}_{a} \hat{\nabla}_b \alpha -\frac{3}{2}\hat{g}^{ab} \hat{\nabla}_a \alpha \hat{\nabla}_{b} \alpha  \bigg], \\
&& \hat{S}_{\text{DE}}[\hat{g}_{ab}, \phi]= \int d^4x \sqrt{-\hat{g}}
\  e^{2 \alpha}\bigg[e^{-\alpha} \hat{X} - V_1(\phi) \bigg], \\
&&\hat{S}_{\text{DM}}[\hat{g}_{ab}, \Pi^{\dagger},\Pi] =
\int d^4x \sqrt{-\hat{g}} \big[ \hat{Y} - V_2(\Pi^{\dagger}\Pi)
\big],
\label{act.dark2}
\eea
\end{subequations}
and $\hat{X}$ and $\hat{Y}$ are the same as $X$ and $Y$ with $g^{ab}$
replaced with $\hat{g}^{ab}$.  We will refer to the hatted frame as the
Jordan frame and the unhatted frame as the Einstein frame, in a slight
generalization of the common terminology.
Note that dark matter has no violations of the WEP of the DD type in the model
(\ref{act.back.3}) satisfying the condition (\ref{simplify}), since from the form of the action
dark matter particles
freely fall along the geodesics of the metric $\hat{g}_{ab}$.  However,
it will have violations of the BD type whenever $\alpha'(\phi) \ne 0$.
Allowing the potential $V_2$ to depend on $\phi$ and $X$ will generically result in
additional violations of the DD type.
The dark energy field $\phi$ then mediates
an extra force on dark matter particles.
In Sec.\ \ref{sec.action} we will express the EFT action in the hatted (generalized Jordan) frame associated
with $\hat{g}_{ab}$.
However, we will conduct most of the following analysis in the unhatted (Einstein) frame.

In what follows we shall neglect the baryonic contributions to the
equations of motion.
The equations of motion derived from the action \eqref{act.back.1},
neglecting baryons, are
the coupled Klein-Gordon equations for the dark energy and the dark matter
fields,
\begin{subequations}
\label{eqm.1}
\bea
&&-\nabla_{a}\nabla^{a} \phi + \nabla_{a}
\Bigg[e^{-2 \alpha} \nabla^{a} \phi V_{2,X}\Bigg]+ \alpha_{,\phi}e^{-\alpha}\Big[Y-2 e^{-\alpha} V_{2}\Big]
+V_{1,\phi}+ e^{-2 \alpha} V_{2, \phi}=0, \\
&&-\nabla^{a} \Big[e^{-\alpha} \nabla_{a} \Pi^{\dagger} \Big]+e^{-2
  \alpha}V_{2, \Pi}=0, \hspace{1cm}-\nabla^{a} \Big[e^{-\alpha}
\nabla_{a} \Pi \Big]+e^{-2 \alpha}V_{2, \Pi^{\dagger}}=0,
\eea
\end{subequations}
and the Einstein's equations
\bea \label{eqm.2}
&& m_p ^2 G_{ab} \equiv m_p ^2 \Big[R_{ab}-\frac{1}{2}R g_{ab} \Big] = T_{ab}^{\text{DE}}
+T_{ab}^{\text{DM-DE}}.
\eea
Here the dark energy and dark matter stress-energy tensors are
\begin{subequations}
\bea\label{stress1}
&&T_{ab}^{\text{DE}}\equiv \frac{-2}{\sqrt{-g}}\frac{\delta S_{\text{DE}}}{\delta g^{ab}} =X_{ab}+ g_{ab}\Big[X
-V_1 \Big],  \\
&&T_{ab}^{\text{DM-DE}} \equiv \frac{-2}{\sqrt{-g}}\frac{\delta S_{\text{DM-DE}}}{\delta g^{ab}} = e^{-\alpha}\Big[ 2 Y_{ab}-e^{-\alpha} V_{2,X} X_{ab}
+g_{ab} \big( Y- e^{-\alpha}V_2\big) \Big].
\eea
\end{subequations}
Also $R_{ab}$ is the Ricci tensor, $X_{ab}\equiv \nabla_{a}\phi \nabla_{b} \phi$,
$Y_{ab}\equiv \nabla_{a}\Pi^{\dagger} \nabla_{b}\Pi$, and we use the notation $A_{,B} \equiv \partial_{B} A$. Note that all indices are raised and lowered using the metric $g_{ab}$.

Note that  by virtue of the Bianchi identity $\nabla^{a} G_{ab}=0$, we have
\be
\nabla^{a} (T_{ab}^{\text{DE}}
+T_{ab}^{\text{DM-DE}}) = 0,
\ee
despite each individual stress energy tensor not being covariantly conserved. However, if $V_2$ did not depend on $\phi$ and $X$,
we could analyse the model in the Jordan frame using the action given in Eqs.\ \eqref{act.back.3} and \eqref{act.back.4}. Then it is not difficult to see that the dark matter stress energy tensor defined with respect to $\hat{g}_{ab}$,
\be
\hat{T}^{\text{DM}} _{ab} \equiv \frac{-2}{\sqrt{-\hat{g}}}\frac{\delta \hat{S}_{\text{DM}}}{\delta \hat{g}^{ab}},
\ee
is covariantly conserved, i.e. $\hat{\nabla}^{a} \hat{T}^{\text{DM}} _{ab}=0$, by virtue of the dark matter equations of motion
derived from the action \eqref{act.back.3}.

\subsection{A study of the minimally coupled dark matter model}\label{sec.min.model}

In the following subsections, we show that the dark matter model
\eqref{act.dark1} admits a fluid
description which comports with two of the main properties of
 cold dark matter, namely that it has negligible equation of state
 parameter and sound speed at the level of the background and the linear perturbations respectively.
To simplify this analysis, we assume that the dark matter field
is minimally coupled to gravity, so that $\alpha=0$ in the action (\ref{act.dark1}).
We also assume that dark energy is non dynamical
and described by a cosmological constant with a value of $\Lambda_0 \sim 10^{-120} m_p ^4$.
We take the dark matter potential to be
\bea \label{potentials}
&& V_2 (\Pi^{\dagger} \Pi) =\pm m^2 \Pi^{\dagger} \Pi + \lambda(\Pi^{\dagger} \Pi)^2,
\eea
where $m$ and $\lambda$ are some constants.
There are several ways of expressing $\Pi$ in terms of two real
scalar fields. The parametrization
\be \label{def-dm}
\Pi \equiv \mathcal{R} e^{i \zeta},
\ee
in terms of the radial and angular variables $\mathcal{R}$ and $\zeta$
makes the presence of the global $U(1)$ symmetry in the dark matter sector manifest, and simplifies the analysis of the background dark matter equations of motion. The dark matter potential $V_2$ in this case becomes
\be
V_2 (\mathcal{R}^2) = \pm m^2 \mathcal{R}^2 + \lambda \mathcal{R}^4.
\ee
Another parametrization is
\be \label{def-dm1}
\Pi \equiv \frac{1}{\sqrt{2}} \big(\varphi_1 + i \varphi_2 \big),
\ee
which is useful for deriving an effective
fluid description beyond the background level (particularly when $\lambda = 0$). We will return to this latter point in Sec.\ \ref{sec.linear.pert}.

\subsubsection{Background dynamics} \label{sec.back.1}
We assume that the background geometry is described by the spatially flat
Friedmann-Robertson-Walker metric
\be \label{bac.metric}
ds^2 = a(\eta)^2 [ -d \eta^2 + dr^2 + r^2 d \Omega^2 ],
\ee
where $a$ is the scale factor, $\eta$ is the conformal
time, and $d \Omega^2$ is the metric on the unit 2-sphere. At the
background level, all fields are
functions of $\eta$ alone. We denote the background values of all
quantities using subscript $0$.

The background Einstein equations are the two Friedmann equations that
are derived from Eqs.\ \eqref{eqm.2} and \eqref{stress1} (set $\alpha
=0$),
\begin{subequations}
\label{friedmann}
\bea
&& 3 m_p ^2 \frac{\mathcal{H}^2}{a^2} = \frac{1}{a^2} \Big(T_{\eta \eta,0}^{\text{DE}} + T_{\eta \eta,0}^{\text{DM}} \Big) =  \rho_{0}^{\text{DE}} + \rho_{0}^{\text{DM}},\\
&& m_p ^2 \bigg(\frac{\mathcal{H}^2}{a^2}- \frac{2 a''}{a^3}\bigg) =
\frac{1}{a^2} \Big(T_{i i,0}^{\text{DE}} + T_{i i,0}^{\text{DM}} \Big)= p_{0}^{\text{DE}} + p_{0}^{\text{DM}},
\eea
\end{subequations}
where $i$ denotes a spatial index, $\mathcal{H} \equiv a'/a$, and
\begin{subequations}
\label{back.press-e}
\bea
&& \rho_0^{\text{DE}} =  \Lambda_0, \hspace{2cm}  p_0^{\text{DE}} = - \Lambda_0, \\
&&\rho_0^{\text{DM}} = \frac{1}{a^2}\bigg[\mathcal{R}_0 ^{\prime 2}+\mathcal{R}_0 ^2 \zeta_{0} ^{\prime 2} \bigg] \pm m^2\mathcal{R}_0 ^2 + \lambda \mathcal{R}_0 ^4,\\
&&p_0^{\text{DM}} = \frac{1}{a^2}\bigg[\mathcal{R}_0 ^{\prime 2}+\mathcal{R}_0 ^2 \zeta_{0} ^{\prime 2} \bigg] \mp m^2 \mathcal{R}_0 ^2 - \lambda \mathcal{R}_0 ^4.
\eea
\end{subequations}
In addition, the following background Klein-Gordon equations are derived from Eq.\ \eqref{eqm.1} for the dark matter fields
\begin{subequations}
\label{klein-min}
\bea
&&\frac{\mathcal{R}_0''}{\mathcal{R}_0}+2 \mathcal{H}\frac{\mathcal{R}_0'}{\mathcal{R}_0}-\zeta_{0}^{\prime 2} \pm a^2 m^2 + 2 a^2 \lambda \mathcal{R}_0 ^2=0,\\
&&2 \mathcal{R}_0 ' \zeta_0 '+ \mathcal{R}_0 \zeta_0 ''
+2 \mathcal{H} \mathcal{R}_0 \zeta_0 '=0.
\label{klein-mit1}
\eea
\end{subequations}
It is also helpful to define the following quantities
\bea \label{w-omega-min}
&&w^{\text{DM}}  \equiv \frac{p_0^{\text{DM}}}{\rho_0^{\text{DM}}},
\hspace{2cm} \Omega^{\text{DM}} \equiv \frac{a^2 \rho_0^{\text{DM}}}{3 m_p ^2 \mathcal{H}^2},
\eea
where the quantity defined on the left is the equation of state
parameter and the one on the right is the density parameter for the background dark matter.

Equation \eqref{klein-mit1} can be integrated to give
\be \label{zeta}
a^2 \mathcal{R}_0 ^2 \zeta_0 ' = c,
\ee
for some constant $c$. The integrability of this equation is
the result of the $U(1)$ symmetry of the dark matter action \footnote{
  Cosmological dark matter models with internal $U(1)$ symmetries have
  previously been previously explored in Ref.\ \cite{spintessence}.}.
The conserved quantity given in Eq.\ \eqref{zeta} can be
thought of as angular momentum in the dark matter field space.
Thus, the background dark matter equations of motion reduce
to a single non-linear ODE given by
\be \label{eqm-dm-min}
\frac{\mathcal{R}_0 ''}{\mathcal{R}_0}+2 \mathcal{H}\frac{\mathcal{R}_0 '}{\mathcal{R}_0}
-\frac{c^2}{a^4 \mathcal{R}_0 ^4} \pm a^2 m^2 + 2 a^2 \lambda \mathcal{R}_0 ^2=0.
\ee

\begin{figure}[h]
\centering
\includegraphics[width=0.7\textwidth]{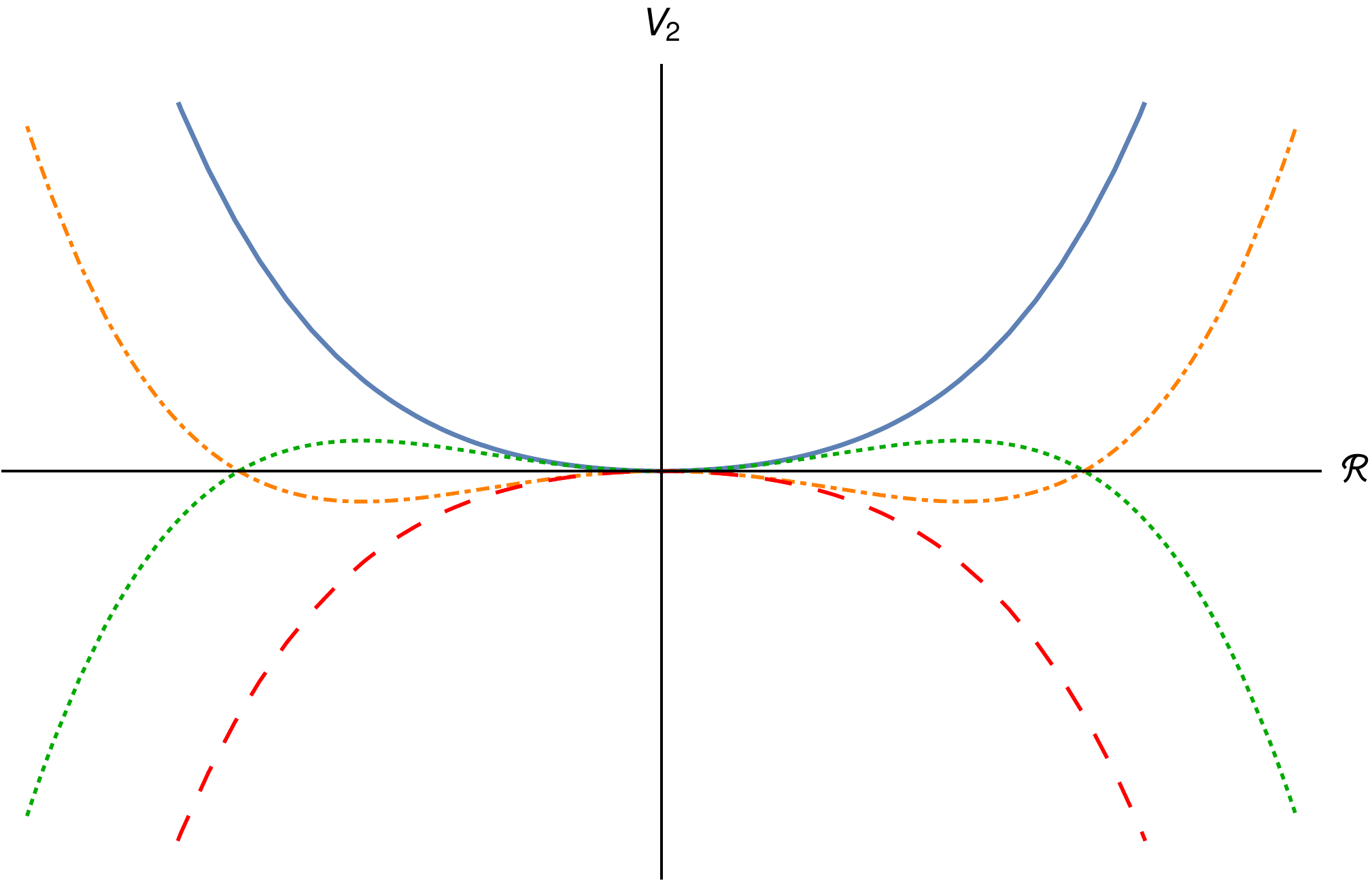}
\caption{Different shapes of the dark matter potential $V_2$ are
drawn here. we have suppressed the angular dependence of the potential.
The solid blue potential corresponds to $m^2 \mathcal{R}^2 +|\lambda| \mathcal{R}^4$ , the dashed red potential to $-m^2 \mathcal{R}^2 -|\lambda| \mathcal{R}^4$, the orange dashed-dotted one to
$-m^2 \mathcal{R}^2 +|\lambda| \mathcal{R}^4$, and the
dotted green one to $m^2\mathcal{R}^2 -|\lambda| \mathcal{R}^4$. }
\label{DMpotentials}
\end{figure}

Prior to conducting a detailed analysis of the background equations \eqref{friedmann} and \eqref{klein-min}, we find it illuminating to
discuss some overall aspects of  the solutions to these equations.
This is largely determined by the form of the dark matter potential
given in Eq.\ \eqref{potentials}. See Fig.\ \ref{DMpotentials}
for the four different possibilities for the dark matter potential.

Since we expect the dark matter energy density to continue to
get diluted into the future, we look for background solutions where
the dark matter field $\mathcal{R}_0$
eventually settles at a minimum of its potential.
Thus, we discard the potential $-m^2 \mathcal{R}^2
-|\lambda| \mathcal{R}^4$ as it has no minima. Also note that
one needs to add the constant term $m^4 / (4 |\lambda|)$ to the
Higgs-like potential $-m^2 \mathcal{R}^2 +|\lambda| \mathcal{R}^4$
in order to set the minimum of the potential to zero, Otherwise
the dark matter's energy density will eventually be dominated by
its non-zero potential, which in turn renders its equation of state
parameter $w^{\text{DM}} \rightarrow -1$.
Adding this constant is compatible with the internal $U(1)$
symmetry of the dark matter action, though doing so is a fine-tuning of the potential.

We first analyse the future asymptotics of Eq.\ \eqref{eqm-dm-min}
for the potentials $m^2 \mathcal{R}^2 \pm |\lambda|\mathcal{R}^4$.
Assuming that the overall background energy density is dominated by $\Lambda_0$
in this limit, we approximate $a$ and $\mathcal{H}$ using
\bea \label{back-scale}
&&a(\eta) = \frac{-\sqrt{3}m_p}{\sqrt{\Lambda_0}\eta} + \mathcal{O}(\eta),
\hspace{2cm} \mathcal{H} = \frac{-1}{\eta}+\mathcal{O}(\eta),
\eea
where $- \infty < \eta < 0 $ and $\eta \rightarrow 0^{-}$ is the future limit.
Moreover, as the dark matter fields roll towards the minimum of their
potential at $\mathcal{R}_0=0$, we ultimately have $\mathcal{R}_0 ^2 \ll m^2/|\lambda|$ (unless $m=0$). Using this approximation and Eq.\eqref{back-scale}, the background dark matter equation \eqref{eqm-dm-min} reduces to
\be \label{eqm.dm.min.1}
\frac{\mathcal{R}_0''}{\mathcal{R}_0}-\frac{2}{\eta} \frac{\mathcal{R}_0'}{\mathcal{R}_0}
-\frac{\Lambda_0 ^2 c^2 \eta^4}{9 m_p ^4 \mathcal{R}_0 ^4} + \frac{3 m_p ^2}{\Lambda_0 \eta^2} m^2 =0.
\ee
If $c \neq 0$ we can find a solution
$\mathcal{R}_0 = \alpha \eta^{3/2}$ where $\alpha$ satisfies
\be \label{rel}
-\frac{9}{4}- \frac{\Lambda_0 ^2 c^2}{9 m_p ^4 \alpha ^4}+\frac{3 m_p ^2 m^2}{\Lambda_0}=0.
\ee
It is easy to check that for this solution, we have
\be
p_0^{\text{DM}} =  \frac{1}{a^2}\bigg[\mathcal{R}_0 ^{\prime 2}+\frac{c^2}{a^4 \mathcal{R}_0 ^2} \bigg] - m^2 \mathcal{R}_0 ^2 = 0.
\ee
This set of solutions therefore correspond to pressureless dark matter
solutions in the asymptotic future.  Although the choice of the value
of $\alpha$ satisfying Eq.\ (\ref{rel}) is a fine tuning, one can
check that linearized perturbations around this solution are stable.
\footnote{On the other hand if $c=0$, the general solution to Eq. \eqref{eqm.dm.min.1} is
\be
\mathcal{R}_0 = \alpha_1 \eta^{\frac{3+\nu}{2}} + \alpha_2 \eta^{\frac{3-\nu}{2}},
\ee
for constants $\alpha_i$, where $\nu = \sqrt{9-12 \frac{m_p ^2 m^2}{\Lambda_0}}$
Using this solution, we compute the background
equation of state parameter for dark matter in this limit to be
\bea \label{eqstate.dm}
&&w^{\text{DM}} \approx \frac{\frac{\mathcal{R}_0 ^{\prime 2}}{a^2}- m^2 \mathcal{R}_0 ^2}{\frac{\mathcal{R}_0 ^{\prime 2}}{a^2}+ m^2 \mathcal{R}_0 ^2} =
 \frac{\eta^{2\nu} \alpha_1 ^2
\bigg[3 \Lambda_0 - 4 m^2 m_p ^2 + \Lambda_0 \nu \bigg]
+ \alpha_2 ^2 \bigg[3 \Lambda_0 - 4 m^2 m_p ^2 - \Lambda_0 \nu \bigg]}{\eta^{2\nu} \alpha_1 ^2
\bigg[3 \Lambda_0 + \Lambda_0 \nu \bigg]
+ 8 m^2 m_p ^2 \alpha_1 \alpha_2 \eta^{\nu}
+\alpha_2 ^2
\bigg[3 \Lambda_0 - \Lambda_0 \nu \bigg]}.
\eea
It can be seen from above that as $\eta\rightarrow 0^{-}$, $w^{\text{DM}} \rightarrow 0  $ if and only if
$m^2 \rightarrow 3 \Lambda_0 / (4 m_p ^2)$. In other words, one
is required to fine-tune the value of the dark matter mass to be on the
order of the Hubble constant today in order to find asymptotically pressureless
solutions in this case.}

Turning to the Higgs-like potential $V_2 = -m^2 \mathcal{R}^2
+|\lambda|\mathcal{R}^4 + m^4/4 |\lambda| $,
Eq. \eqref{eqm-dm-min} in the vicinity of the minima of the potential at $\mathcal{R} = m/\sqrt{2 |\lambda|}$ is approximately
\be
\delta_0 '' -\frac{2}{\eta}\delta_0 ' - \frac{2^{3/2} |\lambda|^{3/2}\Lambda_0 ^2 c^2 \eta^4}{9 m^3 m_p ^4}
+ \frac{6 m^2 m_p ^2 }{\Lambda_0 \eta^2} \delta_0 = 0,
\ee
where we used Eq. \eqref{back-scale} and defined $\delta_0 \equiv \mathcal{R}_0 - m/\sqrt{2 |\lambda|}$. If $c=0$, one arrives at
a conclusion similar to the previous case; namely that
by fine-tuning the value of $m$, one can find asymptotically
pressureless background solutions for the dark matter fields. Alternatively,
if $c \neq 0$, one finds that $\delta_0 \propto \eta^6$. However,
it is not difficult to check that $w^{\text{DM}}\rightarrow 1$ in this case. We therefore disregard the Higgs-like potential
for the following discussions in this section.

\begin{figure}
\centering
\includegraphics[width=1\textwidth]{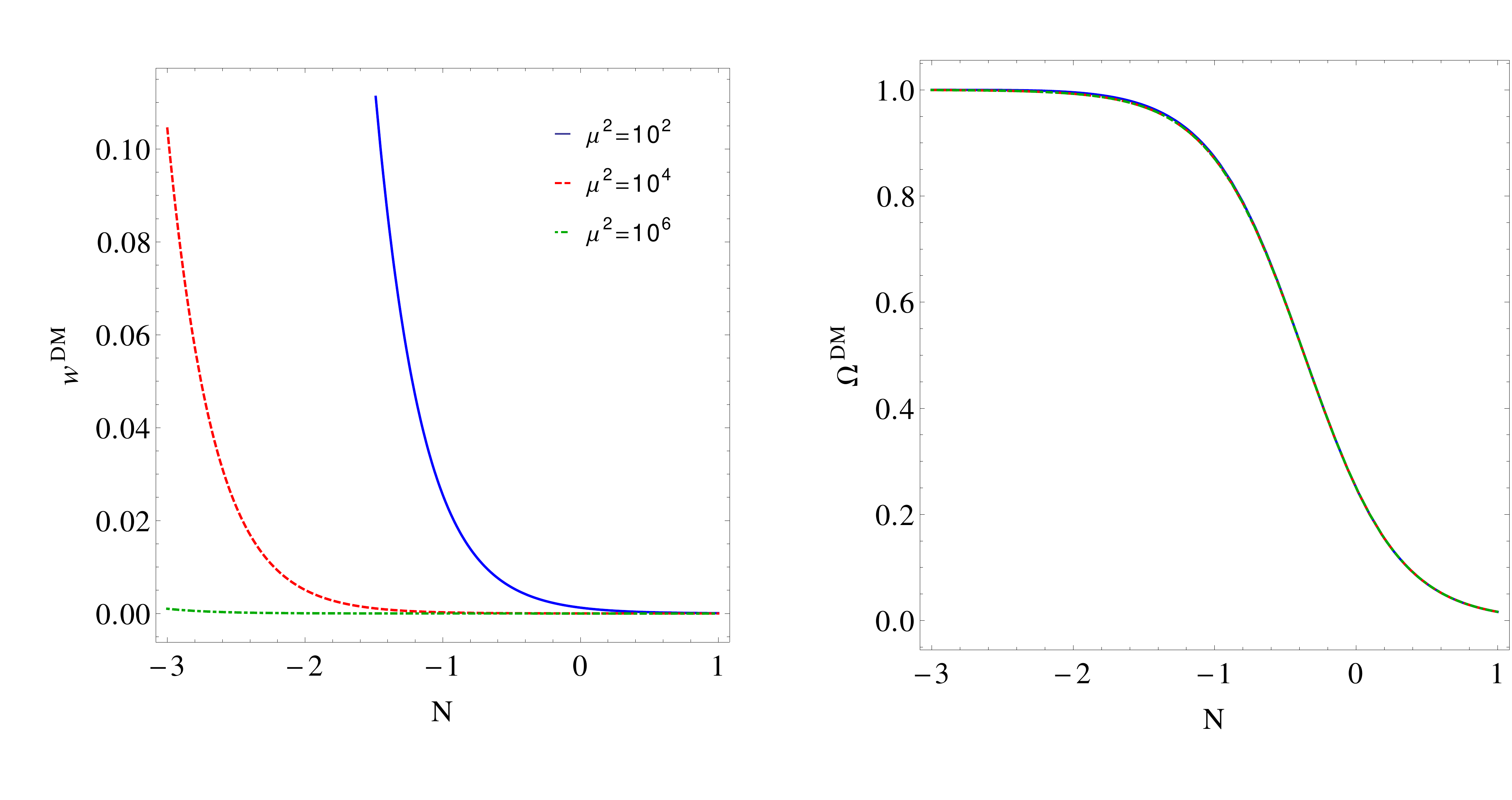}
\caption{We plot the equation of state parameter and the density parameter for the background dark matter when $\lambda=0$ and dark energy
is approximated using the cosmological constant at all times.
N = 0 corresponds to today and N $<0$ corresponds to the past.
In particular, N $ = -3$ roughly corresponds to
the redshift $z \approx 19$. We also defined
$\mu^2 \equiv m^2 m_p ^2 / \Lambda_0$. As can be seen from the plots, the larger the
value of $\mu^2$, the smaller the equation of state parameter will
be as we go to the past.
 Also note that the density parameters almost completely coincide in all
cases.}
\label{back-dyn-min-1}
\end{figure}

\begin{figure}
\centering
\includegraphics[width=1\textwidth]{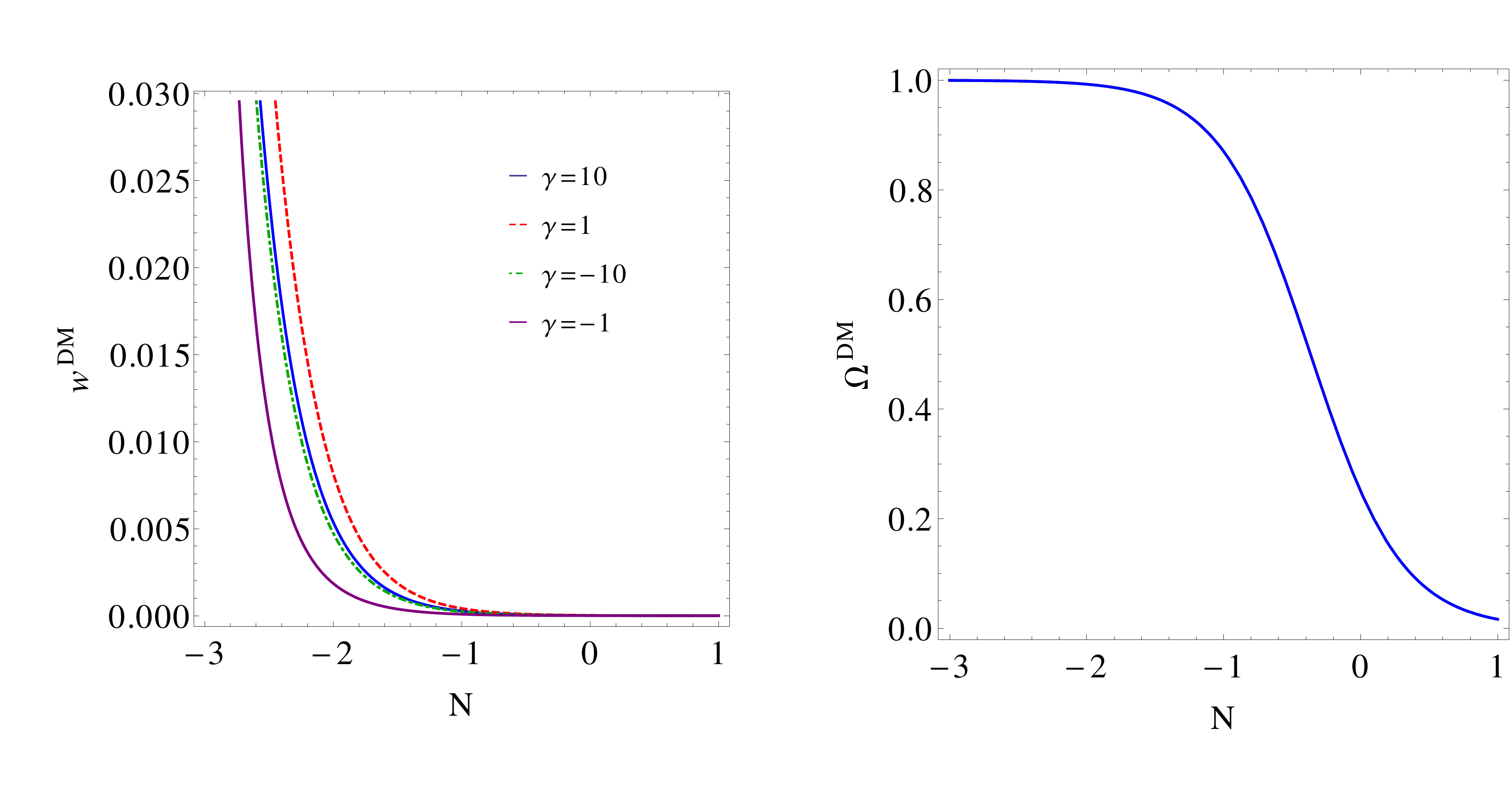}
\caption{Here we plot the equation of state parameter and the density parameter
of the background dark matter for different values of
$\gamma \equiv m^2/(m_p ^2 \lambda)$ when $\lambda \neq 0$
and the dark energy is approximated using the cosmological constant at all times.
While it is evident from the above plot that
the negative values of $\gamma$ result in
a smaller equation of state parameter for the background dark matter,
this observation need not hold asymptotically to the past.}
\label{back-dyn-min-2}
\end{figure}

We now use $\mathcal{R}_0 = \alpha \eta^{3/2}$
as an approximation for $\mathcal{R}_0$ when $\mathcal{R}_0 ^2 \ll m^2 / |\lambda|$. We set $c^2/ \Lambda_0 m_p^2 = 1$ and require $\alpha$ to satisfy \eqref{rel}. We then numerically integrate the background equations given in \eqref{klein-min} away from $\mathcal{R}_0=0$.
We do this for the potentials $V_2 = m^2 \mathcal{R}^2 \pm |\lambda| \mathcal{R}^4 $ for two separate cases:
\begin{itemize}
\item First, we set $\lambda = 0$  while varying $m^2 m_p ^2/\Lambda_0$ (see Fig. \ref{back-dyn-min-1}).
\item Next, we assume $\lambda \neq 0$  and vary $m^2 /(m_p ^2 \lambda)$ (see Fig. \ref{back-dyn-min-2}).
\end{itemize}
Note that all numerical results are plotted as functions of the numberfold N$\equiv \log a/ \log a_0$, where $a_0$ is the value of the scale factor today.

\subsubsection{Dynamics of linear scalar perturbations} \label{sec.linear.pert}

We now turn to analysing the linear scalar
perturbation theory of the minimally coupled
dark matter. We take the dark matter potential to be $V_2 = m^2 \Pi^{\dagger}\Pi$ and show that the sound speed for the dark matter perturbation modes in the linear regime can be averaged to near zero values for sufficiently large values of the dark matter mass $m$.

For the purpose of analysing perturbations, we find it convenient to
express $\Pi$ in terms of $\varphi_1$ and $\varphi_2$ using
Eq. \eqref{def-dm1}. Then the dark matter stress energy tensor can be written as a sum of two separate stress energy tensors, each associated with one of the real scalar fields.
Using Eq. \eqref{stress1} (with $\alpha = 0$) we have
\be
T^{\rm DM}_{ab} [g_{ab},\Pi, \Pi^{\dagger}]= {}^{(1)}T^{\rm DM}_{ab} [g_{ab},\varphi_1]+{}^{(2)}T^{\rm DM}_{ab} [g_{ab},\varphi_2],
\ee
where
\begin{subequations}
\bea
&&{}^{(1)}T^{\rm DM}_{ab} [g_{ab},\varphi_1] = \nabla_a \varphi_1 \nabla_b \varphi_1 + g_{ab} \Big[-\frac{1}{2}\nabla_a \varphi_1 \nabla^a \varphi_1 - \frac{1}{2} m^2 \varphi_1 ^2\Big], \\
&&{}^{(2)}T^{\rm DM}_{ab} [g_{ab},\varphi_2] = \nabla_a \varphi_2
\nabla_b \varphi_2 + g_{ab} \Big[-\frac{1}{2}\nabla_a \varphi_2
\nabla^a \varphi_2 - \frac{1}{2} m^2 \varphi_2 ^2\Big].
\eea
\end{subequations}
The above stress energy tensors can be cast in the perfect
fluid form \footnote{although generally not an isentropic fluid.}. The
energy density, pressure, and four-velocity associated with
$\varphi_1$ are given by \footnote{There are ambiguities with defining
  a comoving frame for the stress energy tensor of a scalar field. For
  instance, if the background solutions $\varphi_{1,0}$  are
  oscillatory, then ${}^{(1)}u_{a,0}$ is discontinuous.  In addition
  it is required that $\nabla_a \varphi_1$ and $\nabla_a \varphi_2$ be
timelike.}
\begin{subequations}
\bea
&&{}^{(1)}\rho = -\frac{1}{2}\nabla_a \varphi_1 \nabla^a \varphi_1 + \frac{1}{2} m^2 \varphi_1 ^2, \\
&&{}^{(1)}p = -\frac{1}{2}\nabla_a \varphi_1 \nabla^a \varphi_1 - \frac{1}{2} m^2 \varphi_1 ^2,\\
&& {}^{(1)}u_a = \frac{\nabla_a \varphi_1}{\sqrt{-\nabla_a \varphi_1 \nabla^a \varphi_1}},
\eea
\end{subequations}
with similar expressions holding for $\varphi_2$. We can then compute the sound speed associated with each dark matter degree of freedom for any given mode $k$ by
\begin{subequations}
\label{soundspeed}
\bea
&&{}^{(1)}c_s ^2 \bigg|_{k}= \frac{{}^{(1)} \delta p}{ {}^{(1)} \delta \rho} \bigg|_{k} =  \frac{\delta \big(-\nabla_c \varphi_1 \nabla^c \varphi_1 - m^2 \varphi_1 ^2\big)}{\delta \big(-\nabla_c \varphi_1 \nabla^c \varphi_1 + m^2 \varphi_1 ^2\big)}\bigg|_{k},\\
&&{}^{(2)}c_s ^2 \bigg|_{k}= \frac{{}^{(2)} \delta p}{ {}^{(2)} \delta \rho} \bigg|_{k} =  \frac{\delta \big(-\nabla_c \varphi_2 \nabla^c \varphi_2 - m^2 \varphi_2 ^2\big)}{\delta \big(-\nabla_c \varphi_2 \nabla^c \varphi_2 + m^2 \varphi_2 ^2\big)}\bigg|_{k}.
\eea
\end{subequations}
Another quantity of interest is the relative boost $\Delta u$ between
the two components of dark matter, which can be parametrized
in terms of
\be \label{rel4}
\Delta u \bigg |_{k} \equiv \sqrt{\bigg|\delta \Big({}^{(1)}u_a {}^{(2)}u^a + 1\Big) \bigg |_{k} }.
\ee

To compute these quantities, we need to track the evolution of the scalar perturbations in the gravity-dark matter system  from the moment that they enter the Hubble horizon up until today \footnote{We ignore radiation and other forms of matter for this analysis.}.
Here we are interested in doing this for the perturbation modes that are in the linear regime, i.e. for the modes with wavenumbers $1 \lesssim k/\mathcal{H}_0 \lesssim 450$, where $\mathcal{H}_0$ is the conformal Hubble constant today.
To begin, recall that the linearly perturbed Friedmann-Robertson-Walker metric in the Newtonian gauge is given by
\be \label{frw-newt}
ds^2 = a(\eta)^2 \Big[ - \big(1+2 \Phi[\eta,x^i]\big) d\eta^2 + \big(1-2 \Psi[\eta,x^i]\big)(dx^{i})^2 \Big],
\ee
where $i$ is an spatial index and $\Phi$ and $\Psi$ are
the metric perturbation functions.
We also perturb the dark matter fields $\varphi_1$ and $\varphi_2$ by writing
\be
\varphi_1(\eta, x^{i}) = \varphi_{1,0}(\eta) + \delta \varphi_1 (\eta, x^{i}), \hspace{2cm} \varphi_2 (\eta, x^{i})= \varphi_{2,0}(\eta) + \delta \varphi_2(\eta, x^{i}),
\ee
where $\varphi_{1,0}$ and $\varphi_{2,0}$ are their background values.

When there is no anisotropic shear in the stress energy tensor of  dark matter, as is the case for the minimally coupled dark matter model under study, it follows from the Einstein equations that $\Phi=\Psi$ \footnote{See any introductory book on the cosmological perturbation theory, e.g. \cite{weinberg}.}.
Therefore, a complete dynamical description of the linear scalar perturbations can be obtained by solving for $\Phi$, $\delta \varphi_1$, and $\delta \varphi_2$. To do this, we linearize the dark matter Klein-Gordon equations given in \eqref{eqm.1} as well as
\be
m_p ^2 \sum_{i=1}^{3} G_{ii} =  \sum_{i=1}^{3} \big(-\Lambda_0 g_{ii} + T^{\text{DM}}_{ii} \big)
\ee
in $\Phi$, $\delta \varphi_1$, and $\delta \varphi_2$. The resultant
equations, after Fourier decomposing the perturbation functions
\footnote{We Fourier decompose a perturbation function using $$\mathcal{F}(\eta,k^i) = \int \frac{d^3 x}{(2 \pi)^{3/2}}\ e^{-i \delta_{ij} x^{i} k^{j}} \mathcal{F}(\eta, x^{i}).$$} are
\begin{subequations}
\label{newt-pert-dm}
\bea
&&\delta \varphi_1 ''+ 2 \mathcal{H} \delta \varphi_1 '+ \delta \varphi_1 \Big[k^2 + m^2 a^2 \Big]+2 m^2 a^2 \varphi_{1,0} \Phi - 4 \varphi_{1,0}' \Phi'=0, \\
&&\delta \varphi_2 ''+ 2 \mathcal{H} \delta \varphi_2 '+ \delta \varphi_2 \Big[k^2 + m^2 a^2 \Big]+2 m^2 a^2 \varphi_{2,0} \Phi - 4 \varphi_{2,0}' \Phi'=0, \\
&&m_p ^2 \Phi''+ 3 m_p ^2 \mathcal{H} \Phi'+ \Phi \Bigg[2 m_p ^2 \bigg(2 \frac{a''}{a} -  \mathcal{H}^2 \bigg)+ \varphi_{1,0}^{\prime 2}+\varphi_{2,0}^{\prime 2} - \frac{1}{2}m^2 a^2 \Big(\varphi_{1,0}^2+\varphi_{2,0} ^2\Big) - a^2 \Lambda_0 \Bigg]\nonumber\\
&&+\frac{1}{2}m^2 a^2 \Big(\varphi_{1,0} \delta \varphi_1 + \varphi_{2,0} \delta \varphi_2\Big) - \frac{1}{2} \Big(\varphi' _{1,0} \delta \varphi'_1 + \varphi' _{2,0} \delta \varphi'_2 \Big) = 0,
\eea
\end{subequations}
where $k$ is the wavenumber associated with a given mode.
We integrate the system of equations \eqref{newt-pert-dm} for a
given mode $k$ from the time it enters the Hubble horizon, i.e.
at the time $\eta_i$ given by $k = \mathcal{H}(\eta_i)$, up until today. If no entropy perturbations were generated during the inflationary period, the initial conditions for the perturbation functions $\Phi$, $\delta \varphi_1$, and $\delta \varphi_2$ at the
time of horizon entry are given by \cite{weinberg}
\begin{subequations}
\label{initial-conditions}
\bea
&&\Phi(k)\big|_{k=\mathcal{H}(\eta_i)} = \mathcal{C}(k)\bigg[-1 + \frac{\mathcal{H}(\eta_i)}{a(\eta_i) ^2}\int_{-\infty}^{\eta_i} a^2 \ d\eta \bigg],\\
&&\delta \varphi_1(k)\big|_{k=\mathcal{H}(\eta_i)} = -\mathcal{C}(k) \frac{\varphi'_{1,0}(\eta_i)}{a(\eta_i) ^2} \int_{-\infty}^{\eta_i} a^2 \ d\eta,\\
&&\delta \varphi_2(k)\big|_{k=\mathcal{H}(\eta_i)} = -\mathcal{C}(k) \frac{\varphi'_{2,0}(\eta_i)}{a(\eta_i) ^2} \int_{-\infty}^{\eta_i} a^2 \ d\eta,
\eea
\end{subequations}
where $\mathcal{C}(k) \approx i 10^{-4} k^{-3/2}$
is the superhorizon value of the so-called comoving curvature perturbation for a single field slow roll model of inflation at the energy scale of $\sim 10^{-2} m_p$ \cite{baumanntasi}.

\begin{figure}
\centering
\includegraphics[width=1\textwidth]{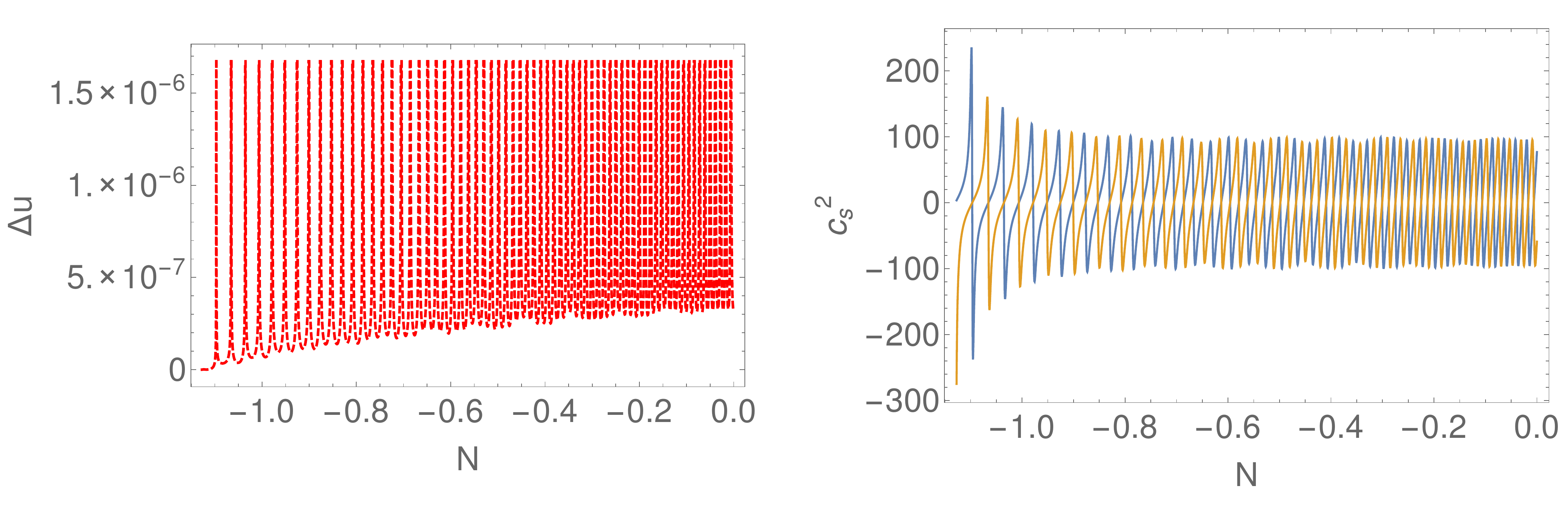}
\caption{This is the plot of $\Delta u$ (left),
${}^{(1)}c_s ^2$ (right-yellow), and ${}^{(2)}c_s ^2$ (right-blue) for the mode  $k/\mathcal{H}_0 = 1$. Here we have taken $m/H_0 = 10^2$. Note that the sound speeds are highly oscillatory with a nearly vanishing mean value throughout. Also, the relative four-velocity between the dark matter components is negligible.}
\label{csu1}
\end{figure}

\begin{figure}
\centering
\includegraphics[width=1\textwidth]{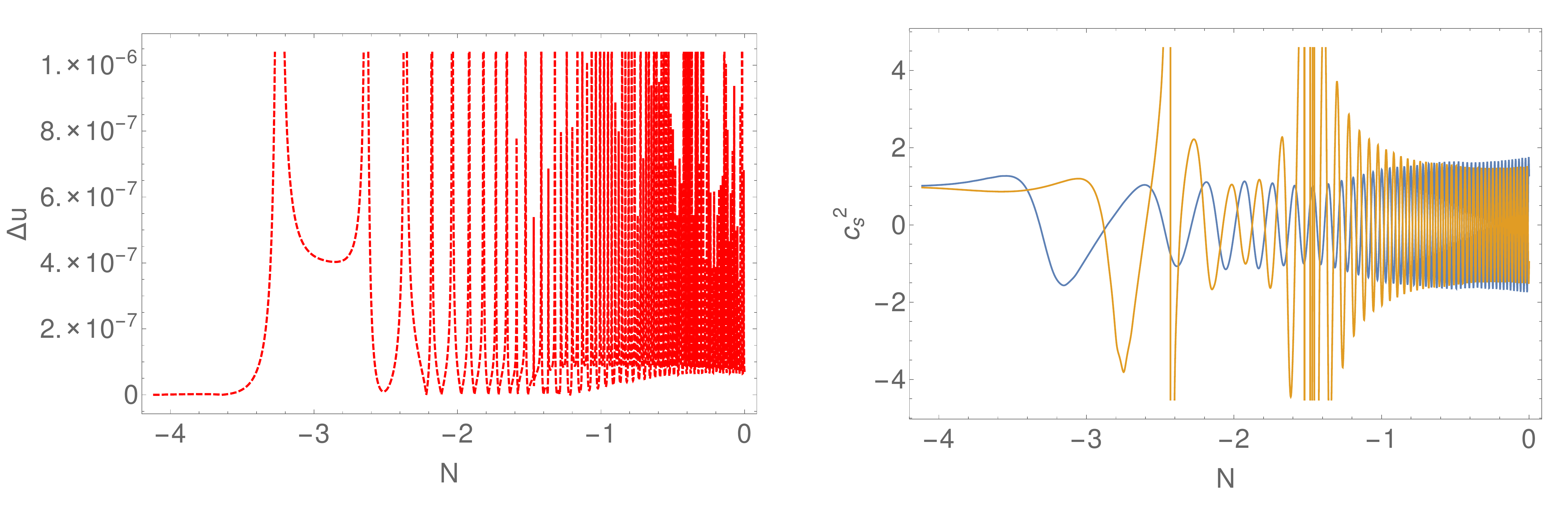}
\caption{This is the plot of $\Delta u$ (left),
${}^{(1)}c_s ^2$ (right-yellow), and ${}^{(2)}c_s ^2$ (right-blue) for the mode  $k/\mathcal{H}_0 = 11$. Here we have taken $m/H_0 = 10^2$. Note that the sound speeds begin with a value of about one, and eventually become highly oscillatory with a nearly vanishing mean value. Also, the relative four-velocity between the dark matter components is negligible.}
\label{csu2}
\end{figure}

Upon solving Eqs. \eqref{newt-pert-dm}, we compute the sound speeds and the relative four-velocity using Eqs. \eqref{soundspeed} and \eqref{rel4}
for any given mode $k$.
Evaluated in the Newtonian gauge, the results are
\bea \label{soundspeedpert}
&&{}^{(1)}c_{s}^2 \bigg| _{k} = \frac{\varphi_{1,0} '\big(\delta \varphi_1 ' - \varphi_{1,0} ' \Phi \big)- m^2 a^2 \varphi_{1,0} \delta \varphi_1 }{\varphi_{1,0} '\big(\delta \varphi_1 ' - \varphi_{1,0} ' \Phi \big)+ m^2 a^2 \varphi_{1,0} \delta \varphi_1 } \bigg|_{k}, \nonumber\\
&&{}^{(2)}c_{s}^2 \bigg| _{k} = \frac{\varphi_{2,0} '\big(\delta \varphi_2 ' - \varphi_{2,0} ' \Phi \big)- m^2 a^2 \varphi_{2,0} \delta \varphi_2 }{\varphi_{2,0} '\big(\delta \varphi_2 ' - \varphi_{2,0} ' \Phi \big)+ m^2 a^2 \varphi_{2,0} \delta \varphi_2 } \bigg|_{k}, \nonumber\\
&& \Delta u \bigg|_k =\sqrt{\frac{k^2}{2} \Bigg(\frac{ \delta \varphi_1}{ \varphi^{\prime}_{1,0}} \bigg| _{k} - \frac{ \delta \varphi_2 }{ \varphi^{\prime}_{2,0}} \bigg| _{k}\Bigg)^2}.
\eea
As an example, we numerically compute the two sound speeds and the relative four-velocity for the modes with $k =\mathcal{H}_0$ and $k =11 \mathcal{H}_0$. We plot the results in Figs. \ref{csu1} and \ref{csu2} respectively. Note that in doing so we have made use of the background solutions found in Sec. \ref{sec.back.1}.

The qualitative features of the plots in Figs. \ref{csu1} and \ref{csu2} can be explained as follows. First, note that the
sound speeds for both modes exhibit highly oscillatory behaviour
as they evolve towards the present time (and in fact, well into the
future). This is the result of the oscillatory behaviour of the
background dark matter fields $\varphi_{1,0}$ and $\varphi_{2,0}$
and their linear perturbations $\delta \varphi_1$ and $\delta \varphi_2$. In fact, for the background solutions that we specialized
to in Sec. \ref{sec.back.1}, it follows from Eqs. \eqref{zeta} and \eqref{rel} that $\varphi_{1,0}$ and $\varphi_{2,0}$ roughly oscillate with frequencies $\zeta_0 '/a = c/(a^3 \mathcal{R}_0 ^2) \sim m $ in the $\eta \rightarrow 0^{-}$ limit. This observation remains valid as long as $p_0^{\rm DM} \approx 0$, as this relation implies $\zeta' _0 / a \sim m$ for $m \gg H_0$. On the other hand, we have already seen in the numerical results presented in the previous section that for values of $m$ that are large compared to the Hubble constant today, the dark matter model under study maintains a near zero background pressure for a longer duration as we evolve the background solutions towards the past \footnote{See Fig. \ref{back-dyn-min-1}.}. Therefore, for sufficiently large values of $m$, we can find background solutions for which  $\varphi_{1,0}$ and $\varphi_{2,0}$ oscillate with frequencies comparable to $m$ and larger than the Hubble parameter from the moment that the shortest mode in the linear regime enters the Hubble horizon up until today.

Similarly, the oscillatory behaviour of the perturbations $\delta \varphi_1$ and $\delta \varphi_2$ in the $\eta \rightarrow 0^{-}$ limit can be surmised from Eq. \eqref{newt-pert-dm}.
By taking the future limit of the latter set of equations, we
arrive at
\begin{subequations}
\label{newt-pert-dm-future}
\bea
&&\delta \varphi_1 ''+ 2 \mathcal{H} \delta \varphi_1 '+ m^2 a^2 \delta \varphi_1 +2 m^2 a^2 \varphi_{1,0} \Phi - 4 \varphi_{1,0}' \Phi'=0, \\
&&\delta \varphi_2 ''+ 2 \mathcal{H} \delta \varphi_2 '+ m^2 a^2 \delta \varphi_2 +2 m^2 a^2 \varphi_{2,0} \Phi - 4 \varphi_{2,0}' \Phi'=0, \\
&&m_p ^2 \Phi''+ 3 m_p ^2 \mathcal{H} \Phi'+ a^2 \Lambda_0 \Phi
+\frac{1}{2}m^2 a^2 \Big(\varphi_{1,0} \delta \varphi_1 +
\varphi_{2,0} \delta \varphi_2\Big) - \frac{1}{2} \Big(\varphi' _{1,0}
\delta \varphi'_1 + \varphi' _{2,0} \delta \varphi'_2 \Big) = 0,
\nonumber \\
\eea
\end{subequations}
where we used the background solutions found in Sec. \ref{sec.back.1}.
It is not difficult to see that in the future limit, $\delta \varphi_1$ and $\delta \varphi_2$ decouple from $\Phi$. Indeed, if we ignore the couplings in Eq. \eqref{newt-pert-dm-future}, then
\begin{subequations}
\bea
&& \delta \varphi_i, \delta \varphi_2 \rightarrow c \eta^{3/2} \eta^{\pm i \frac{m}{H_0}\sqrt{1-\frac{9 H_0 ^2}{4 m^2}}}, \\
&& \Phi \rightarrow c_1 \eta + c_2 \eta^3,
\eea
\end{subequations}
where $c$, $c_1$, and $c_2$ are some constants. Notice the frequency of oscillations for $\delta \varphi_1 $ and $\delta \varphi_2$
which is identical to their background counterparts $\varphi_{1,0}$ and$\varphi_{2,0}$ when $m \gg H_0$ \footnote{Note that $a \propto \eta^{-1} \propto e^{H_0 t}$ in the $\eta \rightarrow 0^{-}$ limit. Thus,
$$
\eta^{\pm i \frac{m}{H_0}\sqrt{1-\frac{9 H_0 ^2}{4 m^2}}} \propto e^{\pm i mt \sqrt{1-\frac{9 H_0 ^2}{4 m^2}} }.
$$
}. To see that this decoupling is consistent with the set of equations \eqref{newt-pert-dm-future}, observe that the corrections due to the $\Phi$ terms to the first two equations become
\be
2 m^2 a^2 \varphi_{1,0} \Phi- 4 \varphi'_{1,0} \Phi' \rightarrow \eta^{1/2} \ \ {\rm or} \ \ \eta^{5/2},
\ee
which induce corrections in $\delta \varphi_1$ and $\delta \varphi_2$  that decay as $\eta^{5/2}$ or $\eta^{9/2}$. Thus, the corrections due to $\Phi$ are subleading
in the $\eta \rightarrow 0^{-}$ limit. On the other hand, the corrections due to $\delta \varphi_1$ and $\delta \varphi_2$  to the
$\Phi$ equation decay at least as $\eta$, which is the same decay form as given by the $\Phi$ terms if $\Phi \rightarrow c_2 \eta^3$ solution is selected. If $\Phi \rightarrow c_1 \eta$ is selected instead, then the $\Phi$ terms in the same equation blow up as $\eta^{-1}$, which implies that the asymptotic dynamics of $\Phi$  is unaffected by its couplings to $\delta \varphi_1$ and $\delta \varphi_2$.
In fact, the leading corrections to the asymptotic solution for
$\Phi$ in this latter case are of $\eta^3$ form.
Numerical solutions to Eq. \eqref{newt-pert-dm} confirm the asymptotic decoupling of the perturbation functions. Also, the asymptotic solutions for $\Phi$ turn out to be of the $\Phi \rightarrow c_1 \eta$ form.

Next, note that the qualitative form of the sound speeds for $k = \mathcal{H}_0$ and $k=11 \mathcal{H}_0$ modes at the onset of horizon
entry are quite different, with the former quickly becoming oscillatory and the latter remaining close to one for roughly an e-fold. Numerical investigations suggest that if $m a \lesssim \mathcal{H}$ at the time of horizon entry for a given mode, then all three sound speeds associated with that mode initially remain close to one, whereas a sharply oscillatory behaviour results for modes with $m a \gg \mathcal{H}$ at the time of their horizon entry. On the other hand, we observe that the sound speeds for both modes end up oscillating with a nearly constant amplitude as they evolve to the present time. This is consistent with the previously found asymptotic forms for the perturbation functions. In fact, using $\Phi \propto \eta$ and $\delta \varphi_{1}, \delta \varphi_{2} \propto \eta^{3/2}$ along with the asymptotic forms of the background solutions, one finds that the numerators and denominators of the sound speeds given
in Eq. \eqref{soundspeedpert} decay as $\eta$ to the leading order.
This implies that the amplitude of oscillations approaches a constant value in the $\eta \rightarrow 0^{-}$ limit. Note that it is more difficult to analytically infer the period of oscillation in this limit, though numerically we have found it to be much shorter than $m^{-1}$.

The other noteworthy feature seen in Figs. \ref{csu1} and
\ref{csu2} is the negligible relative velocity between the
two dark matter fields. This is expected theoretically, because
$\Delta u$ defined in \eqref{rel4} is roughly
$k/(m_p \mathcal{H}_0) \times \delta \varphi \ll k \varphi/(m_p \mathcal{H}_0) \sim k_{\text{phys}}/m$. Therefore, for a
sufficiently large value of dark matter mass $m$, this quantity
remains small for linear perturbations.

To close this discussion, we remind the reader that an averaged out near zero sound speed for the dark matter candidate  is achievable should one select the dark matter mass $m$ to be sufficiently large.
More specifically, in order that the modes with wavenumbers $1 \lesssim k/\mathcal{H}_0 \lesssim 450$ have negligible sound speeds at the present time, one needs $m/H_0 \gtrsim 450$. Requiring
these modes to have negligible sound speeds as early as they become
subhorizon necessitates $m/H_0 \gtrsim \mathcal{H}(\eta_{*})/\mathcal{H}_0 \sim 10^{10}$, where $\eta_{*}$ is the conformal time at which the mode $k/\mathcal{H}_0 = 450$ becomes
subhorizon, assuming that the dark matter candidate generates the CDM background cosmology up until then. This is similar to the conclusion
of a detailed study in \cite{marsh-cosmo} that uses ultralight axions for dark matter.

\section{The effective field theory of interacting dark energy and dark matter} \label{sec.eft}
\subsection{Overview}

In this section we expand the analysis of the preceding section
to allow for more general couplings between the dark matter and dark
energy fields.
We use the techniques that were previously employed in constructing
the EFT of single clock inflation \cite{eft-inflation-1}
to construct an EFT of both dark energy and dark matter\footnote{See \cite{eft-inflation-2} for an extension of the
single clock formalism to multifield models of inflation.
There have also been more recent extensions of this EFT \cite{Ashoorioon:2018uey, Ashoorioon:2018ocr}.
Also see \cite{eft-inflation-weinberg} for a different
EFT formalism for inflation. }. Such
methods have already been applied in constructing
EFTs of dark energy \cite{Gubitosi:2012hu, eft-de-2,eft-de-3,eft-de-4}
\footnote{A different kind EFT valid under certain assumptions about
the UV physics can be found in \cite{eft-de-1}.}.
The successful application of EFT methods in these scenarios is based
on the fact that the degrees of freedom driving
the current and the primordial phases of cosmic acceleration determine
a preferred choice of spacetime foliation.

We now review the basic idea of the construction as applied to the
present context, in a language that does not refer to coordinate choices.
Suppose that we start with a general action
$S = S[g_{ab}, \Pi, \phi]$
for the metric $g_{ab}$, dark energy field $\phi$ and dark matter
field $\Pi$.  This action is diffeomorphism invariant:
\be
S[\psi_* g_{ab}, \psi_* \Pi, \psi_* \phi ] = S[ g_{ab},  \Pi, \phi ]
\ee
where $\psi$ is any smooth diffeomorphism and $\psi_*$ is the pullback operation.
Suppose now that we are given a background solution $g_{ab, 0}, \Pi_0,
\phi_0$.  We choose to use the foliation determined by the dark energy
background solution $\phi_0$, and define a modified action functional
\be
S_1[g_{ab} , \Pi, \delta \phi] = S[g_{ab}, \Pi, \phi_0 + \delta \phi].
\label{s1def}
\ee
which is a functional of the scalar field perturbation $\delta \phi$.
The action $S_1$ is invariant under the transformations
\be
g_{ab} \to \psi_* g_{ab}, \ \ \ \ \Pi \to \psi_* \Pi, \ \ \ \ \delta
\phi \to \psi_* \delta \phi + \psi_* \phi_0 - \phi_0.
\label{symm44}
\ee
The action of these symmetries is linearly realized only for foliation
preserving diffeomorphisms, those for which
\be
\psi_* \phi_0 = \phi_0.
\label{foliationpreserving}
\ee
More general diffeomorphisms are realized nonlinearly, from Eq.\ (\ref{symm44}).

We next define a new action functional of the metric and dark matter
fields by setting $\delta \phi$ to zero:
\be
S_2[g_{ab}, \Pi] = S_1 [g_{ab}, \Pi, 0].
\label{s2def}
\ee
Because of the gauge redundancy in the description of the theory, the
action $S_2$ still contains complete information about the theory.
However it is no longer invariant under the full diffeomorphism group:
the relation
\be
S_2[ \psi_* g_{ab}, \psi_* \Pi] = S_2[ g_{ab}, \Pi]
\ee
is valid only for the foliation preserving diffeomorphisms that satisfy
(\ref{foliationpreserving}).  The action $S_2$ describes the theory
specialized to unitary gauge.

We can reconstruct from the action $S_2$ the fully covariant version
of the theory by performing the so-called
Stueckelberg trick\footnote{See \cite{mgravity} for an excellent
  review of this technique and its applications in gauge theory as
  well as gravitational theories.}.
We define a new action functional that depends on $g_{ab}$, $\Pi$ and
an arbitrary smooth diffeomorphism $\xi: M \to M$ by
\be
S_3[g_{ab}, \Pi, \xi] = S_2[\xi_* g_{ab}, \xi_* \Pi].
\label{s3def}
\ee
This action is invariant under general diffeomorphisms:
\be
S_3[\psi_* g_{ab}, \psi_* \Pi, \psi^{-1} \circ \xi] =S_3[g_{ab}, \Pi,
\xi].
\ee
From the action $S_3$ one can obtain the original action $S$, coming a
full circle, as follows.  We have
\be
S_3[g_{ab}, \Pi, \xi ] = S_2[\xi_* g_{ab}, \xi_* \Pi]
= S_1[\xi_* g_{ab}, \xi_* \Pi, 0],
\ee
from Eqs.\ (\ref{s3def}) and (\ref{s2def}).  Now applying the
invariance (\ref{symm44}) with $\psi = \xi^{-1}$ gives
\be
S_3[g_{ab}, \Pi, \xi ] = S_1[ g_{ab},  \Pi, (\xi^{-1})_* \phi_0 -
\phi_0] = S[g_{ab}, \Pi, \phi].
\ee
Here we have used (\ref{s1def}) and defined $\phi = (\xi^{-1})_*
\phi_0$; the action $S_3$ depends on $\xi$ only through $\phi$.

The key idea of the EFT of inflation/dark energy is to apply the usual
rules of EFT to the unitary gauge form (\ref{s2def}) of the action.
One proceeds by writing down all possible operators that are invariant
under foliation preserving diffeomorphisms, which for perturbations
about FRW solutions includes all time dependent spatial
diffeomorphisms.  This can be done efficiently using the $3+1$ ADM
splitting of the metric, which splits the spacetime metric $g_{ab}$ into a
spatial metric $h_{ij}$, lapse $N$, and shift vector $N^i$ via
\be
ds^2 = h_{ij} (dx^i + N^i dt) (dx^j + N^j dt) - N^2 dt^2.
\ee
Here the surfaces of constant $t$ are chosen to coincide with surfaces of constant $\phi_0$.
Neglecting for the moment the dependence on the dark matter field
$\Pi$, the most general unitary gauge action that is invariant under foliation
preserving diffeomorphisms is
\be
S_2[g_{ab}] = \int d^3 x \int dt \sqrt{h} {\cal L}[t, h_{ij}, N,
K_{ij}, D_i, \partial_t - {\cal L}_{\vec N}],
\label{eq:e4}
\ee
where $D_i$ is the 3D covariant derivative associated with $h_{ij}$
and $K_{ij} = ({\dot h}_{ij} - D_i N_j - D_j N_i)/(2N)$ is the
extrinsic curvature, with $N_i = h_{ij} N^j$.
This process is more efficient than other approaches
since the field $\phi$ does not appear anywhere in the action (\ref{s2def}). After all the
appropriate operators have been identified, one can always return to
the fully covariant form of the theory using the Stueckelberg trick.

The action (\ref{eq:e4}) can depend on the three dimensional Riemann
tensor ${}^{(3)} R_{ijkl}$ through the derivative $D_i$.
It is convenient to re-express this dependence and the dependence on
the lapse and the extrinsic curvature in terms of the perturbations
\begin{eqnarray}
\delta N &\equiv& N - \alpha(t), \\
\delta K_{ij} &\equiv& K_{ij} -\beta(t) h_{ij},\\
\delta {}^{(3)}R_{ijkl} &\equiv& {}^{(3)}R_{ijkl} - \gamma(t) \left[
  h_{ik} h_{jl} - h_{jk} h_{il}
-  h_{il} h_{jk} + h_{jl} h_{ik}\right].
\end{eqnarray}
Here the functions $\alpha(t)$, $\beta(t)$ and $\gamma(t)$ parametrize the
background solution, and are regarded as fixed functions of time,
which are allowed in the action (\ref{s2def}).  Finally the dependence
on three dimensional curvature can be re-expressed in terms of a
dependence on four dimensional curvature and on the extrinsic curvature
using the Gauss-Codazzi equations \cite{eft-inflation-1}.

In the following subsections we follow this approach, but generalize
previous treatments by including the dark matter field $\Pi$.
For simplicity, we disregard baryonic matter.  In Sec. \ref{sec.operators} we identify all the
relevant (dimension $\leq4$) and marginally relevant (dimension $=4$)
operators in the dark matter sector \footnote{Irrelevant operators
  (dimension $>4$) could be important if one
is interested in probing the dynamics of energy scales that are
comparable to the cutoff scale of the EFT. See Sec. \ref{sec.scale}
for a discussion.}.
Just as in the EFT of dark energy, we only include the geometric
objects that are either spacetime tensors or tensors intrinsic to the
surfaces of constant dark energy field.
We find it convenient to implement the Stueckelberg trick from the
beginning by expressing all terms in the action in their covariant
forms. We present the EFT action in Sec. \ref{sec.action}, discuss the
EFT regime of validity in Sec. \ref{sec.scale}, and derive the
relevant equations of motion in Sec. \ref{sec.eft.eqm}. Our present
work is intended to complement previous attempts to formulate a
generalized interacting theory of dark energy and dark matter
\cite{Gleyzes:2013ooa, eft-interactingDE-1,eft-interactingDE-2, eft-interactingDE-3, eft-interactingDE-4, mariana} \footnote{See \cite{Tsujikawa:2015upa, DeFelice:2016ucp} for studies on some aspects of the effective field theory of dark energy in presence of matter.}.

\subsection{The effective field theory operators for the dark matter sector}\label{sec.operators}

In this section we identify all the operators of $ \text{dim} \leq 4$ that
appear in the dark matter effective action up to the second order in
perturbations. The goal is to replace the action \eqref{act.back.1} with a more general action.
Following \cite{eft-inflation-1} we perform a field redefinition
\be
\phi \to {\bar \phi} = {\bar \phi}(\phi)
\ee
so that in the new coordinates on field space the background solution
is just ${\bar \phi}_0(t) =t$.
We denote the perturbation to the scalar field in these field space
coordinates by
\be \label{phibar}
\tau(t,x^i) = \delta {\bar \phi}(t,x^i) = {\bar \phi}(t,x^i) - {\bar \phi}_0(t) = {\bar \phi}(t,x^i)
- t.
\ee
As discussed above, the allowed operators can be constructed from
scalar functions such as the dark matter field $\Pi$,
spacetime tensors such as $g_{ab}$ and $R_{abcd}$, and foliation
dependent tensors such as the unit normal $n^a$, induced metric $h_{ab}$, and extrinsic curvature
tensor $K_{ab}$. After covariantizing using the Stueckelberg trick, these
quantities can be written as
\begin{eqnarray}
n^a &=& \frac{g^{ab}\phi_{,b}}{\sqrt{-g^{c d}{\bar \phi}_{,c}
    {\bar \phi}_{,d}}}, \\
h_{ab}&=&g_{ab}+n_{a}n_{b},\\
K_{ab} &=& h_{a} \ ^{c} h_{b} \ ^{d} \nabla_{c} n_{d}.
\end{eqnarray}
The independent operators in the dark matter sector are given in Table \ref{table.operator}. Some of the operators are
relevant for both the background and the linear perturbation theory as discussed
in the subsequent sections. [See Appendix \ref{sec.operator} for a more detailed  discussion of all possible operators.]

\begin{table}[h]
\centering
\begin{tabular}{cc|c|c|c|c}
\cline{3-5}
& & Background & Perturbations & dimension  \\ \cline{1-5}
\multicolumn{1}{|}{}&$\Pi^{\dagger} \Pi$&  $\checkmark$ & $\checkmark$ & 2 &  \\ \cline{1-5}
\multicolumn{1}{|}{}&$\delta f \ \Pi^{\dagger} \Pi $&  $\text{\ding{55}}$ & $\checkmark$ & 2 &  \\ \cline{1-5}
\multicolumn{1}{|}{}&$n^{a}\nabla_{a}\delta f \ \Pi^{\dagger} \Pi$&  $\text{\ding{55}}$ & $\checkmark$ & 3 &  \\ \cline{1-5}
\multicolumn{1}{|}{}&$\delta K \ \Pi^{\dagger} \Pi $&  $\text{\ding{55}}$ & $\checkmark$ & 3 &  \\ \cline{1-5}
\multicolumn{1}{|}{}&$(\Pi^{\dagger} \Pi)^2$ & $\checkmark$ & $\checkmark$ & 4 &     \\ \cline{1-5}
\multicolumn{1}{|}{}& $g^{ab} \nabla_a \Pi^{\dagger} \nabla_b \Pi$ & $\checkmark$ & $\checkmark$ & 4 &  \\ \cline{1-5}
\multicolumn{1}{|}{}& $h^{ab} \nabla_a \Pi^{\dagger} \nabla_b \Pi$ & $\text{\ding{55}}$ & $\checkmark$ & 4 &  \\ \cline{1-5}
\multicolumn{1}{|}{}&$g^{ab} \nabla_a \nabla_b \delta f \ \Pi^{\dagger} \Pi$&  $\text{\ding{55}}$ & $\checkmark$ & 4 &  \\ \cline{1-5}
\multicolumn{1}{|}{}&$h^{ab} \nabla_a \nabla_b \delta f \ \Pi^{\dagger} \Pi$&  $\text{\ding{55}}$ & $\checkmark$ & 4 &  \\ \cline{1-5}
\multicolumn{1}{|}{}&$n^a \nabla_a \delta K \ \Pi^{\dagger} \Pi$&  $\text{\ding{55}}$ & $\checkmark$ & 4 &  \\ \cline{1-5}
\multicolumn{1}{|}{}&$\delta K^2 \ \Pi^{\dagger} \Pi$&  $\text{\ding{55}}$ & $\checkmark$ & 4 &  \\ \cline{1-5}
\multicolumn{1}{|}{}&$ \delta \big[K_{ab} K^{ab} \big] \ \Pi^{\dagger} \Pi$&  $\text{\ding{55}}$ & $\checkmark$ & 4 &  \\ \cline{1-5}
\multicolumn{1}{|}{}&$h^{ab} \delta R_{ab} \Pi^{\dagger} \Pi$&  $\text{\ding{55}}$ & $\checkmark$ & 4 &  \\ \cline{1-5}
\multicolumn{1}{|}{}&$ R \ \Pi^{\dagger} \Pi$&  $\checkmark$ & $\checkmark$ & 4 &  \\ \cline{1-5}
\end{tabular}
\caption{Operators that are relevant for the effective linear perturbation theory of interacting dark energy-dark matter.
Here we have defined $f \equiv g^{ab} {\bar \phi}_{,a} {\bar \phi}_{,b}$. Also, for any object $X$, $\delta X \equiv X - X_0$ with $X_0$ being its background value.}
\label{table.operator}
\end{table}

\subsection{The effective field theory action}\label{sec.action}

We are now ready to express the EFT action.
This action provides a complete picture for the dynamics of the background
and the linear scalar perturbation theory of interacting dark energy and
dark matter. The general EFT action in the ``hatted frame'' \footnote{This is a slight generalization of the commonly used Jordan frame. For a general interacting theory of dark energy and dark matter, there could be no conformal frame in which all couplings between the two sectors are eliminated. The hatted frame here coincides with the standard Jordan frame if we set $\alpha_i = \beta_i = 0$.} with metric $\hat{g}_{ab}$ is given by
[see Eqs. (\ref{trans}) and (\ref{act.back.3})]
\be  \label{eft.act}
\hat{S}[\hat{g}_{ab},{\bar \phi},\Pi^{\dagger},\Pi,\psi]=\hat{S}_{\text{gravity}}[\hat{g}_{ab}, {\bar \phi}]+\hat{S}_{\text{DE}}[\hat{g}_{ab}, {\bar \phi}]
+\hat{S}_{\text{DM-int}}[\hat{g}_{ab}, {\bar \phi}, \Pi^{\dagger},\Pi] + S_b[e^\alpha
{\hat g}_{ab},\psi].
\ee
Here $\hat{S}_{\text{gravity / DE}}$ are the effective gravitational /
dark energy actions that were worked out in Ref. \citep{Gubitosi:2012hu, eft-de-2}, and
$\hat{S}_{\text{DM-int.}}$ is the effective dark matter action which
includes its interactions with the dark energy field and gravity. Since our focus in
this paper is on the dynamics of dark matter,
we choose the following simple forms for the effective gravitational and dark energy actions
\bea \label{eft.act.de}
&&\hat{S}_{\text{gravity}}[\hat{g}_{ab}, \bar{\phi}] = \int d^4 x  \sqrt{-\hat{g}} \ \frac{e^{\alpha}m_p ^2}{2}  \ \Bigg[ \hat{R} - 3 \hat{g}^{ab} \hat{\nabla}_a \hat{\nabla}_b
\alpha -\frac{3}{2} \hat{g}^{ab} \hat{\nabla}_a \alpha \hat{\nabla}_b \alpha \Bigg],\nonumber\\
&&\hat{S}_{\text{DE}}[\hat{g}_{ab}, \bar{\phi}] = \int d^4 x  \sqrt{-\hat{g}} \ e^{2 \alpha} \ \Bigg[ \frac{-e^{- \alpha}}{2}\Lambda_0  \hat{f} - \tilde{\Lambda} \Bigg],
\eea
where
\be
{\hat f} = {\hat g}^{ab} {\hat \nabla}_a {\bar \phi} {\hat \nabla}_b {\bar \phi}
\ee
and $\tilde{\Lambda} \equiv \Lambda_0 (1+ \beta_1)$
for some function $\beta_1$ of $\bar{\phi}$. This is essentially
the same action for gravity and dark energy that was given in Eq. \eqref{act.back.4}.
Using the operators listed in Table \ref{table.operator},
the effective dark matter action takes the
following form
\bea \label{eft.act.dm}
&& \hat{S}_{\text{DM-int}}[\hat{g}_{ab}, \bar{\phi}, \Pi^{\dagger},\Pi] = \int d^4x \sqrt{-\hat{g}}\ \Bigg[- \hat{g}^{ab} \hat{\nabla}_a \Pi^{\dagger} \hat{\nabla}_b \Pi +  \alpha_1  \hat{h}^{ab} \hat{\nabla}_a \Pi^{\dagger} \hat{\nabla}_b \Pi  - \tilde{\lambda} (\Pi^{\dagger} \Pi)^2 \nonumber\\
&&+ \Big( \mp \tilde{m}^2 + \tilde{\gamma} \hat{R}
+ \mu^2 \alpha_2 \delta \hat{f} + \mu \alpha_3  \hat{n}^a
\hat{\nabla}_a \delta \hat{f} + [\alpha_4 \hat{g}^{ab} + \alpha_5 \hat{h}^{ab}]
\hat{\nabla}_a \hat{\nabla}_b \delta \hat{f}+\mu \alpha_6 \delta \hat{K} \nonumber\\
&&+ \alpha_7 \hat{n}^{a} \hat{\nabla}_a \delta \hat{K}+ \alpha_8 \delta [\hat{K}^2] + \alpha_{9} \delta[\hat{K}_{ab} \hat{K}^{ab}]+ \alpha_{10} \hat{h}^{ab}\delta \hat{R}_{ab} \bigg) \Pi^{\dagger} \Pi \Bigg],
\eea
where $\alpha_i$ are some functions of $\bar{\phi}$ and $\mu$ is some
constant of mass dimension one. Also, $\tilde{m}^2 \equiv m^2 (1+ \beta_2)$,
$\tilde{\lambda} \equiv \lambda (1+ \beta_3)$, and $\tilde{\gamma} \equiv \gamma (1+ \beta_4)$ for some constants $m^2$, $\lambda$, and $\gamma$ and some functions $\beta_i$ of $\bar{\phi}$.

The action \eqref{eft.act.dm} incorporates different possibilities for 
dark matter interactions. The term proportional to $\alpha_1$ is associated
with local Lorentz violation for the dark matter perturbations. The term
proportional to $\tilde{\gamma}$ encodes non-minimal coupling to gravity. Note that this term does not result in the
violation of the WEP as long as $\beta_4=0$.
All terms proportional to $\alpha_i$
result in the violation of the WEP by virtue of being foliation dependent.

Finally, note that expressing $\tilde{\Lambda}$, $\tilde{m}$, $\tilde{\lambda}$, and $\tilde{\gamma}$ in terms of constant terms
and the $\beta_i$ functions is a gauge dependent procedure. To compare the predictions of this model with observations, one would need to fix the values of the $\beta_i$ functions at a fiducial redshift. 

\subsection{The effective field theory regime of validity}\label{sec.scale}

We now determine the domain of validity of the EFT
constructed in the previous section. We first examine
the bounds on the UV cutoff of the EFT by estimating the magnitude
of the dark matter irrelevant operators and requiring them to be suppressed compared to the relevant operators. We then complement our estimates by demanding radiative stability for the coefficients of the main operators, namely the masses and the leading
coupling constants of the dark matter and dark energy fields.
Here we do not speculate on the nature of the UV physics from which this EFT results in the infrared limit.

\begin{figure}
\centering
\includegraphics[width=1\textwidth]{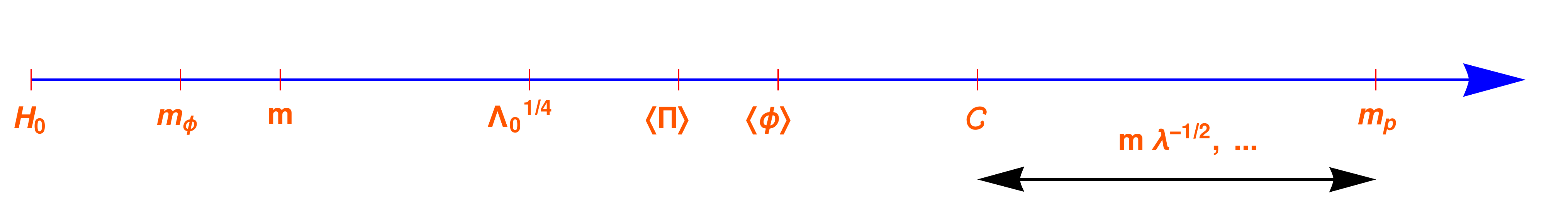}
\caption{An illustration of the various energy scales $E$ in the
  effective field theory, where $H$ is the Hubble parameter. Here
  $\langle \Pi \rangle$ and $\langle \phi \rangle$ denote the present
  values of the dark matter and dark energy fields respectively. The
  UV cutoff $\mathcal{C}$ of the EFT is generically expected to be
  larger than $\langle \Pi \rangle$ and $\langle \phi \rangle$ and the
  masses $m$ and $m_{\phi}$ of the dark matter and dark energy
  fields. Radiative stability for the parameters of the dark
  matter-dark energy Lagrangian places an upper bound on the cutoff
  $\mathcal{C}$. One such upper bound is $m \lambda^{-1/2}$ which
  comes from the one loop contribution of $\lambda (\Pi^{\dagger}
  \Pi)^2$ to the mass $m$ of the dark matter fields.}
\label{energy-axis}
\end{figure}

Since we have multiple degrees of freedom in the EFT, it is possible that the scales at which different sectors of the EFT become strongly coupled are substantially different. In this work, our estimates for the range of the UV cutoff are based on the dark matter sector,
including dark matter-dark energy interactions. Our main conclusion
is that in order to minimize the amount of fine tuning for the parameters of the EFT Lagrangian,
\begin{itemize}
\item the hierarchy between the dark matter mass $m$
and the dark energy mass $m_{\phi}$ should not be more than
a few orders of magnitude,
\item the dark matter mass should be many orders of magnitude
larger than the Hubble parameter today, and
\item the UV cutoff $\mathcal{C}$ should be several orders of magnitude larger than $\Lambda_0^{1/4} \sim 10^{-3}$eV.
\end{itemize}
See Fig. \ref{energy-axis} for an illustration of these conclusions.

\subsubsection{Cutoff estimation based on the dark matter potential}

Let us begin by ignoring the dark matter interactions with dark energy and gravity.
Focusing on the operators with dim $\leq 4$, the effective dark matter Lagrangian that we formulated in Sec. \ref{sec.action} has two parameters, namely the mass $m$ of the dark matter fields and the dark
matter coupling constant $\lambda$.
As a first step,
it is reasonable to demand that the UV cutoff ${\cal C}$  be within the range $m \ll {\cal C} \ll m_p$.
 Note that $m \gg H_0$ as the dark matter mass is required to be several orders of magnitude heavier than the Hubble constant
\footnote{See the analysis given in Sec. \ref{sec.linear.pert}. Also, the lightest dark matter candidates that
can account for the entire CDM content of the Universe
have $m \gtrsim 10^{-24} {\rm eV} \sim 10^9 H_0$ \cite{marsh-cosmo}.}.

It is necessary in any EFT that
irrelevant operators be suppressed compared to relevant and marginally
relevant operators in the regime where EFT is expected to be predictive.
To examine this more closely, consider non-derivative self-interactions in the dark matter potential which we parametrize as
\be \label{dm.pot.series}
\sum_{n \geq 1} \sigma_{n} \frac{(\Pi^{\dagger}\Pi)^{n} }{{\cal C}^{2n-4}},
\ee
where $\sigma_n$ are some dimensionless constants.The dark matter mass $m$
and coupling constant $\lambda$ are expressed as $\sqrt{\sigma_1} {\cal C}$ and $\sigma_2$ respectively. The condition $m \lesssim {\cal C}$ requires $\sigma_1$ to be small compared to unity.
If we assume that dark matter is weakly interacting with itself, then
\be
\lambda (\Pi^{\dagger}\Pi)^2 \sim \lambda \frac{\Lambda_0 ^2}{m^4} \ll \Lambda_0,
\ee
where we used $m^2 \Pi^{\dagger} \Pi \sim \Lambda_0$. This implies that $\lambda \ll m^4/\Lambda_0$. If $m \lesssim \Lambda_0 ^{1/4}$, then $\lambda$ must be suppressed compared to unity  \footnote{Recall that our analysis in Sec. \ref{sec.back.1} indicated that smaller values of $\lambda$ result in $w^{\rm DM} \ll 1$ for higher redshifts and therefore are better suited for reproducing the background $\Lambda$CDM cosmology.}.
Setting $\lambda$ to values much smaller than unity is a fine tuning if no assumptions regarding the UV physics have been made.

Turning to terms in Eq. \eqref{dm.pot.series} with $n \geq 3$ (irrelevant operators), we have
\be
\sigma_n \frac{(\Pi^{\dagger}\Pi)^n}{{\cal C}^{2n-4}} \sim \sigma_n \frac{\Lambda_0 ^n}{m^{2n} {\cal C}^{2n-4}} \ll \Lambda_0
\ee
which using $\Lambda_0 \sim m_p^2 H_0^2 \sim (10^{-3} \, {\rm eV})^4$
requires the cutoff to be bounded below by
\be \label{cutoff.lb}
{\cal C} \gg \sigma_n ^{\frac{1}{2n-4}} \sqrt{m_p H_0} \left( \frac{m}{\sqrt{m_p H_0}} \right)^{- \frac{n}{n-2}}.
\ee
Assuming $\sigma_n \sim \mathcal{O}(1)$ this gives a lower bound below $10^{-3} \, {\rm eV}$ for $m \gtrsim 10^{-3} \, {\rm eV}$,
but a more stringent lower bound above $10^{-3} \, {\rm eV}$ for larger $m$.
If the dark matter fields are similar to the QCD axions, then
$m \gtrsim 10^{-6} {\rm eV}$ \footnote{See \cite{preskill-axion, ABBOTT1983133, Dine:1982ah, steinhardt-axion}. The lower bound on the mass of the QCD axions can be significantly lowered if one allows for fine tuning of its parameters \cite{hertzberg-axion,wantz-axion}. } which implies ${\cal C} \gtrsim \mathcal{O}(1){\rm MeV}$. Note that $m \gtrsim 10^{-13} {\rm eV}$ as
${\cal C}$ cannot exceed $m_p$.

If one allows $\sigma_n \ll 1$, then smaller values for $m$ are feasible at the expense of fine tuning. This may be necessary if one
is interested in significant interactions between dark matter and
dark energy fields when the dark energy field has a mass $m_{\phi}\sim H_0$ \footnote{Ultralight dark matter fields can interact significantly with dark energy if the hierarchy between their respective masses is not too large \cite{kaloper}.}.
We can rearrange Eq. \eqref{cutoff.lb} to derive the following upper bound on $\sigma_n$
\be \label{bound-sigma}
\sigma_n \ll \frac{H_0^2}{m_p^2}
\left(\frac{{\cal C}}{m_p}\right)^{2n-4} \left(\frac{m}{H_0}\right)^{2n}.
\ee
We expect $m/H_0 \geq 10^9$
if the dark matter candidate accounts for all of dark matter in the
Universe \cite{marsh-cosmo}. For $n=3$ and using ${\cal C} \lesssim
m_p$ this gives $\sigma_3 \ll 10^{-68}$, which is an extreme fine
tuning. If we lower the cutoff scale to $\mathcal{C} \sim
{\rm eV}$, the fine tuning of $\sigma_3$ becomes more
severe, where now it becomes bounded above by $10^{-120}$. Overall,
the smaller the hierarchy between the cutoff scale and the dark matter
mass, and the lighter the dark matter fields, more severe fine tunings
should be expected.
However, this is not particularly surprising because fine tuning
is a generic problem of dynamical dark energy models which has now
become more severe as a result of dark energy coupling to dark matter.
This level of fine tuning may not harm the self consistency of the EFT
model as long as the EFT parameters are stable under radiative
corrections. Whether extreme fine tuning of parameters can prevent
the EFT from admitting a well defined UV completion is a relevant
question which we will not address in the present work.

\subsubsection{Bounds on the cutoff from radiative corrections} \label{sec.radiative}

Unless prohibited or strongly constrained by a symmetry, radiative corrections can induce large changes in the parameters of a Lagrangian.
Such corrections, which are cutoff dependent, are then absorbed by carefully tuned ``bare'' Lagrangian parameters in order to produce ``physical'' Lagrangian parameters, which are cutoff independent \footnote{We are using brute-force cutoff as a method of regularization.}.
This process requires fine tuning of bare Lagrangian parameters.
If this fine tuning is undesirable, then the
cutoff needs to be suitably lowered in order to keep radiative corrections under control. By demanding radiative stability at one loop order for a number of primary EFT parameters, we derive
an upper bound for the EFT cutoff $\mathcal{C}$.

For simplicity, we ignore direct couplings of gravity to dark energy
and dark matter \footnote{In our case, these direct couplings take the forms $e^{\alpha} R$ and $\tilde{\gamma} R \Pi^{\dagger}\Pi$ that appear in Eqs. \eqref{eft.act.de} and \eqref{eft.act.dm} respectively.}. We restrict attention to modes with
momenta $k \gg \mathcal{H}$ for which we can neglect the influence
of background geometry on the mode dynamics.
The EFT  action that we formulated in Sec. \ref{sec.action} can then
be written as (ignoring irrelevant operators) \footnote{We ignore foliation dependent operators as they result in local Lorentz violation. Note, however, that we have no direct evidence that dark matter or dark energy respect local Lorentz symmetry. Here we do so  to simplify our analysis.}
\bea\label{large-k-action}
&&S^{k \gg \mathcal{H}}_{\text{DE-DM}}[\phi,\Pi^{\dagger},\Pi] =
\int d^4x \ \bigg[-\frac{1}{2}\partial_a \phi \partial^a \phi
-V(\phi) -\partial_a \Pi^{\dagger}\partial^a \Pi -m^2 \Pi^{\dagger}\Pi
-\lambda (\Pi^{\dagger}\Pi)^2\nonumber\\
&& - \mu \phi \Pi^{\dagger}\Pi - \epsilon \phi^2 \Pi^{\dagger}\Pi \bigg],
\eea
where $\mu$ and $\epsilon$ are constants of dimension one and zero
respectively.  The dark energy potential $V(\phi)$ is assumed to be
sufficiently flat with a current value of approximately $\Lambda_0$. If we take this potential to be $V(\phi) = m_{\phi}^2 \phi^2 /2$,
we have
\be \label{de-pot}
V(\langle \phi \rangle) = \frac{1}{2} m_{\phi}^2 \langle \phi \rangle ^2 \approx \Lambda_0,
\ee
where $\langle \phi \rangle$ is the present value of the dark energy
field. Potential flatness then requires $m_{\phi} \lesssim \Lambda_0 ^{1/4} \lesssim \langle \phi \rangle$ \ \footnote{For the dark energy  potential to be flat, we require
$
|V'/V^{3/4}|, |V''/V^{1/2}| \ll 1
$.}. The interaction terms in action \eqref{large-k-action}
result from expanding $\tilde{m}^2$ in Eq. \eqref{eft.act.dm}
in powers of $\phi$. If $\tilde{m}^2$ has an analytic dependence on $\phi$, we can write
\be \label{mtilde}
\tilde{m}^2(\phi) = m^2 \bigg[1+ \sum_{i=1}^{\infty} \xi_i \bigg(\frac{\phi}{{\cal C}}\bigg)^i\bigg],
\ee
for some constants $\xi_i$. In the EFT framework,
the strong coupling limit occurs
when $\phi \sim \mathcal{C}/\xi_i ^{1/i}$.
This parametrization gives $\mu \equiv \xi_1 m^2/\mathcal{C}$ and $\epsilon \equiv \xi_2 m^2/\mathcal{C}^2$. Note that we should require $\epsilon, \lambda \ll 1$ in order to rely on perturbation methods for computing radiative corrections.
This latter requirement implies in particular that $\xi_2 \ll \mathcal{C}^2/m^2$, which is not a stringent constraint given that
$\mathcal{C} \gtrsim m$.

\begin{figure}
\centering
\includegraphics[width=1\textwidth]{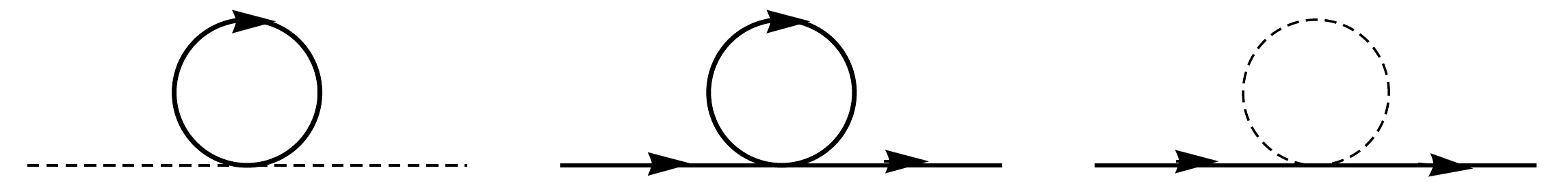}
\caption{One loop corrections to $m_{\phi}$ (left) and $m$ (right)
from $\phi^2 \Pi^{\dagger}\Pi$ operator, as well as the
one loop correction to $m$ (center) from $(\Pi^{\dagger}\Pi)^2$.}
\label{feynman-1}
\end{figure}

\begin{figure}
\centering
\includegraphics[width=0.6\textwidth]{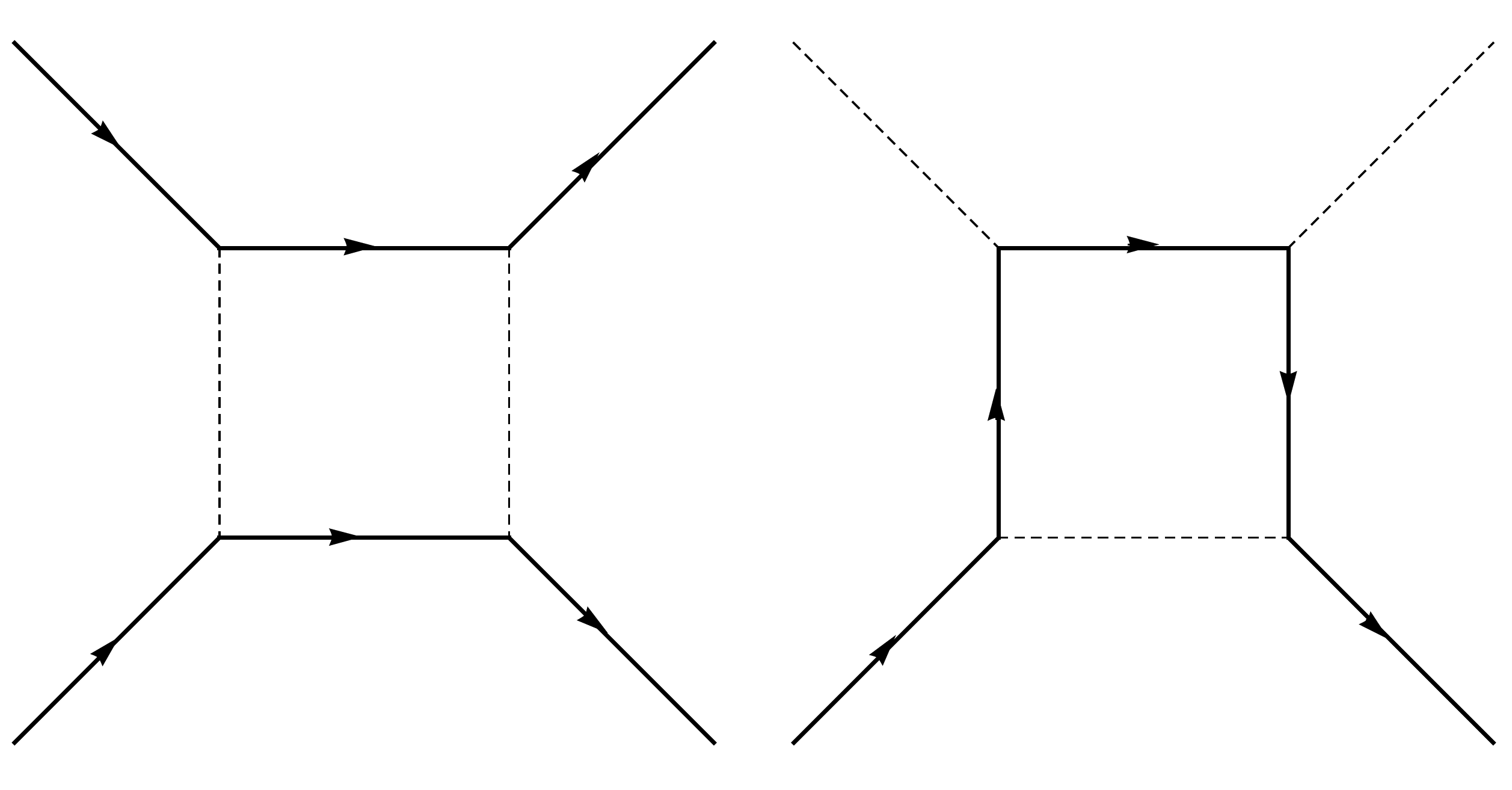}
\caption{One loop corrections to $\lambda$ (left) and $\epsilon$ (right)
from $\phi \Pi^{\dagger}\Pi$ operator.}
\label{feynman-2}
\end{figure}

Our objective here is to keep the one loop radiative corrections to
$m_{\phi}$, $m$, $\mu$, $\lambda$, and $\epsilon$ not large compared to their bare values. The estimates for radiative corrections to these parameters are provided in Table \ref{table.rc}.
\begin{table}[h]
\centering
\begin{tabular}{cc|c|c|}
\cline{3-4}
& & DM & DE  \\ \cline{1-4}
\multicolumn{1}{|}{}&$\delta m^2$&  $\sim \lambda \ \mathcal{C}^2$ & $\sim \epsilon \ \mathcal{C}^2$ \\ \cline{1-4}
\multicolumn{1}{|}{}&$\delta m_{\phi}^2$&  $\sim \epsilon \ \mathcal{C}^2$ &  \\ \cline{1-4}
\multicolumn{1}{|}{}&$\delta \mu$&  $\sim \lambda \mu \  \log{\big(\frac{\mathcal{C}}{m}\big)}$ &  \\ \cline{1-4}
\multicolumn{1}{|}{}&$\delta \lambda$&  $\sim \lambda^2 \  \log{\big(\frac{\mathcal{C}}{m}\big)}$  & $\sim \epsilon^2 \  \log{\big(\frac{\mathcal{C}}{m_{\phi}}\big)}$ \\ \cline{1-4}
\multicolumn{1}{|}{}&$\delta \epsilon$&  $\sim \lambda \epsilon \  \log{\big(\frac{\mathcal{C}}{m}\big)}$ &  \\ \cline{1-4}
\end{tabular}
\caption{Leading one loop corrections to $m_{\phi}$, $m$, $\mu$, $\epsilon$, and $\lambda$ due to the dark matter (DM) and dark energy (DE) loops based on the interactions terms that appear in Eq. \eqref{large-k-action}. One loop corrections that are inversely proportional to $\mathcal{C}$ are omitted.}
\label{table.rc}
\end{table}
Among these parameters, $m$ and $m_{\phi}$ are most sensitive to the UV cutoff where $\delta m^2, \delta m_{\phi}^2 \propto \mathcal{C}^2$.
The most stringent bounds on $\mathcal{C}$ result from these corrections (See Fig. \ref{feynman-1} for the Feynman diagrams associated with these loop corrections).
Requiring $\delta m^2 \lesssim m^2$ and $\delta m_{\phi} ^2 \lesssim m_{\phi} ^2$ gives
\be \label{cutoff-ub-1}
\mathcal{C} \lesssim \text{min} \bigg\{\frac{m}{\sqrt{\lambda}},
\frac{m}{\sqrt{\epsilon}}, \frac{m_{\phi}}{\sqrt{\epsilon}} \bigg\}.
\ee
Several remarks should be made regarding the above relation.
First, using the definition given in Eq. \eqref{mtilde} for
$\epsilon$ we have
\be
\frac{\text{min}\{m,m_{\phi}\}}{\sqrt{\xi_2} m} \gtrsim 1,
\ee
which necessitates fine tuning of $\xi_2$ to values smaller
than order unity unless $m_{\phi} \gtrsim m$. Additionally, while
lowering $\lambda$ increases $m \lambda^{-1/2}$
classically, the quantum corrections to $\lambda$ from dark energy loops require $\lambda_{\rm phys} \equiv \lambda _{\rm bare}+\delta \lambda \gtrsim \epsilon^2 \log{(\mathcal{C}/m_{\phi})}$. if we assume
$m_{\phi}\sim m$, this lower bound on $\lambda_{\rm phys}$ gives
\be
\mathcal{C} \lesssim m \times \text{min}\Bigg\{\frac{1}{\sqrt{\epsilon}}, \frac{1}{\epsilon \sqrt{\log{\big(\frac{\epsilon^2}{2 }\big)}}} \Bigg\} = \frac{m}{\sqrt{\epsilon}}
\ee
for $\epsilon \ll 1$. Finally, note that $\epsilon$ too is not
protected from quantum corrections even if we classically
set $\epsilon  = 0$. Indeed, one loop corrections to $\lambda$
and $\epsilon$ from $\phi \Pi^{\dagger}\Pi$ operator are
roughly $\mu^4/\mathcal{C}^4$, which using the definition of $\mu$
become roughly $\xi_1 ^4 m^8/\mathcal{C}^8$ (See Fig. \ref{feynman-2} for the relevant Feynman diagrams).
This gives
\be
\mathcal{C} \lesssim \frac{m}{\sqrt{\epsilon_{\rm phys}}} = \frac{\mathcal{C}^4}{\xi_1 ^2 m^3} \Rightarrow \mathcal{C} \gtrsim \xi_1^{2/3} m,
\ee
which, while consistent with our expectations that $m \lesssim \mathcal{C}$ for $\xi_1 \sim \mathcal{O}(1)$, does not provide
any upper bounds on $\mathcal{C}$. Therefore, the UV cutoff
cannot be constrained using Eq. \eqref{cutoff-ub-1} if $\lambda$ and $\epsilon$ are tuned to zero classically and $\phi \Pi^{\dagger} \Pi$ is the only interaction operator. In that case, an upper bound
for ${\cal C}$ is obtained by requiring the one loop correction $\delta m_{\phi} ^2$ to the mass of the dark energy field to be
bounded by $m_{\phi}^2$, which gives $\log{(\mathcal{C}/m)} \lesssim  m_{\phi}^2/\mu^2$. In particular, if ${\cal C}$ is larger than $m$ by at least a few orders of magnitude, then $\mu^2 \lesssim m_{\phi} ^2$.

\section{The effective field theory dynamical equations} \label{sec.eft.eqm}

In this section we derive the effective dynamical
equations for the gravitational and the dark matter
fields by varying the action \eqref{eft.act} with respect to $g_{ab}$, $\Pi^{\dagger}$, and $\Pi$. Given the parameterization \eqref{phibar} of the dark energy field as $\bar{\phi}$, we find it more convenient to derive the dark energy background and linear perturbation equations separately in Secs. \ref{sec.background.eft} and \ref{sec.eqm.scalar}
respectively.

The Einstein equations are (we drop the hat from the operators in Sec. \ref{sec.action})
\be \label{eft.einstein}
\mathcal{G}_{ab} =  T_{ab} ^{\text{DE}} + T^{\text{DM}} _{ab}+ T^{\text{int}} _{ab}
\ee
where the gravitational tensor $\mathcal{G}_{ab}$ is defined as
\bea \label{eft.einstein.tensor}
&& \mathcal{G}_{ab} \equiv \frac{2}{\sqrt{-g}}\frac{\delta S_{\text{gravity}}}{\delta g^{ab}}= e^{\alpha}m_p ^2\bigg[ G_{ab} + \frac{1}{2}g_{ab} \Big(2 g^{ab} \nabla_a \nabla_b \alpha + \frac{1}{2}g^{ab} \nabla_a \alpha \nabla_b \alpha \Big) - \nabla_a \nabla_b \alpha \nonumber\\
&&+ \frac{1}{2} \nabla_a \alpha \nabla_b \alpha \bigg],
\eea
the effective dark energy stress energy tensor $T_{ab} ^{\text{DE}}$ is given by
\bea \label{eft.de.tensor}
&&T_{ab} ^{\text{DE}} \equiv \frac{-2}{\sqrt{-g}}\frac{\delta S_{\text{DE}}}{\delta g^{ab}} =  e^{\alpha}  \Lambda_0 f_{ab} + e^{2\alpha} g_{ab} \Big( -\frac{1}{2} e^{-\alpha}\Lambda_0 f - \tilde{\Lambda} \Big),
\eea
the effective dark matter stress energy tensor $T_{ab} ^{\text{DM}}$ is given by
\bea \label{eft.dm.tensor.1}
&& T_{ab} ^{\text{DM}}\equiv \frac{-2}{\sqrt{-g}}\frac{\delta S_{\text{DM-int}}}{\delta g^{ab}} \bigg|_{\alpha_i = \beta_i= \gamma = 0}= 2 \nabla_{(a}\Pi^{\dagger} \nabla_{b)}\Pi + g_{ab} \Big[- g^{cd} \nabla_{c} \Pi^{\dagger} \nabla_d \Pi  \mp m^2  \Pi^{\dagger} \Pi \nonumber\\
&& - \lambda (\Pi^{\dagger} \Pi)^2 \Big],
\eea
and the effective stress energy tensor $T_{ab} ^{\text{int}}$ for the dark matter interactions with dark energy and gravity is given by
\bea\label{eft.dm.tensor.2}
&&T_{ab} ^{\text{int}}\equiv \frac{-2}{\sqrt{-g}}\frac{\delta S_{\text{DM-int}}}{\delta g^{ab}} - T^{\text{DM}}_{ab}= -2 \alpha_1 \nabla_{(a}\Pi^{\dagger} \nabla_{b)}\Pi + g_{ab}\Big[ \alpha_1 h^{cd} \nabla_{c} \Pi^{\dagger} \nabla_d \Pi  + ( \mp m^2 \beta_2 + \tilde{\gamma} R) \nonumber\\
&& \times \Pi^{\dagger} \Pi - \lambda \beta_3 (\Pi^{\dagger} \Pi)^2 - 2 g^{cd} \nabla_c \nabla_d (\tilde{\gamma} \Pi^{\dagger} \Pi) \Big] - 2 \tilde{\gamma} R_{ab} \Pi^{\dagger} \Pi - 2\alpha_1 n^c \Big[ \nabla_c \Pi^{\dagger} \ \nabla_{(a}\Pi \ n_{b)} + \nabla_c \Pi \nonumber\\
&&\times \nabla_{(a}\Pi^{\dagger} \ n_{b)}\Big]+2 \alpha_1 J_{ ab} \ n^c n^d \nabla_c \Pi^{\dagger} \nabla_d \Pi  + 2 \nabla_a \nabla_b (\tilde{\gamma} \Pi^{\dagger} \Pi )+\Bigg [\mu^2 \alpha_2 \Big \{g_{ab} \delta f - 2 \delta f_{ab}\Big \}+ \mu \alpha_3 \Big \{g_{ab} \nonumber\\
&&\times n^c \nabla_c \delta f -2 n_{(a}\nabla_{b)} \delta f  +J_{ab} n^c \nabla_c \delta f \Big \} + \alpha_5 \Big \{ g_{ab} h^{cd} \nabla_c \nabla_d \delta f -2\nabla_{(a} \nabla_{b)} \delta f -4 n^c n_{(a} \nabla_{b)} \nabla_c \delta f   \nonumber\\
&&+2  J_{ab} n^c n^d \nabla_c \nabla_d \delta f \Big \}+ \mu \alpha_6 \Big \{g_{ab}\delta K - 2 \delta \Big(K_{ab} - \frac{1}{2}J_{ab} K\Big)\Big \}+ \alpha_7 \Big \{ g_{ab} n^c \nabla_c \delta K  + 2 \delta \Big[K \Big(K_{ab} \nonumber\\
&&-\frac{1}{2} J_{ab} K \Big) \Big]\Big \} + \alpha_8 \Big \{g_{ab}\delta K^2 - 4 \delta \Big[K \Big(K_{ab}-\frac{1}{2} J_{ab} K \Big) \Big] \Big\}   + \alpha_9 \Big \{ g_{ab}\delta \big(K_{cd} K^{cd}\big) -4 \delta \Big(K^{c} \ _{(a} K_{b)c} \nonumber\\
&&-\frac{1}{2} J_{ab} K^{cd} K_{cd}\Big)\Big \} + \alpha_{10} \Big \{g_{ab} h^{cd}\delta  R_{cd} - 2  \Big(\delta R_{ab} + 2 n^c \delta R_{c(a} n _{b)} - n^c n^d \delta R_{cd} J_{ab}\Big) \Big\} \Bigg] \Pi^{\dagger} \Pi - 2 \delta f_{ab}  \nonumber\\
&&\times \Big[ - \mu \nabla_c (\alpha_3 n^c \Pi^{\dagger} \Pi) +  g^{cd} \nabla_c \nabla_d (\alpha_4 \Pi^{\dagger} \Pi) + \nabla_c \nabla_d ( \alpha_5 h^{cd} \Pi^{\dagger} \Pi)\Big]- g_{ab} g^{cd} \nabla_c \delta f \nabla_d [\alpha_4 \Pi^{\dagger} \Pi]  \nonumber\\
&&+ 2 \nabla_{(a} \big[\alpha_4 \Pi^{\dagger} \Pi \big] \nabla_{b)}\delta f   - g^{cd} \nabla_d \Big[ \alpha_5 h_{ab} \nabla_c \delta f \Pi^{\dagger} \Pi \Big]+ 2 \nabla_c \Big[ \alpha_5 \nabla_{(a} \delta f \ h^{c} \ _{b)} \Pi^{\dagger} \Pi \Big] + 2 \delta \bigg \{ \nabla_c \bigg[ \Big( n_{(a} h_{b)} \ ^{c}  \nonumber\\
&&- \frac{1}{2} h_{ab} n^c \Big) \bigg \} \Big( [\mu \alpha_6 - \alpha_7 K] \Pi^{\dagger} \Pi - n^c \nabla_c [\alpha_7 \Pi^{\dagger} \Pi]\Big) \bigg]-2 \nabla_c \bigg[\alpha_7  \Big(n_{(a}h_{b)} \ ^c
- \frac{1}{2} h_{ab} n^c \Big) \delta K \Pi^{\dagger} \Pi \bigg]   \nonumber\\
&&+ 2\nabla_c [\alpha_7 \Pi^{\dagger} \Pi] \bigg[ n^c \delta \Big(K_{ab} - \frac{1}{2}J_{ab} K \Big) + \delta K
\Big(\delta^c _{(a} \ n_{b)} - \frac{1}{2} n^c J_{ab}\Big) \bigg] + 4 \delta \bigg \{\nabla_c \bigg [ K \Big( n_{(a} h_{b)}\ ^{c} - \frac{1}{2}h_{ab} n^c \Big) \bigg \}\nonumber\\
&&\times  \alpha_8 \Pi^{\dagger} \Pi \bigg]  + 4 \delta \bigg \{ \nabla_c \bigg [ \Big(n_{(a} K_{b)} \ ^{c} - \frac{1}{2} n^c K_{ab} \Big) \bigg \} \alpha_9 \Pi^{\dagger} \Pi \bigg]
+ \delta \bigg \{ 2 \nabla_c \nabla_{(b} \bigg \} \Big[\alpha_{10} h_{a)} \ ^{c} \Pi^{\dagger} \Pi \Big]- \delta \bigg \{ g^{cd}\nabla_c \nabla_{d} \bigg \}  \nonumber\\
&&\times \Big[\alpha_{10} h_{ab} \Pi^{\dagger} \Pi \Big] - \delta \bigg \{ g_{ab}\nabla_c \nabla_{d} \bigg \} \Big[ \alpha_{10} h^{cd}   \Pi^{\dagger} \Pi \Big].
\eea 
In Eqs. \eqref{eft.einstein.tensor}, \eqref{eft.de.tensor}, \eqref{eft.dm.tensor.1}, and \eqref{eft.dm.tensor.2} we defined
$f_{ab} \equiv \bar{\phi}_{,a} \bar{\phi}_b$, $\delta f_{ab} \equiv (\bar{\phi}_{,a} \bar{\phi}_{,b} - a^2 \delta_{\eta a} \delta_{\eta b})$, and $J_{ab} \equiv \bar{\phi}_{,a} \bar{\phi}_{,b}/(\bar{\phi}_{,c} \bar{\phi}^{,c})$. We also used
the notation $X_{(a} Y_{b)} \equiv (X_a Y_b + X_b Y_a)/2$.

For much of the subsequent analysis we use the  dark matter
parameterization in terms of $\mathcal{R}$ and $\zeta$ defined in Eq. \eqref{def-dm}. We therefore find it convenient to express the dark matter and interactions stress energy tensors as a sum of stress energy tensors each associated with $\mathcal{R}$ and $\zeta$ \footnote{The $\zeta$ pieces of both stress energy tensors actually include factors of $\mathcal{R}^2$. This is expected because $\zeta$ is dimensionless.}. The dark matter stress energy tensor given in Eq. \eqref{eft.dm.tensor.1} is written as
\be
T^{\text{DM}}_{ab} = {}^{\mathcal{R}}T_{ab} ^{\text{DM}} + {}^{\zeta}T_{ab} ^{\text{DM}}
\ee
where
\bea \label{eft.stress.dm.1.r&z}
&&{}^{\mathcal{R}}T_{ab} ^{\text{DM}} = 2 \nabla_a \mathcal{R} \nabla_b \mathcal{R} + g_{ab} \big[- \nabla_c \mathcal{R} \nabla^c \mathcal{R}  \mp m^2 \mathcal{R}^2 - \lambda \mathcal{R}^4 \big], \nonumber\\[10pt]
&&{}^{\zeta}T_{ab} ^{\text{DM}} = \mathcal{R}^2 \Big[ 2 \nabla_a \zeta \nabla_b \zeta - g_{ab} \nabla_c \zeta \nabla^c \zeta \Big].
\eea
Similarly, we write the interactions stress energy tensor given in Eq. \eqref{eft.dm.tensor.2} as
\be
T^{\text{int}}_{ab} = {}^{\mathcal{R}}T_{ab} ^{\text{int}} + {}^{\zeta}T_{ab} ^{\text{int}}
\ee
where
\bea \label{eft.stress.dm.2.r&z}
&&{}^{\zeta}T_{ab} ^{\text{int}}= \mathcal{R}^2 \alpha_1 \Big[ -2 \nabla_a \zeta \nabla_b \zeta - 4 n^c \nabla_c \zeta \nabla_{(a} \zeta \ n_{b)}
+ 2  J_{ab} n^c n^d \nabla_c \zeta \nabla_d \zeta +  g_{ab} h^{cd} \nabla_c \zeta \nabla_d \zeta \Big], \nonumber\\[10pt]
&& {}^{\mathcal{R}}T_{ab} ^{\text{int}}= -2 \alpha_1 \nabla_a \mathcal{R} \nabla_b \mathcal{R} - 4 \alpha_1 n^c \nabla_c \mathcal{R} \nabla_{(a} \mathcal{R} \ n_{b)} + 2 \alpha_1 J_{a b} n^c n^d \nabla_c \mathcal{R} \nabla_d \mathcal{R}- 2 \tilde{\gamma}  R_{ab}\mathcal{R}^2 \nonumber\\
&&  + 2 \nabla_a \nabla_b [\tilde{\gamma} \mathcal{R}^2] + g_{ab} \big[ \alpha_1 h^{cd} \nabla_c \mathcal{R} \nabla_d \mathcal{R} + (\mp m^2 \beta_2 + \tilde{\gamma} R) \mathcal{R}^2 - \lambda \beta_3 \mathcal{R}^4  - 2  \nabla^c \nabla_c (\tilde{\gamma} \mathcal{R}^2)\big] \nonumber\\
&&+ {}^{\text{non-cov}}T_{ab} ^{\text{int}}\bigg|_{\Pi^{\dagger}\Pi= \mathcal{R}^2}.\nonumber\\
\eea
In the above equation,  ${}^{\text{non-cov}}T_{ab} ^{\text{int}}$ refers to
the operators in the interaction stress energy tensor that are expressed in
the non-covariant format.

Lastly, the dark matter equations of motion are
\bea \label{eft.eqm.dm}
&& \nabla^a \nabla_a \Pi^{\dagger} - \nabla_a \Big[ \alpha_1 h^{ab} \nabla_b \Pi^{\dagger} \Big] + \Big[\mp \tilde{m}^2 -2 \tilde{\lambda}[\Pi^{\dagger} \Pi]  + \tilde{\gamma} R + \mu^2 \alpha_2 \delta f + \mu \alpha_3 n^a \nabla_a \delta f \nonumber\\
&&+ \big(\alpha_4 g^{ab} + \alpha_5 h^{ab}\big)\nabla_a \nabla_b \delta f + \mu \alpha_6 \delta K + \alpha_7 n^a \nabla_a \delta K + \alpha_8 \delta K^2
+ \alpha_9 \delta \big(K_{ab} K^{ab}\big) \nonumber\\
&&+ \alpha_{10} h^{ab}\delta R_{ab} \Big] \Pi^{\dagger} = 0,
\eea
along with its complex conjugate.
Expressed in terms of the radial and angular fields
$\mathcal{R}$ and $\zeta$, they become
\bea \label{eft.eqm.dm.r&z}
&&\big(g^{ab}-\alpha_1 h^{ab}\big)\nabla_a \nabla_b \mathcal{R}-\nabla_a \big(\alpha_1 h^{ab} \big)\nabla_b \mathcal{R} - \mathcal{R} \big(g^{ab}-\alpha_1 h^{ab}\big)\nabla_a \zeta \nabla_b \zeta + \Big[\mp \tilde{m}^2 -2 \tilde{\lambda}\mathcal{R}^2 \nonumber\\
&& + \tilde{\gamma} R + \mu^2 \alpha_2 \delta f + \mu \alpha_3 n^a \nabla_a \delta f
+ \big(\alpha_4 g^{ab} + \alpha_5 h^{ab}\big)\nabla_a \nabla_b \delta f + \mu \alpha_6 \delta K + \alpha_7 n^a \nabla_a \delta K + \alpha_8 \delta K^2 \nonumber\\
&&+ \alpha_9 \delta \big(K_{ab} K^{ab}\big)+ \alpha_{10} h^{ab} \delta R_{ab} \Big] \mathcal{R} = 0, \nonumber\\[10pt]
&& \nabla_a \Big(\mathcal{R}^2 \big[g^{ab} - \alpha_1 h^{ab}\big] \nabla_b \zeta\Big)=0.
\eea

\subsection{Background equations revisited}\label{sec.background.eft}
The EFT action that was worked out in Sec.\ref{sec.action} subsumes the original action of Sec. \ref{sec.intro-model}. Due to the dark energy parametrization, the effective action's dependence on the dark energy field is partly disguised in our choices for the EFT functions $\alpha$, $\alpha_i$, and $\beta_i$. This dark energy parametrization significantly simplifies the study of the background dynamics. We now use the results
of the previous subsection to derive the effective background equations.

The Einstein equations \eqref{eft.einstein} evaluated on the flat FRW
background reduce to the following two Friedmann equations
\bea \label{eft.friedman}
&&\frac{1}{a^2} \mathcal{G}_{\eta \eta,0} = \frac{1}{a^2} T_{\eta \eta,0}^{\text{DE}} + \frac{1}{a^2} T_{\eta \eta,0} ^{\text{DM}} +  \frac{1}{a^2} T_{\eta \eta,0} ^{\text{int}} = \rho^{\text{DE}} _0 + \rho_0 ^{\text{DM}}+ \rho^{\text{int}}_0\nonumber\\
&& \Rightarrow e^{\alpha} m_p ^2 \bigg( 3 \frac{\mathcal{H}^2}{a^2} + \frac{3\alpha ^{\prime 2}}{4a^2}
+ 3 \frac{\mathcal{H} \alpha'}{a^2} \bigg) =  e^{2 \alpha}  \Big(\frac{e^{-\alpha}}{2}\Lambda_0 + \tilde{\Lambda}\Big)
+ \frac{\mathcal{R}_0 ^{\prime 2}}{a^2} + \mathcal{R}_0 ^2 \frac{\zeta_0 ^{\prime 2}}{a^2} -  \frac{6\mathcal{H}}{a^2}\big(\tilde{\gamma} \mathcal{R}_0 ^2 \big)'
 \nonumber\\
&&+ \bigg(\pm \tilde{m}^2+ \tilde{\lambda} \mathcal{R}_0 ^2 -  \frac{6\mathcal{H}^2}{a^2} \tilde{\gamma} \bigg)\mathcal{R}_0 ^2, \nonumber\\[10pt]
&&\frac{1}{a^2}  \mathcal{G}_{i i,0} = \frac{1}{a^2}  T_{ii,0}^{\text{DE}} + \frac{1}{a^2} T_{ii,0} ^{\text{DM}}+ \frac{1}{a^2} T_{ii,0} ^{\text{int}} = p_0^{\text{DE}} + p_0^{\text{DM}}+ p_0^{\text{int}}\nonumber\\
&&\Rightarrow  e^{\alpha} m_p ^2 \bigg( \frac{\mathcal{H}^2}{a^2} - \frac{2 a''}{a^3}
- \frac{\alpha ''}{a^2} - \frac{\mathcal{H} \alpha'}{a^2}- \frac{\alpha^{\prime 2}}{4 a^2} \bigg) = e^{2 \alpha}  \Big(\frac{e^{-\alpha}}{2}\Lambda_0 - \tilde{\Lambda}\Big) +\frac{\mathcal{R}_0 ^{\prime 2}}{a^2} + \mathcal{R}_0 ^2 \frac{\zeta_0 ^{\prime 2}}{a^2} + \frac{2}{a^2} \big( \tilde{\gamma} \mathcal{R}_0 ^2 \big)'' \nonumber\\
&& + \frac{2 \mathcal{H}}{a^2} \big( \tilde{\gamma} \mathcal{R}_0 ^2 \big)'
+ \bigg(\mp \tilde{m}^2 - \tilde{\lambda} \mathcal{R}_0 ^2  + \Big[4 \frac{a''}{a^3} - 2 \frac{\mathcal{H}^2}{a^2}\Big]\tilde{\gamma} \bigg) \mathcal{R}_0 ^2,
\eea
where $i$ is a spatial index and we used
the dark matter parametrization in terms of $\mathcal{R}$ and $\zeta$ given in Eq. \eqref{def-dm}.

The background dark matter equations of motion evaluated using Eq. \eqref{eft.eqm.dm} are
\bea \label{eft.back.dm}
&& \frac{\mathcal{R}_0 ''}{\mathcal{R}_0} + 2 \mathcal{H} \frac{\mathcal{R}_0 '}{\mathcal{R}_0} - \zeta_0 ^{\prime 2} + \bigg(\pm a^2 \tilde{m}^2 +  2 a^2 \tilde{\lambda} \mathcal{R}_0 ^2 - 6 \frac{a''}{a} \tilde{\gamma}  \bigg)=0
\eea
and the same equation as in \eqref{klein-min} for $\zeta_0$.

Finally, to derive the background equation associated with the dark energy field
we expand the effective action \eqref{eft.act} to linear order in the dark energy
perturbation function $\tau$, then vary the resultant expression with respect
to $\tau$. Doing this we get
\bea \label{eft.back.de}
&&\frac{m_p ^2}{2 a} e^{\alpha} \alpha' \bigg(6 \frac{a''}{a^3} + 3 \frac{\alpha''}{a^2}+\frac{3 \alpha{\prime 2}}{2 a^2}+6 \frac{\mathcal{H} \alpha'}{a^2}\bigg) - e^{2 \alpha}\Big(\frac{e^{-\alpha}}{2 a}\Lambda_0 [\alpha' + 6 \mathcal{H}] +  \frac{\tilde{\Lambda}}{a} \Big[ 2 \alpha' + \frac{\tilde{\Lambda}'}{\tilde{\Lambda}}\Big]\Big) \nonumber\\
&&+ \Big(\mp m^2 \beta_2 ' - \lambda \beta_3 ' \mathcal{R}_0 ^2 + 6 \gamma \beta_4' \frac{a''}{a^3}  \Big) \frac{\mathcal{R}_0 ^2}{a} + 2 \mu^2 a^3 \alpha_2 \mathcal{R}_0 ^2 \Big(\frac{\alpha_2 ' }{\alpha_2} +3 \mathcal{H} + 2 \frac{\mathcal{R}_0 '}{\mathcal{R}_0}\Big)\nonumber\\
&&+2 \mu a^2 \alpha_3 \mathcal{R}_0 ^2 \Big(3 \mathcal{H}^2 + 10 \mathcal{H} \frac{\mathcal{R}_0 '}{\mathcal{R}_0} + 2 \frac{\mathcal{R}_0 ^{\prime 2}}{\mathcal{R}_0 ^2} + 5 \mathcal{H} \frac{\alpha_3 '}{\alpha_3} + 4 \frac{\mathcal{R}_0 ' \alpha_3 '}{\mathcal{R}_0 \alpha_3}  + 3 \frac{a''}{a} +  2 \frac{\mathcal{R}_0 ''}{\mathcal{R}_0} + \frac{\alpha_3 ''}{\alpha_3}\Big)\nonumber\\
&& -2 a \alpha_4 \mathcal{R}_0 ^2 \Big(6 \mathcal{H} \frac{\mathcal{R}_0 ^{\prime 2}}{\mathcal{R}_0 ^2}+12 \mathcal{H} \frac{\mathcal{R}_0 ' \alpha_4 '}{\mathcal{R}_0 \alpha_4} + 6 \frac{\mathcal{R}_0 ^{\prime 2} \alpha_4 '}{\mathcal{R}_0^2 \alpha_4}+4 \frac{a'' \mathcal{R}_0 '}{a \mathcal{R}_0} + 2 \frac{a''\alpha_4 '}{a \alpha_4}+6 \mathcal{H} \frac{\mathcal{R}_0 ''}{\mathcal{R}_0} + 6 \frac{\mathcal{R}_0 ' \mathcal{R}_0 ''}{\mathcal{R}_0 ^2}   \nonumber\\
&&+6 \frac{\alpha_4'  \mathcal{R}_0''}{\alpha_4 \mathcal{R}_0}+3 \mathcal{H} \frac{\alpha_4 ''}{\alpha_4} + 6 \frac{\mathcal{R}_0 ' \alpha_4 ''}{\mathcal{R}_0 \alpha_4} + 2 \frac{\mathcal{R}_0 '''}{\mathcal{R}_0}+ \frac{\alpha_4 '''}{\alpha_4}\Big) + 24 a \alpha_5 \mathcal{R}_0 ^2 \Big(-\mathcal{H}^3 + 2 \mathcal{H}^2 \frac{\mathcal{R}_0 '}{\mathcal{R}_0} + \mathcal{H}^2 \frac{\alpha_5 '}{\alpha_5 }\nonumber\\
&& + 2 \mathcal{H} \frac{a''}{a}\Big)=0.
\eea
As expected, the above equation is non-dynamical, rather it enforces
a constraint that is to be satisfied  by the background and the EFT functions. It can be seen from the above derivations that
we have a total of five background equations
for a total of three background functions which are the FRW scale factor $a$ and
the dark matter fields $\mathcal{R}_0$ and $\zeta_0$, as well as all the EFT functions in the action \eqref{eft.act} except $\alpha_1$,  $\alpha_6$,  $\alpha_7$,  $\alpha_8$,  $\alpha_9$, and $\alpha_{10}$.
Therefore, thirteen of fourteen EFT functions can be freely specified.

Lastly, we find it worthwhile to relate the results of our EFT formalism to the phenomenological models for interacting dark matter and dark energy, where such interactions are typically represented via the non-conservation of the dark matter stress energy tensor \cite{maartens1,wands1,wands2,wands3}.
To derive the background continuity equation for the dark matter,
we begin by noting that
\bea
&&\nabla_{a} T^{\text{DM} \ a} \ _{\eta} =  \nabla_a  \mathcal{G}^a \ _{\eta} - \nabla_{a}  T^{\text{DE} \ a} \ _{\eta} - \nabla_{a}  T^{\text{int} \ a} \ _{\eta} \ {\rm (evaluated \ on \ background)}\nonumber\\
&&\Rightarrow \rho_0^{\text{DM} \ \prime } + 3 \mathcal{H} (\rho_0^{\text{DM}} +p_0^{\text{DM}})
=-\rho^{\text{int} \ \prime } _0 - 3 \mathcal{H} (\rho^{\text{int}} _0+p^{\text{int}}_0)
- \rho^{\text{DE} \ \prime } _0 - 3 \mathcal{H} (\rho^{\text{DE}} _0+p^{\text{DE}} _0) \nonumber\\
&& -  (\mathcal{G}^{\eta} \ _{\eta,0})' - 3 \mathcal{H} ( \mathcal{G}^{\eta} \ _{\eta,0}- \mathcal{G}^{i} \ _{i,0}).
\eea
Using Eqs. \eqref{eft.friedman} and \eqref{eft.back.de} we find
\bea\label{eft.back.continuity}
&&\rho_0^{\text{DM} \ \prime }  + 3 \mathcal{H} (\rho_0^{\text{DM}} +p_0^{\text{DM}})=
-2 \mathcal{R}_0 \mathcal{R}_0 ' \Big(\pm m^2 \beta_2 + 2 \lambda \beta_3 \mathcal{R}_0 ^2 - 6 \tilde{\gamma} \frac{a''}{a^3}\Big)-2 \mu^2 a^4 \alpha_2 \mathcal{R}_0 ^2 \Big(\frac{\alpha_2 ' }{\alpha_2} +3 \mathcal{H}\nonumber\\
&& + 2 \frac{\mathcal{R}_0 '}{\mathcal{R}_0}\Big)-2 \mu a^3 \alpha_3 \mathcal{R}_0 ^2 \Big(3 \mathcal{H}^2 + 10 \mathcal{H} \frac{\mathcal{R}_0 '}{\mathcal{R}_0} + 2 \frac{\mathcal{R}_0 ^{\prime 2}}{\mathcal{R}_0 ^2} + 5 \mathcal{H} \frac{\alpha_3 '}{\alpha_3} + 4 \frac{\mathcal{R}_0 ' \alpha_3 '}{\mathcal{R}_0 \alpha_3}  + 3 \frac{a''}{a} +  2 \frac{\mathcal{R}_0 ''}{\mathcal{R}_0} + \frac{\alpha_3 ''}{\alpha_3}\Big)\nonumber\\
&& +2 a^2 \alpha_4 \mathcal{R}_0 ^2 \Big(6 \mathcal{H} \frac{\mathcal{R}_0 ^{\prime 2}}{\mathcal{R}_0 ^2}+12 \mathcal{H} \frac{\mathcal{R}_0 ' \alpha_4 '}{\mathcal{R}_0 \alpha_4} + 6 \frac{\mathcal{R}_0 ^{\prime 2} \alpha_4 '}{\mathcal{R}_0^2 \alpha_4}+4 \frac{a'' \mathcal{R}_0 '}{a \mathcal{R}_0} + 2 \frac{a''\alpha_4 '}{a \alpha_4}+6 \mathcal{H} \frac{\mathcal{R}_0 ''}{\mathcal{R}_0} + 6 \frac{\mathcal{R}_0 ' \mathcal{R}_0 ''}{\mathcal{R}_0 ^2}   \nonumber\\
&&+6 \frac{\alpha_4'  \mathcal{R}_0''}{\alpha_4 \mathcal{R}_0}+3 \mathcal{H} \frac{\alpha_4 ''}{\alpha_4} + 6 \frac{\mathcal{R}_0 ' \alpha_4 ''}{\mathcal{R}_0 \alpha_4} + 2 \frac{\mathcal{R}_0 '''}{\mathcal{R}_0}+ \frac{\alpha_4 '''}{\alpha_4}\Big) - 24 a^2 \alpha_5 \mathcal{R}_0 ^2 \Big(-\mathcal{H}^3 + 2 \mathcal{H}^2 \frac{\mathcal{R}_0 '}{\mathcal{R}_0} + \mathcal{H}^2 \frac{\alpha_5 '}{\alpha_5 }\nonumber\\
&& + 2 \mathcal{H} \frac{a''}{a}\Big).
\eea
As expected from our choice of frame in the effective action \eqref{eft.act}, the background dark matter continuity equation deviates
from the standard CDM result\footnote{Note that for CDM one has $p^{\text{DM}}_0\ll \rho^{\text{DM}}_0$. Here we did not omit pressure from the continuity equation. However, in Sec. \ref{sec.back.1} we showed that pressureless solutions can be found.} by terms that either break the WEP or are associated with non-minimal gravitational interactions.

In Ref. \cite{santos}, the current value of the right hand side
of Eq. \eqref{eft.back.continuity} is roughly parametrized
as $\xi \mathcal{H}_0 \Lambda_0$, where the CMB anisotropy maps together with the matter power spectrum are used to show that $|\xi|\ll 0.1$.
We apply this constraint individually to the terms appearing on the
right hand side of Eq. \eqref{eft.back.continuity}, upon which
we arrive at the following set of constrains:
\bea \label{cosmo-constraints}
&& |\tilde{\gamma}| \ll 0.1 \frac{m^2}{H_0 ^2}, \nonumber\\
&& |\beta_2| \ll 0.1, \ |\beta_3| \ll 0.1\frac{m^4}{\lambda \Lambda_0}, \nonumber\\
&& |\alpha_2|\ll 0.1 \frac{m^2}{\mu^2}, \ |\alpha_3|\ll 0.1 \frac{m^2}{\mu H_0}, \ |\alpha_4|, |\alpha_5| \ll 0.1 \frac{m^2}{H_0 ^2},
\eea
where we estimated $m^2 \mathcal{R}_0 ^2 \sim \Lambda_0$, $\partial_{\eta}^n\mathcal{R}_0 \sim \mathcal{H}^n \mathcal{R}_0$, $\partial^n _{\eta}\alpha_i \sim \mathcal{H}^n \alpha_i$, and set $a_0=1$. Of the above constraints, only the ones on $\beta_2$ and possibly $\alpha_2$ are stringent.
Indeed, following the arguments provided in Sec. \ref{sec.scale},
we have
\be
H_0 \ll m, \ \ \frac{m^4}{\lambda \Lambda_0} \gg 1,
\ee
which implies that the above upper bounds on $\tilde{\gamma}$, $\beta_3$, $\alpha_4$, and $\alpha_5$ are large compared to unity. Furthermore, as we argued
in Sec. \ref{sec.scale}, we require $\mu \lesssim m_{\phi}$. Given that a small hierarchy between $m$ and $m_{\phi}$ is preferred based
on the fine tuning arguments previously outlined, we can have $\mu \lesssim m$. This latter
implies that while the above upper bound on $\alpha_3$ is not stringent at all, the upper bound on $\alpha_2$ could be somewhat stringent if $m^2 / \mu^2 \lesssim 10$.

\subsection{Linear perturbation equations in the Newtonian gauge} \label{sec.eqm.scalar}

We now derive the linearly perturbed Einstein equations \eqref{eft.einstein}, the dark matter equations \eqref{eft.eqm.dm.r&z}, and the dark energy equation in the Newtonian gauge. The latter is derived by expanding the EFT action \eqref{eft.act} to second order in the scalar perturbations and then varying the resultant expression with respect to $\tau$.

We express the components of the Einstein equations in the following way:

\bea \label{einstein-newt-1}
&&\delta(\mathcal{G}_{ab} n^a n^b)=\delta(T^{\text{DE}}_{ab} n^a n^b + T^{\text{DM}}_{ab} n^a n^b + T^{\text{int}}_{ab} n^a n^b)\nonumber\\
&&\Rightarrow e^{\alpha} m_p ^2 \bigg[\Phi \Big(-6 \frac{\mathcal{H}^2}{a^2}-6\frac{\mathcal{H} \alpha'}{a^2}  -\frac{3}{2} \frac{\alpha^{\prime 2}}{a^2}\Big) + 3 \tau \Big(\frac{\mathcal{H}\alpha^{\prime 2}}{2a^3}+\frac{\alpha^{\prime 3}}{4 a^3}
+\frac{\mathcal{H} \alpha''}{a^3}+\frac{\alpha' \alpha''}{2 a^3}\Big)\nonumber\\
&&-\frac{\alpha'}{a^3}\partial^2 \tau + \frac{2}{a^2}\partial^2 \Psi
+3\Big( \frac{\mathcal{H} \alpha'}{a^3} + \frac{\alpha^{\prime 2}}{2a^3}\Big)\tau'-3\Big(2\frac{\mathcal{H}}{a^2}+\frac{\alpha'}{a^2}\Big)\Psi'\bigg] = \delta \rho^{\text{DE}}+\delta \rho^{\text{DM}} +\delta \rho^{\text{int}},\nonumber\\
&&
\eea
\bea \label{einstein-newt-2}
&&\delta(\mathcal{G}_{ab} h^{ab})=\delta(T^{\text{DE}}_{ab} h^{ab} + T^{\text{DM}}_{ab} h^{ab} + T^{\text{int}}_{ab} h^{ab})\nonumber\\
&&\Rightarrow e^{\alpha}m_p ^2 \bigg[\Phi \Big(-6\frac{\mathcal{H}^2}{a^2}
+6 \frac{\mathcal{H} \alpha'}{a^2} + \frac{3\alpha^{\prime 2}}{2a^2}
+12 \frac{a''}{a^3} + 6 \frac{\alpha''}{a^2}\Big)+3 \tau \Big(- \frac{\mathcal{H} \alpha^{\prime 2}}{2 a^3}-\frac{\alpha^{\prime 3}}{4 a^3}\nonumber\\
&&- \frac{\alpha' a''}{a^4}+\frac{\mathcal{H} \alpha''}{a^3}-\frac{3\alpha' \alpha''}{2 a^3}
-\frac{\alpha'''}{a^3}\Big)+2\frac{\alpha'}{a^3}\partial^2 \tau
+\frac{2}{a^2}\partial^2 [\Phi-\Psi]+3\Big(\frac{\mathcal{H}\alpha'}{a^3}
-\frac{\alpha^{\prime 2}}{2 a^3}-2 \frac{\alpha''}{a^3}\Big)\tau'\nonumber\\
&&+3 \Big(\frac{\alpha'}{a^2}+2 \frac{\mathcal{H}}{a^2}\Big)[\Phi'+2 \Psi']-3 \frac{\alpha'}{a^3}\tau''+\frac{6}{a^2}\Psi''\bigg] =  3 (\delta \mathfrak{p}^{\text{DE}}+\delta p^{\text{DM}}
+ \delta \mathfrak{p}^{\text{int}}),\nonumber\\
&&
\eea
\bea \label{einstein-newt-3}
&&\delta(\mathcal{G}_{ab} h^{(a} \ _{i} n^{b)})=\delta(T^{\text{DE}}_{ab} h^{(a} \ _{i} n^{b)} + T^{\text{DM}}_{ab} h^{(a} \ _{i} n^{b)} + T^{\text{int}}_{ab} h^{(a} \ _{i} n^{b)})\nonumber\\
&&\Rightarrow \frac{e^{\alpha} m_p ^2}{2}\bigg[\Big(4 \frac{\mathcal{H}^2}{a^2}-2 \frac{a''}{a^3}\Big)\partial_i \tau -\Big( \frac{2\mathcal{H}+\alpha'}{a}\Big)\partial_i \Phi+\frac{\alpha'}{a^2}\partial_i \tau' - \frac{2}{a}\partial_i \Psi' \bigg]= \Big(\partial_i \tau - \frac{a \partial_i \delta \mathcal{R}}{\mathcal{R}_0 '}\Big) [{}^{\mathcal{R}}\rho_0^{\text{DM}} \nonumber\\
&&+{}^{\mathcal{R}}p_0^{\text{DM}}+{}^{\mathcal{R}}\rho^{\text{int}} _0+{}^{\mathcal{R}}p_0^{\text{int}}]+\Big(\partial_i \tau - \frac{a \partial_i \delta \zeta}{\zeta'}\Big) [{}^{\zeta}\rho_0^{\text{DM}}+{}^{\zeta}p_0^{\text{DM}}]+\frac{1}{4}{}^{\mathcal{R}}\mathfrak{C}^{\text{int}}_{i} ,\nonumber\\
&&
\eea
\bea \label{einstein-newt-4}
&&\delta(\mathcal{G}_{ab} h^a \ _{i} h^{b} \ _{i})-\frac{1}{3}\delta(\mathcal{G}_{ab} h^{ab} h_{ii})= \delta(T^{\text{DE}}_{ab} h^a \ _{i} h^{b} \ _{i} + T^{\text{DM}}_{ab} h^a \ _{i} h^{b} \ _{i} + T^{\text{int}}_{ab} h^a \ _{i} h^{b} \ _{i})\nonumber\\
&&-\frac{1}{3}\delta(T_{ab}^{\text{DE}} h^{ab} h_{ii}+T_{ab}^{\text{DM}} h^{ab} h_{ii}+T_{ab}^{\text{int}} h^{ab} h_{ii})\nonumber\\
&&\Rightarrow \frac{e^{\alpha} m_p ^2}{6}  \bigg[\partial^2 \Big(\Phi-\Psi +\frac{\alpha'}{a}\tau\Big)-3\partial^2 _{i} \Big(\Phi-\Psi + \frac{\alpha'}{a} \tau \Big) \bigg] = {}^{\mathcal{R}}\Sigma ^{\text{int}}_{ii}, \nonumber\\[20 pt]
&&\delta(\mathcal{G}_{ab} h^a \ _{i} h^{b} \ _{j})-\frac{1}{3}\delta(\mathcal{G}_{ab} h^{ab} h_{ij})= \delta(T^{\text{DE}}_{ab} h^a \ _{i} h^{b} \ _{j} + T^{\text{DM}}_{ab} h^a \ _{i} h^{b} \ _{j} + T^{\text{int}}_{ab} h^a \ _{i} h^{b} \ _{j})\nonumber\\
&&-\frac{1}{3}\delta(T_{ab}^{\text{DE}} h^{ab} h_{ij}+T_{ab}^{\text{DM}} h^{ab} h_{ij}+T_{ab}^{\text{int}} h^{ab} h_{ij})\nonumber\\
&&\Rightarrow -\frac{e^{\alpha}m_p ^2}{2}\partial
_i \partial_j\Big[\Phi-\Psi+\frac{\alpha '}{a} \tau\Big]= {}^{\mathcal{R}}\Sigma ^{\text{int}}_{ij},
\eea
where we used Eq. \eqref{eft.einstein.tensor} and the results
of the last subsection. The functions $\delta \rho$, $\delta p$,
$\delta \mathfrak{p}$, $\mathfrak{C}_i$, and $\Sigma_{ij}$
are the energy density perturbation, the pressure perturbation,
the ``generalized pressure'' perturbation, the ``heat flow'' covector,
and the ``anisotropic stress'' tensor for the dark matter, dark energy, and interaction sectors. We have borrowed this terminology from  imperfect relativistic fluids, which we
briefly review in App. \ref{sec.imp.fluid}. The complete definition
for these quantities and their expressions in the Newtonian gauge is provided in App. \ref{sec.pert.stress}.

Similarly, we derive the linearized dark matter equations using
Eq. \eqref{eft.eqm.dm.r&z}. For the radial dark matter perturbation
$\delta \mathcal{R}$ we find
\bea \label{dm-newt-r}
&&-\frac{1}{a^2} (\partial_{\eta} ^2-[1-\alpha_1] \partial^2)\delta \mathcal{R}-\frac{\mathcal{H}}{a^2}\delta \mathcal{R}'(2+9 \alpha_1)
-\delta \mathcal{R} \Big(\pm \tilde{m}^2 + 6 \tilde{\lambda} \mathcal{R}_0 ^2 - 6 \tilde{\gamma} \frac{a''}{a^3} - \frac{\zeta_0 ^{\prime 2}}{a^2}\Big)\nonumber\\
&&+2 \frac{\mathcal{R}_0\zeta_0 '}{a^2} \delta \zeta' + \Phi \Big(2 \mu^2 \mathcal{R}_0 + 6 \frac{\mathcal{H}^2}{a^2}\mathcal{R}_0 [\alpha_7 - 3 \alpha_8 - \alpha_9 - \alpha_{10}]+3\frac{\mathcal{H}}{a} \mu \alpha_6 \mathcal{R}_0 - 2 \mathcal{R}_0 \frac{\zeta_0 ^{\prime 2}}{a^2} \nonumber\\
&&+2 \frac{\mathcal{R}_0 ''}{a^2}+2 \frac{\mathcal{H} \mathcal{R}_0 '}{a^2} (2+9 \alpha_1)-3 \mathcal{R}_0 \frac{a''}{a^3}[4 \tilde{\gamma}+\alpha_7 + 2 \alpha_{10}]\Big)- 6\alpha_{10} \mathcal{R}_0 \Big(\frac{\mathcal{H}^2}{a^2}+\frac{a''}{a^3}\Big)\Psi \nonumber\\
&& +\frac{\tau}{a}\Big(\mp m^2 \mathcal{R}_0 \beta_2 ' - 2 \lambda \mathcal{R}_0 ^3 \beta_3 ' -9 \mathcal{R}_0 '\alpha_1 ' \frac{\mathcal{H}}{a^2} + 6 \gamma \mathcal{R}_0 \beta_4 ' \frac{a''}{a^3}\Big)+ \frac{\Phi'}{a} \Big(-2 \mu \alpha_3 \mathcal{R}_0+\frac{\mathcal{R}_0 '}{a} \nonumber\\
&&- \frac{\mathcal{H}}{a} \mathcal{R}_0 [6 \tilde{\gamma}  + 4 \alpha_4  +6 \alpha_5 +3 \alpha_7 +3 \alpha_{10}]\Big)+ \frac{\mathcal{R}_0}{a}\Psi' \Big(3 \mu \alpha_6
+3 [1+3 \alpha_1]  \frac{\mathcal{R}_0 '}{a \mathcal{R}_0}\nonumber\\
&& +\frac{\mathcal{H}}{a}[-18 \tilde{\gamma}+3 \alpha_7  -18 \alpha_8 -6 \alpha_9 -15 \alpha_{10}]\Big)+\frac{\mathcal{R}_0}{a}\tau'\Big(-6 \frac{\mathcal{H}^2}{a^2}\alpha_5 - 2 \mu ^2 \alpha_2 - 2 \alpha_4 \frac{a''}{a^3} \Big) \nonumber\\
&&+\frac{\mathcal{R}_0}{a^2}\tau'' (2 \mu \alpha_3 + 6 \alpha_5)-2 \frac{\mathcal{R}_0}{a^2}\alpha_4 \Phi'' -3 \frac{\mathcal{R}_0}{a^2}(2 \tilde{\gamma}+\alpha_7+\alpha_{10})\Psi'' +\frac{\mathcal{R}_0}{a^2}\partial^2 \tau \Big(\alpha_1 \frac{\mathcal{R}_0 '}{a\mathcal{R}_0} \nonumber\\
&&+\mu \alpha_6 +2 \alpha_7- 6 \alpha_8 - 2\alpha_9 \Big)  -\frac{\mathcal{R}_0}{a^2} \partial^2 \Phi (2 \tilde{\gamma} -2 \alpha_4 - 2 \alpha_5 + \alpha_{10} ) + 4 \frac{\mathcal{R}_0}{a^2}\partial^2 \Psi (\tilde{\gamma}+\alpha_{10})\nonumber\\
&& + 2 \frac{\mathcal{R}_0}{a^3}\alpha_4 \tau''' - \frac{\mathcal{R}_0}{a^3}\partial^2 \tau' (2 \alpha_4 + 2 \alpha_5 + \alpha_7) =0,\nonumber\\
\eea
and for the angular dark matter perturbation $\delta \zeta$ we find
\bea \label{dm-newt-z}
&&-\frac{\mathcal{R}_0 ^2}{a^2}(\partial^2 _{\eta}-[1-\alpha_1]\partial^2)\delta \zeta - 2 \frac{\mathcal{R}_0 ^2}{a^2}\Big(\mathcal{H}+\frac{\mathcal{R}_0 '}{\mathcal{R}_0}\Big)\delta \zeta ' -2 \frac{\mathcal{R}_0 \zeta_0 '}{a^2}\delta \mathcal{R}'+2\frac{\mathcal{R}'_0 \zeta_0 '}{a^2}\delta \mathcal{R} + \frac{\mathcal{R}_0 ^2 \zeta_0 '}{a^2} (\Phi'+3 \Psi') \nonumber\\
&&+\frac{\mathcal{R}_0 ^2 \alpha_1 \zeta_0 '}{a^3}\partial^2 \tau=0, \nonumber\\
\eea
where we used Eq. \eqref{zeta}.

Finally, we derive the linear perturbation equation for the dark energy field by varying the perturbed effective action with respect to $\tau$. In doing so, for simplicity and due to our focus being on the dark energy interactions with dark matter, we only keep the second order perturbation terms in $S_{\text{DM-int}}$ that depend on the dark matter perturbations. Other terms that are either quadratic in $\tau$ or the metric perturbations, or are products of $\tau$ and the metric perturbations, are discarded as they would be generated by the EFT of dark energy. Doing this we find

\bea \label{de-newt}
&&4\frac{e^{\alpha}}{a^2}\Big(2 \Lambda_0 - \frac{3 m_p ^2 \alpha^{\prime 2}}{a^2}\Big)[\tau''-\partial^2 \tau]+4 \frac{e^{\alpha}}{a^2} \Big(2 \Lambda_0 [2\mathcal{H} +  \alpha']-3 \frac{m_p ^2}{a^2}\Big[\alpha^{\prime 3}+2\alpha' \alpha''\Big] \Big)\tau' \nonumber\\
&&+2 \frac{e^{\alpha}}{a^2} \Big(\Lambda_0\big[10 \mathcal{H} \alpha' + 2 \alpha^{\prime 2}+2 \alpha '' -4 e^{\alpha}\{ \mathcal{H} \beta_1 ' -4 \alpha' \beta_1 '-\beta_1 ''\}\big]-8 e^{\alpha}\tilde{\Lambda}\big[\mathcal{H} \alpha' - 2 \alpha^{\prime 2}-\alpha''\big] \nonumber\\
&&+3\frac{m_p ^2}{a^2}\Big[4 \mathcal{H}^2 \alpha^{\prime 2}-\mathcal{H} \alpha^{\prime 3}-\alpha^{\prime 4}+4 \mathcal{H} \alpha' \frac{a''}{a}
-2 \alpha^{\prime 2}\frac{a''}{a}-2\mathcal{H} \alpha' \alpha ''-5 \alpha^{\prime 2}\alpha '' -4 \alpha '' \frac{a''}{a}-2 \alpha ^{\prime \prime 2} \nonumber\\
&&-2 \alpha ' \alpha ''' \Big]\Big)\tau + \frac{e^{\alpha}}{a}\Big(\Lambda_0 [-24 \mathcal{H} - 4 \alpha']+16 \tilde{\Lambda}e^{\alpha} \alpha' + 8 e^{\alpha}\Lambda_0 \beta_1 '
+ 6 \frac{m_p ^2}{a^2}\Big[2 \mathcal{H} \alpha^{\prime 2} + \alpha^{\prime 3} + 2 \alpha' \frac{a''}{a} \nonumber\\
&& + 4 \alpha' \alpha '' +2 \alpha '''\Big]\Big)\Phi+ 3\frac{e^{\alpha}}{a}\Big(\Lambda_0 [-24 \mathcal{H} - 4 \alpha']-16 \tilde{\Lambda}e^{\alpha} \alpha' - 8 e^{\alpha}\Lambda_0 \beta_1 '
+ 6 \frac{m_p ^2}{a^2}\Big[2 \mathcal{H} \alpha^{\prime 2} + \alpha^{\prime 3} \nonumber\\
&&+ 2 \alpha' \frac{a''}{a}+ 4 \alpha' \alpha '' +2 \alpha '''\Big]\Big)\Psi+ \frac{32\mathcal{R}_0}{a}\bigg(\mu^2 \alpha_2\Big\{-3 \mathcal{H}-\frac{\mathcal{R}_0 '}{\mathcal{R}_0}-\frac{\alpha_2 '}{\alpha_2}\Big\}+\mu \alpha_3\Big\{-3\frac{\mathcal{H}^2}{a}-5\frac{\mathcal{H} \mathcal{R}_0 '}{a \mathcal{R}_0}\nonumber\\
&&-5 \frac{\mathcal{H}\alpha_3'}{a \alpha_3}-2\frac{\mathcal{R}_0 '\alpha_3'}{a\mathcal{R}_0 \alpha_3}-3\frac{a''}{a^2}-\frac{\mathcal{R}_0 ''}{a \mathcal{R}_0}-\frac{\alpha_3 ''}{a \alpha_3}\Big\}+\alpha_4 \Big\{ 2\frac{a'' \mathcal{R}_0 '}{a^3 \mathcal{R}_0}+3 \frac{\mathcal{H} \mathcal{R}_0 ''}{a^2 \mathcal{R}_0}+\frac{\mathcal{R}_0 '''}{a^2 \mathcal{R}_0}+\frac{6 \mathcal{H} \mathcal{R}_0 ' \alpha_4 '}{a^2 \mathcal{R}_0 \alpha_4}\nonumber\\
&&+2\frac{a'' \alpha_4 '}{a^3 \alpha_4}+3 \frac{\mathcal{R}_0 '' \alpha_4 '}{a^2\mathcal{R}_0 \alpha_4}+3 \frac{\mathcal{H} \alpha_4 ''}{a^2 \alpha_4}+3 \frac{\mathcal{R}_0 ' \alpha_4 ''}{a^2 \mathcal{R}_0 \alpha_4}+\frac{\alpha_4 '''}{a^2 \alpha_4}\Big\}+\alpha_5\Big\{3 \frac{\mathcal{H}^3}{a^2}-3\frac{\mathcal{H}^2 \mathcal{R}_0 '}{a^2 \mathcal{R}_0}-6\frac{\mathcal{H} a''}{a^3}-6\frac{a'' \mathcal{R}_0 '}{a^3 \mathcal{R}_0}\nonumber\\
&&-3\frac{\mathcal{H} \mathcal{R}_0 ''}{a^2 \mathcal{R}_0}-3\frac{a'''}{a^3}-3\frac{\mathcal{H}^2 \alpha_5 '}{a^2 \alpha_5}-6\frac{\mathcal{H} \mathcal{R}_0 ' \alpha_5 '}{a^2 \mathcal{R}_0 \alpha_5}-6 \frac{a'' \alpha_5 '}{a^3 \alpha_5}-3\frac{\mathcal{H} \alpha_5 ''}{a^2 \alpha_5}\Big\}+\frac{m^2 \beta_2 '}{2}+\lambda \beta_3 ' \mathcal{R}_0 ^2 - 3 \gamma \beta_4 ' \frac{a''}{a^3}\bigg)\delta \mathcal{R} \nonumber\\
&& + \frac{16\mathcal{R}_0}{a^2}\bigg(-\alpha_1 \frac{\mathcal{R}_0 '}{a \mathcal{R}_0}+\alpha_4\Big\{-2\frac{\mathcal{H}}{a}-2\frac{\mathcal{R}_0 '}{a \mathcal{R}_0}-2\frac{\alpha_4'}
{a\alpha_4}\Big\}+\alpha_5\Big\{-2\frac{\mathcal{H}}{a}-2\frac{\mathcal{R}_0 '}{a \mathcal{R}_0}-2\frac{\alpha_5'}{a \alpha_5} \Big\}-\mu \alpha_6 \nonumber\\
&& -2 \alpha_7 \frac{\mathcal{H}}{a}+6 \alpha_8 \frac{\mathcal{H}}{a}+\alpha_9\Big\{ \frac{\mathcal{H}}{a} -  \frac{\mathcal{R}_0 '}{a \mathcal{R}_0}-\frac{\alpha_9 '}{a \alpha_9}\Big\}\bigg)\partial^2 \delta \mathcal{R}-16 \alpha_1 \frac{\mathcal{R}_0 ^2 \zeta_0 '}{a^3}\partial^2 \delta \zeta+8 e^{\alpha}m_p ^2 \frac{\alpha'}{a^3}\partial^2[\Phi-3\Psi] \nonumber\\
&&+32 \frac{\mathcal{R}_0}{a}\bigg(-\mu^2 \alpha_2 + \mu \alpha_3\Big\{-5 \frac{\mathcal{H}}{a}-2\frac{\mathcal{R}_0 '}{a \mathcal{R}_0}-2\frac{\alpha_3 '}{a \alpha_3}\Big\}+ \alpha_4 \Big\{6 \mathcal{H} \frac{\mathcal{R}_0'}{a^2 \mathcal{R}_0}+2\frac{a''}{a^3}+3\frac{\mathcal{R}_0''}{a^2 \mathcal{R}_0}+6 \frac{\mathcal{H} \alpha_4 '}{a^2 \alpha_4} \nonumber\\
&&+6\frac{\mathcal{R}_0 ' \alpha_4 '}{a^2 \mathcal{R}_0 \alpha_4}+3\frac{\alpha_4''}{a^2 \alpha_4}\Big\}+\alpha_5 \Big\{ - 3  \frac{\mathcal{H}^2}{a^2} -6\frac{\mathcal{H} \mathcal{R}_0 '}{a^2 \mathcal{R}_0}-6\frac{a''}{a^3}-6\frac{\mathcal{H} \alpha_5 '}{a^2 \alpha_5}\Big\} \bigg) \delta \mathcal{R}'+ \frac{e^{\alpha}}{a}\Big(-8 \Lambda_0 + 12 \frac{m_p ^2}{a^2} \nonumber\\
&&\times [\mathcal{H} \alpha' + \alpha^{\prime 2} + \alpha'']\Big) (\Phi'+3 \Psi')-16 \frac{\mathcal{R}_0}{a^3}(2 \alpha_4 + 2 \alpha_5 +  \alpha_9) \partial^2 \delta \mathcal{R}' +32 \frac{\mathcal{R}_0}{a^2}\bigg(-\mu \alpha_3 + \alpha_4 \Big\{3\frac{\mathcal{H}}{a}\nonumber\\
&&+3\frac{\mathcal{R}_0'}{a \mathcal{R}_0}+3\frac{\alpha_4 '}{a \alpha_4}\Big\} -3 \alpha_5 \frac{\mathcal{H}}{a}\bigg) \delta \mathcal{R}''+24 e^{\alpha} m_p ^2 \frac{\alpha'}{a^3}\Psi'' +32 \alpha_4 \frac{\mathcal{R}_0}{a^3}\delta \mathcal{R}'''=0.
\eea
We will be using these equations in the next section to compute
the effective Newton's constant for dark matter in this EFT.

\section{Effective Newton's constant for the weak equivalence principle violating dark matter} \label{sec.observables}

In order to compute the effective Newton's constant $G^{\text{DM}}_{\text{eff}}$
for dark matter, we first take the Newtonian limit of the Einstein equations \eqref{einstein-newt-1} and \eqref{einstein-newt-2}, the dark energy equation of motion \eqref{de-newt}, as well as the angular dark matter equation of motion \eqref{dm-newt-z} by omitting all
terms that include a perturbation functions with conformal time derivatives
\footnote{In the Newtonian limit, the gravitational potentials
$\Phi$ and $\Psi$ are assumed to be nearly time independent. The situation for the dark matter perturbation functions is less obvious. However, a numerical study shows that terms like $\delta \mathcal{R}''$ and $\delta \zeta''$ are suppressed compared to terms $k^2 \delta \mathcal{R}$ and $k^2 \delta \zeta$.} and dropping all background function of dimension two \footnote{e.g. $\mathcal{H}^2, a''/a,\mathcal{R}_0 ''/\mathcal{R}_0,...$}
in favour of $k^2$, where $k$ is the wavenumber of a given mode. We also disregard all components of the stress energy tensors
for the dark energy, dark matter, and interactions besides the energy
densities. In addition, we make two simplifying assumptions regarding the EFT functions, namely that $\{\alpha, \beta_1\} \rightarrow 0$ and  all terms involving the EFT functions with two or more conformal time derivatives are vanishingly small. The former is assumed in order to disregard the Universal corrections to the Newton's constant \footnote{i.e. corrections that are not specific to dark matter.}, and the latter is done assuming that the EFT functions are slowly varying with time \footnote{in consistency with approximate de Sitter invariance of the spacetime.}.

After implementing the above approximations, the Einstein equations \eqref{einstein-newt-1} and \eqref{einstein-newt-2}, the dark energy equation of motion \eqref{de-newt}, and the angular dark matter equation of motion \eqref{dm-newt-z} reduce to (we set $\partial^2 \rightarrow - k^2$)
\bea \label{eqm-newtonian}
&& m_p ^2 \bigg[-6 \frac{\mathcal{H}^2}{a^2}{}^{\text{N}}\Phi - \frac{2 k^2}{a^2}{}^{\text{N}}\Psi \bigg] \approx {}^{\text{N}}\delta \rho^{\text{DE}}+{}^{\text{N}}\delta \rho^{\text{DM}} +{}^{\text{N}}\delta \rho^{\text{int}},\nonumber\\
&&{}^{\rm N} \Phi \approx {}^{\rm N} \Psi,\nonumber\\
&&\frac{8 \Lambda_0 k^2}{a^2}{}^{\text{N}}\tau - \frac{24\Lambda_0 \mathcal{H}}{a}({}^{\text{N}}\Phi+ 3 {}^{\text{N}}\Psi)+ \frac{32\mathcal{R}_0}{a}\bigg[\frac{\mathcal{R}_0' k^2}{2a^2 \mathcal{R}_0}\alpha_1 +\mu^2 \alpha_2\Big\{-3 \mathcal{H}-\frac{\mathcal{R}_0 '}{\mathcal{R}_0}-\frac{\alpha_2 '}{\alpha_2}\Big\}+\mu \alpha_3 \nonumber\\
&&\times\Big\{-3\frac{\mathcal{H}^2}{a}-5\frac{\mathcal{H} \mathcal{R}_0 '}{a \mathcal{R}_0}-5 \frac{\mathcal{H}\alpha_3'}{a \alpha_3}-2\frac{\mathcal{R}_0 '\alpha_3'}{a\mathcal{R}_0 \alpha_3}-3\frac{a''}{a^2}-\frac{\mathcal{R}_0 ''}{a \mathcal{R}_0}\Big\} +\frac{k^2}{a^2}\alpha_4 \Big\{\mathcal{H}+ \frac{\mathcal{R}_0'}{\mathcal{R}_0}+\frac{\alpha_4 '}{\alpha_4}\Big\}\nonumber\\
&&+\frac{k^2}{a^2}\alpha_5\Big\{\mathcal{H}+\frac{\mathcal{R}_0 '}{\mathcal{R}_0}+\frac{\alpha_5 '}{ \alpha_5}\Big\}+\frac{k^2}{2a}\mu \alpha_6+\frac{k^2\mathcal{H}}{a^2}\alpha_7-3\frac{k^2 \mathcal{H}}{a^2}\alpha_8- \frac{ k^2}{2a^2}\alpha_9\Big\{ \mathcal{H} -  \frac{\mathcal{R}_0 '}{ \mathcal{R}_0}-\frac{\alpha_9 '}{ \alpha_9}\Big\} \nonumber\\
&&\pm\frac{m^2 \beta_2 '}{2}+\lambda \beta_3 ' \mathcal{R}_0 ^2 - 3 \gamma \beta_4 ' \frac{a''}{a^3} \bigg]{}^{\text{N}}\delta \mathcal{R}+16 \alpha_1 \frac{\mathcal{R}_0 ^2 \zeta_0 ' k^2}{a^3}{}^{\text{N}}\delta \zeta \approx 0, \nonumber\\
&&-\frac{\mathcal{R}_0 ^2 k^2}{a^2}[1-\alpha_1]\ {}^{\text{N}}\delta \zeta -2\frac{\mathcal{R}_0}{a^2}\Big(2 \mathcal{H} \zeta_0 '+ \frac{\mathcal{R}_0 ' \zeta_0 '}{\mathcal{R}_0} + \zeta_0 '' \Big) {}^{\text{N}}\delta \mathcal{R}-\frac{\mathcal{R}_0 ^2 k^2 \alpha_1 \zeta_0 '}{a^3} \ {}^{\text{N}}\tau \approx 0,
\eea
where the left superscript ``N'' on the perturbation functions
stands for ``Newtonian''. In this limit, the perturbed energy densities for the dark energy, dark matter, and interactions become
\bea \label{set-newtonian}
&& {}^{\text{N}}\delta \rho^{\text{DE}} \approx -\Lambda_0 {}^{\rm N}\Phi,\nonumber\\
&&{}^{\text{N}}\delta \rho^{\text{DM}}={}^{\text{N},\mathcal{R}}\delta \rho^{\text{DM}}+{}^{\text{N},\zeta}\delta \rho^{\text{DM}} \approx
2 \mathcal{R}_0 {}^{\rm N}\delta \mathcal{R}\Big[\pm m^2 + 2 \lambda \mathcal{R}_0 ^2 + \frac{\zeta_0 ^{\prime 2}}{a^2}\Big] -2 \Big[\frac{\mathcal{R}_0 ^{\prime 2}}{a^2}+\mathcal{R}_0 ^2 \frac{\zeta_0 ^{\prime 2}}{a^2}\Big]{}^{\rm N}\Phi,\nonumber\\
&&{}^{\text{N}}\delta \rho^{\text{int}} \approx 2 \mathcal{R}_0 \bigg[\pm m^2 \beta_2+ 2 \lambda \beta_3 \mathcal{R}_0 ^2-\frac{2 k^2}{a^2} \tilde{\gamma} \bigg] {}^{\rm N}\delta \mathcal{R} + \mathcal{R}_0 ^2  \bigg[\frac{12}{a^2} \tilde{\gamma} \Big( \mathcal{H}^2 + 2\frac{\mathcal{H} \mathcal{R}_0 '}{ \mathcal{R}_0} +\frac{\mathcal{H}\tilde{\gamma}'}{\tilde{\gamma}}\Big)  -2 \mu^2 \alpha_2 \nonumber\\
&&  +2\frac{k^2}{a^2}\alpha_5-\frac{6}{a} \mu \mathcal{H}\alpha_6 + \alpha_7 \Big(-\frac{6 \mathcal{H}^2}{a^2}+\frac{3 a''}{a^3}\Big)+\frac{6\mathcal{H}^2}{a^2}(3\alpha_8+\alpha_9)+ 6 \alpha_{10} \Big(\frac{\mathcal{H}^2}{a^2}+\frac{a''}{a^3}\Big) \bigg] {}^{\rm N}\Phi\nonumber\\
&& +2 \frac{\mathcal{R}_0 ^2 k^2}{a^2} (2\tilde{\gamma}+3\alpha_{10}) {}^{\rm N}\Psi+\frac{\mathcal{R}_0 ^2}{a}  \bigg[\pm m^2 \beta_2 ' +\lambda \beta_3 ' \mathcal{R}_0 ^2  -\frac{k^2}{a^2} \Big\{2\tilde{\gamma}' + 4 \alpha_7 \Big(\frac{\mathcal{R}_0 '}{ \mathcal{R}_0}+\mathcal{H}+\frac{\alpha_7 '}{\alpha_7}\Big) \nonumber\\
&&-2 \mathcal{H}(3\alpha_8+\alpha_9 )
\Big\}\bigg]  {}^{\rm N}\tau,
\eea
where we used Eqs. \eqref{eft.set.imp.de}, \eqref{eft.set.imp.zeta},   \eqref{eft.set.imp.r}, and \eqref{eft.rho.imp.int}.

Using the above equations  we find
the following Poisson equation for the dark matter perturbations,
\be \label{dm-poisson}
-\frac{k^2}{a^2}{}^{\rm N}\Psi = 4 \pi G^{\text{DM}}_{\text{eff}}\ {}^{\text{N}}\delta \rho^{\text{DM}},
\ee
where the effective Newton's constant $G^{\text{DM}}_{\text{eff}}$
is given by
\be \label{geff}
\frac{G^{\text{DM}}_{\text{eff}}}{G}= 1+ \frac{\Delta_{\gamma} + \Delta_{\gamma = 0}}{m^2 + \frac{\zeta_0 ^{\prime 2}}{a^2}+2 \lambda \mathcal{R}_0 ^2} + \mathcal{O}(2).
\ee
Here $\mathcal{O}(2)$ stands for terms that are second order
or higher in the EFT functions, and we defined
\bea \label{I}
&& \Delta_{\gamma}  = -2 \frac{k^2}{a^2} \tilde{\gamma} + \mathcal{O}(k^0),\nonumber\\
&& \Delta_{\gamma=0} = m^2 \beta_2 + 2 \mathcal{R}_0 ^2 \lambda \beta_3
-8 \pi G \mathcal{R}_0 ^2 \bigg(m^2 + \frac{\zeta_0 ^{\prime 2}}{a^2}+2 \lambda \mathcal{R}_0 ^2\bigg) (\alpha_5 + 3 \alpha_{10}) + \mathcal{O}(k^{-2}),
\eea
where $\Delta_{\gamma}$ vanishes when $\gamma\rightarrow 0$ and
$\Delta_{\gamma=0}$ has no dependence on $\gamma$.
The above linearization in the EFT functions is based on the assumption that the violation of the WEP and the non-minimal coupling between the spacetime curvature and the dark matter fields are small effects that can be incorporated as perturbations to the standard minimally coupled dark matter. This assumption is reasonable if
dark matter has similar gravitational / dark energy interactions to baryonic matter.

As can be seen from Eq. \eqref{I}, only five of the nearly dozen EFT operators introduced in Sec. \ref{sec.operators} contribute significantly in the Newtonian limit. We now use the results of
the analysis performed in Ref. \cite{yangbai}
to derive individual constraints on the five EFT functions $\tilde{\gamma},\beta_2,\beta_3,\alpha_5, \alpha_{10}$. \footnote{In  Ref. \cite{yangbai}, the authors use the induced changes to the location of the acoustic peaks of the CMB anisotropy spectra to constrain the difference between the inertial and gravitational masses of dark matter. Note that the constraints are valid for the modes with physical wavelengths $\lambda_{\rm phys} \gtrsim \mathcal{O}(1)$Mpc.}
From Ref. \cite{yangbai} we have
\be
\frac{G_{\rm eff} ^{\rm DM}}{G} - 1 \lesssim 10^{-5},
\ee
which gives
\bea \label{constraints-eft-func}
&& |\tilde{\gamma}|\lesssim \frac{m^2}{k_{\rm phys} ^2} 10^{-5},\nonumber\\
&&|\beta_2|\lesssim 10^{-5}, \hspace{1cm} |\beta_3| \lesssim \frac{m^4}{\Lambda_0 \lambda} 10^{-5}, \nonumber\\
&&|\alpha_5|,|\alpha_{10}| \lesssim \frac{m^2}{H_0 ^2} 10^{-5}.
\eea
Here $k_{\rm phys}$ denotes the current value of the wavenumber $k$ and we used the approximations $m^2 + \zeta^{\prime 2}/a^2 + 2 \lambda \mathcal{R}_0 ^2 \sim m^2$ and $m^2 \mathcal{R}_0 ^2 \sim \Lambda_0$.
Given that $m \gtrsim 10^{-24} {\rm eV}$ and $k_{\rm phys} \lesssim 10^{-30} {\rm eV}$ for the CMB modes, the functions $\tilde{\gamma}$,
$\alpha_5$, and $\alpha_{10}$ are not strongly constrained. On the other hand, the above constraint on $\beta_2$ is quite strong. Likewise, unless $m^4/(\lambda \Lambda_0) \gtrsim 10^5$, $\beta_3$ is tightly bounded. Note that the constraints given in Eq. \eqref{constraints-eft-func}
 on $\tilde{\gamma}$,
$\beta_2$, $\beta_3$, and $\alpha_5$ are far more stringent than
the ones obtained from Eq. \eqref{cosmo-constraints}.

\section{Concluding remarks}

Our main objective in this paper was to incorporate dark matter in the EFT of dark energy. Unfortunately, our current understanding of dark matter physics does not provide us with a preferred model for dark matter. We chose to model dark matter using a complex scalar field, the coherent excitations of which corresponds to the usual notion of dark matter particles. In this respect, the dark matter model under study overlaps  with axion dark matter models.

The formulated EFT has a total of fourteen relevant and marginally relevant operators in the dark matter sector, ten of which only
contribute to the linear perturbation theory while the other four  contribute to the background dynamics as well. The former set of operators necessarily violate the WEP in the dark matter sector.
We used the results of two cosmological studies to constrain the coefficients of eight of these operators. However, the most meaningful constraint only applied to one of these coefficients, namely $\beta_2 (\phi)$ which enters the dark matter effective action via $m^2 \beta_2(\phi) \Pi^{\dagger}\Pi$. The most stringent bound on this quantity was obtained from a study of the CMB acoustic peaks \cite{yangbai}, which places an upper bound of $10^{-5}$ on its magnitude. This is the best observational evidence constraining the WEP violation in the dark matter sector.
It should be noted, however, that the majority of the EFT operators
are not constrained or at least well constrained by the current
observational data.

It was noted that the astrophysical and cosmological signatures of the WEP violation, primarily the effective Newton's constant $G^{\rm DM} _{\rm eff}$ for dark matter, is degenerate with the effects of non-minimal coupling between dark matter and gravity. This coupling is negligible for background cosmology as $R \sim H_0 ^2 \ll m^2$. However, its contribution to $G^{\rm DM} _{\rm eff}$ is proportional
to $k_{\rm phys} ^2 /m^2$, which becomes significant on scales shorter
than the Compton wavelength of the dark matter particles. Such a coupling could significantly modify structure formation on sub-galactic scales, particularly for ultralight dark matter fields.
Moreover, since the dark matter perturbations in this model are expected to have a significant sound speed on scales shorter than their Compton wavelengths \cite{marsh-axion-cosmology}, it would be interesting to explore through N-body simulations how non-minimal couplings to gravity alter structure formation on small scales as predicted by this model.

Finally, as we noted in Sec. \ref{sec.scale}, the EFT constructed here
requires a certain degree of fine tuning to both successfully reproduce the background $\Lambda$CDM cosmology and to be sensible
as a radiatively stable EFT. Discovering any connections with UV theories from which such an EFT can result would be of primary interest.

\section{Acknowledgments}
We thank Mariana Carrillo Gonzalez, \'Eanna Flanagan, Maxim Perelstein, Mark Trodden, and Yu-dai Tsai for helpful discussions.
We especially thank \'Eanna Flanagan for very valuable comments on the manuscript. This work was supported in part by the NSF grants PHY-1707800, PHY-1404105, PHY-1505411, the Eberly research funds of Penn State, the John and David Boochever Prize Fellowship in Fundamental Theoretical Physics, and the Urania Stott fund of Pittsburgh Foundation. 
\appendix

\section{Some background geometric tensors} \label{sec.back.tensors}

We find it useful to reminder the reader of the values of some of the
geometric tensors evaluated on the spatially flat FRW background that we used in the construction of the EFT given in Sec. \ref{sec.eft}. We gather this information in Table \ref{table.back}.

\begin{table}[h]
\centering
\begin{tabular}{ccc|c|c}
\cline{4-4}
& & & Background Value  \\ \cline{2-4}
& \multicolumn{1}{|c}{1}& \multicolumn{1}{|c|}{$ \bar{\phi}$} & $t$ &  \\ \cline{2-4}
& \multicolumn{1}{|c}{2}& \multicolumn{1}{|c|}{$\vec{n}$} & $-\frac{1}{a} \vec{\partial}_{\eta}$ &  \\ \cline{2-4}
& \multicolumn{1}{|c}{3}& \multicolumn{1}{|c|}{$\nabla_c n^c$} & $-3 \frac{\mathcal{H}}{a}$ &  \\ \cline{2-4}
& \multicolumn{1}{|c}{4}& \multicolumn{1}{|c|}{$n^c \nabla_c n^d$} & $0$ &  \\ \cline{2-4}
& \multicolumn{1}{|c}{5}& \multicolumn{1}{|c|}{$h_{cd}$} & $a^2 \delta_{cd}$ &  \\ \cline{2-4}
& \multicolumn{1}{|c}{6}& \multicolumn{1}{|c|}{$K_{cd}$} & $-a' \delta_{cd}$ &  \\ \cline{2-4}
& \multicolumn{1}{|c}{7}& \multicolumn{1}{|c|}{$R_{cd} dx^c dx^d$} & $3\Big(\mathcal{H}^2 - \frac{a''}{a} \Big)d \eta^2 + \Big(\mathcal{H}^2 + \frac{a''}{a} \Big)dx^i dx_i$ &  \\ \cline{2-4}
& \multicolumn{1}{|c}{8}& \multicolumn{1}{|c|}{$R$} & $6 \frac{a''}{a^3}$ &  \\ \cline{2-4}
\end{tabular}
\caption{The background values for some of the geometric tensors used in the EFT formalism.}
\label{table.back}
\end{table}

\section{The stress energy tensor for imperfect relativistic fluids}\label{sec.imp.fluid}

Here we review the basic formalism and definitions for the stress energy
tensor of imperfect relativistic fluids. For a more comprehensive discussion, one may consult \cite{weinberg-gr}.

For imperfect fluids, one typically expresses the stress energy tensor
using the following ansatz
\be
T_{ab} = \mathfrak{e} u_a u_b - \mathfrak{C}_{(a} \ u_{b)} + \mathfrak{p} h_{ab} + \Sigma_{ab},
\ee
where $u_a$ is the unit normal vector field associated with the fluid's comoving
frame, $\mathfrak{e}\equiv T_{ab} u^a u^b$ is the locally measured energy
density in the fluid's comoving frame, $\mathfrak{C}_a \equiv T_{cd} h^{(c} \ _{a} u^{b)}$ is regarded as the heat flow covector for fluids in thermal equilibrium, $h_{ab} \equiv g_{ab} + u_a u_b$ is the spatial metric intrinsic to the surfaces of constant time in the fluid's comoving frame, $\mathfrak{p} \equiv T_{ab} h^{ab} /3 $ is the generalized pressure for the fluid, and $\Sigma_{ab} \equiv T_{cd} h^c _{a} h^d _{b} - T_{cd} h^{cd} h_{ab} /3$ is the anisotropic
stress tensor for the fluid. The generalized pressure $\mathfrak{p}$ is typically
decomposed in the following way
\be
\mathfrak{p} \equiv p - \nu \nabla_a u^a,
\ee
where $p$ is regarded as the pressure and $\nu >0$ is called the bulk
viscosity. For fluids in thermal equilibrium, $\mathfrak{C}_a$ is proportional
to the temperature gradient, though its interpretation for fluids that
are not in thermal equilibrium is somewhat obscure. Also, note that $\Sigma_{ab}$ is symmetric and traceless by definition.

\subsection{Stress energy tensors in comoving frames} \label{sec.pert.stress}

Here we compute the dark energy, dark matter, and interactions stress energy tensors up to linear order in the scalar perturbations in the Newtonian gauge defined in Eq. \eqref{frw-newt}. To begin, we express each stress energy tensor
in terms of its components in the corresponding comoving frame by writing \footnote{The stress energy tensors are analogous to that of an imperfect relativistic fluid. The analogy, however, is superficial in that a general interacting field theory does not admit a fluid description.}
\bea \label{stress.imperfect}
&& T_{ab} ^{\text{DE}} = \mathfrak{e}^{\text{DE}} n_a n_b - n_{(a}\mathfrak{C}^{\text{DE}} _{b)}
+\mathfrak{p}^{\text{DE}} h_{ab} + \Sigma_{ab} ^{\text{DE}}, \nonumber\\
&&{}^{\mathcal{R}}T_{ab} ^{\text{DM}} = {}^{\mathcal{R}}\mathfrak{e}^{\text{DM}} \ {}^{\mathcal{R}}u_a \ {}^{\mathcal{R}}u_b -  {}^{\mathcal{R}}u_{(a}{}^{\mathcal{R}}\mathfrak{C}^{\text{DM}} _{b)}
+{}^{\mathcal{R}}\mathfrak{p}^{\text{DM}} \  {}^{\mathcal{R}}H_{ab} +  {}^{\mathcal{R}}\Sigma_{ab} ^{\text{DM}}, \nonumber\\
&&{}^{\mathcal{R}}T_{ab} ^{\text{int}} = {}^{\mathcal{R}}\mathfrak{e}^{\text{int}} \ {}^{\mathcal{R}}u_a \ {}^{\mathcal{R}}u_b -  {}^{\mathcal{R}}u_{(a}{}^{\mathcal{R}}\mathfrak{C}^{\text{int}} _{b)}
+{}^{\mathcal{R}}\mathfrak{p}^{\text{int}} \  {}^{\mathcal{R}}H_{ab} +  {}^{\mathcal{R}}\Sigma_{ab} ^{\text{int}}, \nonumber\\
&&{}^{\zeta}T_{ab} ^{\text{DM}} = {}^{\zeta}\mathfrak{e}^{\text{DM}} \ {}^{\zeta}u_a \ {}^{\zeta}u_b -  {}^{\zeta}u_{(a}{}^{\zeta}\mathfrak{C}^{\text{DM}} _{b)}
+{}^{\zeta}\mathfrak{p}^{\text{DM}} \  {}^{\zeta}H_{ab} +  {}^{\zeta}\Sigma_{ab} ^{\text{DM}},\nonumber\\
&&{}^{\zeta}T_{ab} ^{\text{int}} = {}^{\zeta}\mathfrak{e}^{\text{int}} \ {}^{\zeta}u_a \ {}^{\zeta}u_b -  {}^{\zeta}u_{(a}{}^{\zeta}\mathfrak{C}^{\text{int}} _{b)}
+{}^{\zeta}\mathfrak{p}^{\text{int}} \  {}^{\zeta}H_{ab} +  {}^{\zeta}\Sigma_{ab} ^{\text{int}}.
\eea
For each stress energy tensor above, $\mathfrak{e}$ is the locally measured energy density in the comoving frame of the given fluid component,
 $\mathfrak{p}$ is the generalized pressure, and $\mathfrak{C}^a$ and $\Sigma_{ab}$
are the analogous heat flow vector and anisotropic shear stress tensor for an imperfect fluid.  We also define the four velocities
and the spatial metrics associated with the surfaces of constant $\mathcal{R}$ and $\zeta$ using
\bea
&& {}^{\mathcal{R}}u_a \equiv \frac{\nabla_a \mathcal{R}}{\sqrt{-g^{cd} \nabla_c \mathcal{R} \nabla_d \mathcal{R}}}, \hspace{2cm} {}^{\zeta}u_a \equiv \frac{\nabla_a \zeta}{\sqrt{-g^{cd} \nabla_c \zeta \nabla_d \zeta}}, \nonumber\\
&& {}^{\mathcal{R}}H_{ab} \equiv g_{ab} + {}^{\mathcal{R}}u_a {}^{\mathcal{R}}u_b, \hspace{2.5cm} {}^{\zeta}H_{ab} \equiv g_{ab} + {}^{\zeta}u_a {}^{\zeta}u_b.
\eea
Our objective then comes down to computing $\mathfrak{e}$, $\mathfrak{p}$, $\mathfrak{C}_{a}$ and $\Sigma_{ab}$ for each of the above stress energy tensors up to linear order in the scalar perturbations. A summary of the contribution of the EFT operators to the stress energy tensors is given in Table. \ref{table.operator-stress}.
\begin{table}[h]
\centering
\begin{tabular}{cc|c|c|c|c|c}
\cline{3-6}
& & energy density & generalized pressure & heat flow & anisotropic stress \\ \cline{1-6}
\multicolumn{1}{|}{}&$\beta_1$&  $\dagger$ & $\dagger$ &  & &  \\ \cline{1-6}
\multicolumn{1}{|}{}&$\beta_2$&  $\checkmark$ & $\checkmark$ &  & &  \\ \cline{1-6}
\multicolumn{1}{|}{}&$\beta_3$&  $\checkmark$ & $\checkmark$ &  &  \\ \cline{1-6}
\multicolumn{1}{|}{}&$\tilde{\gamma} $& $\checkmark$ & $\checkmark$ & $\checkmark$ & $\checkmark$ &  \\ \cline{1-6}
\multicolumn{1}{|}{}&$\alpha$ &$\dagger$  &$\dagger$  &  & &     \\ \cline{1-6}
\multicolumn{1}{|}{}&$\alpha_1$ &  &  &  & &     \\ \cline{1-6}
\multicolumn{1}{|}{}& $\alpha_2$ & $\checkmark$ & $\checkmark$ & $\checkmark$ & &  \\ \cline{1-6}
\multicolumn{1}{|}{}& $\alpha_3$ & $\checkmark$ &  & $\checkmark$ & &  \\ \cline{1-6}
\multicolumn{1}{|}{}&$\alpha_4$ & $\checkmark$ & $\checkmark$ & $\checkmark$ &  \\ \cline{1-6}
\multicolumn{1}{|}{}&$\alpha_5$ & $\checkmark$ & $\checkmark$ & $\checkmark$ &  \\ \cline{1-6}
\multicolumn{1}{|}{}&$\alpha_6$&  $\checkmark$ & $\checkmark$ &  & &  \\ \cline{1-6}
\multicolumn{1}{|}{}&$\alpha_7$&  $\checkmark$ & $\checkmark$ & $\checkmark$ & &  \\ \cline{1-6}
\multicolumn{1}{|}{}&$ \alpha_8$ & $\checkmark$ & $\checkmark$ & $\checkmark$ &  \\ \cline{1-6}
\multicolumn{1}{|}{}&$\alpha_9$&   $\checkmark$ & $\checkmark$ &  $\checkmark$ &  $\checkmark$& \\ \cline{1-6}
\multicolumn{1}{|}{}&$\alpha_{10}$&  $\checkmark$ & $\checkmark$ &  $\checkmark$ &  $\checkmark$&  \\ \cline{1-6}
\end{tabular}
\caption{Above is a summary of the contribution of the EFT functions and their associated operators to the dark energy (denoted with $\dagger$) and the radial interactions (denoted with $\checkmark$) stress energy tensors.}
\label{table.operator-stress}
\end{table}

For the dark energy stress energy tensor we find
\bea \label{eft.set.imp.de}
&&\mathfrak{e}^{\text{DE}} = \rho^{\text{DE}} _{0} + \delta \rho ^{\text{DE}}  = e^{2 \alpha}\bigg[\frac{e^{-\alpha}}{2}\Lambda_0 + \tilde{\Lambda} \bigg] + e^{\alpha} \bigg[\Lambda_0 \Big(-\Phi + \frac{\alpha '}{2 a} \tau +  \frac{\tau '}{a}\Big)+ \frac{e^{\alpha}}{a} \tilde{\Lambda} \tau \Big(2 \alpha' + \frac{\tilde{\Lambda}'}{\tilde{\Lambda}} \Big)  \bigg], \nonumber\\
&& \mathfrak{p}^{\text{DE}} = p^{\text{DE}} _{0} + \delta p ^{\text{DE}} = e^{2 \alpha}\bigg[\frac{e^{-\alpha}}{2}\Lambda_0 - \tilde{\Lambda} \bigg] +  e^{\alpha} \bigg[\Lambda_0 \Big(-\Phi + \frac{\alpha '}{2 a} \tau + \frac{\tau '}{a} \Big)-  \frac{e^{\alpha}}{a} \tilde{\Lambda} \tau \Big(2 \alpha' + \frac{\tilde{\Lambda}'}{\tilde{\Lambda}} \Big)  \bigg], \nonumber\\
&&\mathfrak{C}^{\text{DE}}_{i} =  \Sigma^{\text{DE}}_{ab} =0,
\eea
and for the angular dark matter and interactions stress energy tensors we find
\bea\label{eft.set.imp.zeta}
&&{}^{\zeta}\mathfrak{p}^{\text{DM}}=
 {}^{\zeta}\mathfrak{e}^{\text{DM}} =
{}^{\zeta}\rho^{\text{DM}}_0+ {}^{\zeta}\delta \rho^{\text{DM}}
=\mathcal{R}_0 ^2 \frac{\zeta_0 ^{\prime 2}}{a^2} + 2 \mathcal{R}_0 \bigg[\big(\delta \mathcal{R} - \mathcal{R}_0 \Phi\big) \frac{\zeta_0 ^{\prime 2}}{a^2} + \mathcal{R}_0 \frac{\zeta_0 ^{\prime} \delta \zeta '}{a^2}\bigg], \nonumber\\
&& {}^{\zeta}\mathfrak{C}^{\text{DM}}_i = {}^{\zeta}\Sigma_{ab} ^{\text{DM}}= 0,
\eea
and
\be
{}^{\zeta}\mathfrak{p}^{\text{int}}= {}^{\zeta}\mathfrak{e}^{\text{int}}
={}^{\zeta}\mathfrak{C}_a ^{\text{int}} = {}^{\zeta}\Sigma_{ab} ^{\text{int}}=0,
\ee
where $i$ denotes a spatial index.
For the radial dark matter stress energy tensor we find
\bea \label{eft.set.imp.r}
&& {}^{\mathcal{R}}\mathfrak{e}^{\text{DM}} =
{}^{\mathcal{R}}\rho^{\text{DM}}_0+ {}^{\mathcal{R}}\delta \rho^{\text{DM}}  = \frac{\mathcal{R}_0 ^{\prime 2}}{a^2}
+ \mathcal{R}_0 ^2 \Big[\pm m^2 +\lambda \mathcal{R}_0 ^2 \Big]
+ 2 \mathcal{R}_0 \delta \mathcal{R} \Big[\pm m^2 + 2 \lambda \mathcal{R}_0 ^2 \Big]\nonumber\\
&& - \frac{2\mathcal{R}_0 ^{\prime 2}}{a^2} \Phi  + \frac{2 \mathcal{R}'_0 \delta \mathcal{R}' }{a^2}, \nonumber\\
&&{}^{\mathcal{R}}\mathfrak{p}^{\text{DM}} =
{}^{\mathcal{R}}p^{\text{DM}}_0+ {}^{\mathcal{R}}\delta p^{\text{DM}}= \frac{\mathcal{R}_0 ^{\prime 2}}{a^2} + \mathcal{R}_0 ^2 \Big[\mp m^2 - \lambda \mathcal{R}_0 ^2\Big] + 2 \mathcal{R}_0 \delta \mathcal{R} \Big[\mp m^2 - 2 \lambda \mathcal{R}_0 ^2\Big]\nonumber\\
&&- \frac{2\mathcal{R}_0 ^{\prime 2}}{a^2} \Phi +\frac{2\mathcal{R}'_0 \delta \mathcal{R}'}{a^2},\nonumber\\
&&{}^{\mathcal{R}}\mathfrak{C}_i ^{\text{DM}}= {}^{\mathcal{R}}\Sigma_{ab} ^{\text{DM}} = 0.
\eea
As expected, the dark matter stress energy tensor is that of a perfect fluid.

Finally, for the radial interactions stress energy we find
\bea\label{eft.rho.imp.int}
&&{}^{\mathcal{R}}\mathfrak{e}^{\text{int}}
={}^{\mathcal{R}}\rho_0 ^{\text{int}} + {}^{\mathcal{R}}\delta \rho ^{\text{int}}=
\mathcal{R}_0 ^2 \bigg[\pm m^2 \beta_2  + \lambda \beta_3 \mathcal{R}_0 ^2 - \frac{6}{a^2} \tilde{\gamma} \Big(\mathcal{H}^2+\frac{\mathcal{H}\tilde{\gamma}'}{\tilde{\gamma}}+2 \frac{\mathcal{H} \mathcal{R}_0'}{ \mathcal{R}_0}\Big)\bigg]+2 \mathcal{R}_0 \delta \mathcal{R} \nonumber\\
&& \times \bigg[\pm m^2 \beta_2+ 2 \lambda \beta_3 \mathcal{R}_0 ^2-\frac{6}{a^2} \tilde{\gamma} \Big( \mathcal{H}^2 + \frac{\mathcal{H}\mathcal{R}_0 '}{ \mathcal{R}_0}+\frac{\mathcal{H}\tilde{\gamma}'}{\tilde{\gamma}} \Big)\bigg]+ \mathcal{R}_0 ^2 \Phi \bigg[\frac{12}{a^2} \tilde{\gamma} \Big( \mathcal{H}^2 + 2\frac{\mathcal{H} \mathcal{R}_0 '}{ \mathcal{R}_0} +\frac{\mathcal{H}\tilde{\gamma}'}{\tilde{\gamma}}\Big)\nonumber\\
&& -2 \mu^2 \alpha_2
-\frac{6}{a} \mu \mathcal{H}\alpha_6 + \alpha_7 \Big(-\frac{6 \mathcal{H}^2}{a^2}+\frac{3 a''}{a^3}\Big)+\frac{6\mathcal{H}^2}{a^2}(3\alpha_8+\alpha_9) + 6 \alpha_{10} \Big(\frac{\mathcal{H}^2}{a^2}+\frac{a''}{a^3}\Big) \bigg]-6 \frac{\mathcal{R}_0 ^2}{a^2} \alpha_{10} \Psi\nonumber\\
&&\times \bigg(2\mathcal{H}^2 + \frac{2 \mathcal{H} \mathcal{R}_0 '}{\mathcal{R}_0} + \frac{\mathcal{H}\alpha_{10} '}{ \alpha_{10}}\bigg)+\frac{\mathcal{R}_0 ^2}{a} \tau  \bigg[\pm m^2 \beta_2 ' +\lambda \beta_3 ' \mathcal{R}_0 ^2 -\frac{6}{a^2} \tilde{\gamma}' \Big(2\frac{\mathcal{H} \mathcal{R}_0 '}{ \mathcal{R}_0}+\frac{\mathcal{H}\tilde{\gamma}''}{\tilde{\gamma}'} \Big) \bigg]+\frac{4 \tilde{\gamma} \mathcal{R}_0}{a^2}\partial^2 \delta \mathcal{R} \nonumber\\
&&+\frac{\mathcal{R}_0 ^2}{a^3}\partial^2 \tau \bigg[2 \tilde{\gamma}' +4 \alpha_7 \Big(\frac{\mathcal{R}_0 '}{ \mathcal{R}_0}+\mathcal{H}+\frac{\alpha_7 '}{\alpha_7}\Big)-2 \mathcal{H}(3\alpha_8+\alpha_9 )
\bigg] -2 \frac{\mathcal{R}_0 ^2}{a^2}\bigg[\alpha_5 \partial^2 \Phi+  (2\tilde{\gamma}+3\alpha_{10}) \partial^2 \Psi \bigg] \nonumber\\
&&  -\frac{12 \tilde{\gamma} \mathcal{H} \mathcal{R}_0}{a^2}\delta \mathcal{R}'-\frac{2  \mathcal{R}_0 ^2}{a} \tau' \bigg[\frac{3}{a^2}\tilde{\gamma} ' \mathcal{H}
+ \mu^2 \alpha_2 + \frac{2}{a} \mu \alpha_3 \Big(2 \frac{\mathcal{R}_0 '}{\mathcal{R}_0} + 3 \mathcal{H} + \frac{\alpha_3 '}{\alpha_3}\Big)-\frac{2}{a^2} \alpha_4 \Big(2\frac{\mathcal{R}_0 ^{\prime 2}}{ \mathcal{R}_0 ^2}+2\frac{\mathcal{R}_0 ''}{ \mathcal{R}_0}\nonumber\\
&&+5\frac{\mathcal{H}\mathcal{R}_0 '}{ \mathcal{R}_0} + \frac{\alpha_4 '}{\alpha_4} \Big\{\frac{5\mathcal{H}}{2}+\frac{4\mathcal{R}_0 '}{ \mathcal{R}_0}\Big\} + \frac{\alpha_4 ''}{\alpha_4}\Big)+\frac{3}{a^2} \alpha_5 \Big(3 \mathcal{H}^2+\frac{4 \mathcal{H}\mathcal{R}_0'}{ \mathcal{R}_0} + \frac{2 a''}{a}+ \frac{2 \mathcal{H}\alpha_5 '}{\alpha_5}\Big) \bigg]+ \frac{\mathcal{R}_0 ^2}{a^2}\bigg[\Phi'\nonumber\\
&&\times \Big(2\alpha_4\Big\{\frac{2\mathcal{R}_0 '}{ \mathcal{R}_0}+\frac{\alpha_4 '}{\alpha_4}\Big\}+3 \mathcal{H}(-2\alpha_5 + \alpha_7)\Big) + 3\Psi' \Big(2 \tilde{\gamma}  \Big\{2 \mathcal{H}  + 2 \frac{\mathcal{R}_0 '}{ \mathcal{R}_0} + \frac{\tilde{\gamma}'}{\tilde{\gamma}}\Big\}+\alpha_7\Big\{\frac{4 \mathcal{R}_0 '}{ \mathcal{R}_0}+ 5\mathcal{H}  \nonumber\\
&&+2 \frac{\alpha_7 '}{\alpha_7}\Big\} - 2 \mathcal{H}(3\alpha_8 +  \alpha_9) - \alpha_{10} \Big\{2 \mathcal{H} + \frac{2 \mathcal{R}'_0}{ \mathcal{R}_0}+\frac{\alpha_{10} '}{\alpha_{10}}\Big\}\Big)\bigg] + \frac{\mathcal{R}_0 ^2}{a^3} (2 \alpha_5 + \alpha_7) \partial^2 \tau'  +  \frac{\mathcal{R}_0 ^2}{a^2} \bigg[\frac{\tau''}{a^3} \Big\{6 \mathcal{H} \alpha_5 \nonumber\\
&& -2\alpha_4\Big(2 \frac{\mathcal{R}_0 '}{ \mathcal{R}_0} + \frac{\alpha_4 '}{\alpha_4}\Big)\Big\} + 3 \alpha_7 \Psi'' \bigg],
\eea
\bea
&&{}^{\mathcal{R}}\mathfrak{p}^{\text{int}}
={}^{\mathcal{R}}p_0 ^{\text{int}} + {}^{\mathcal{R}}\delta \mathfrak{p}^{\text{int}}= \mathcal{R}_0 ^2 \bigg[\mp m^2 \beta_2 -\lambda \beta_3 \mathcal{R}_0 ^2 + \frac{2}{a^2} \tilde{\gamma} \Big(-\mathcal{H}^2+2\frac{\mathcal{H} \mathcal{R}_0 '}{ \mathcal{R}_0}+2 \frac{\mathcal{R}_0 ^{\prime 2}}{ \mathcal{R}_0 ^2}+ \frac{\mathcal{H} \tilde{\gamma}'}{\tilde{\gamma}}\nonumber\\
&&+4\frac{\mathcal{R}_0 ' \tilde{\gamma}'}{\mathcal{R}_0 \tilde{\gamma}}+2 \frac{a''}{a} + 2 \frac{\mathcal{R}_0 ''}{ \mathcal{R}_0 } + \frac{\tilde{\gamma} ''}{\tilde{\gamma}} \Big) \bigg]+
2 \mathcal{R}_0 \delta \mathcal{R} \bigg[\mp m^2 \beta_2 -2 \lambda \beta_3 \mathcal{R}_0 ^2 + \frac{2}{a^2} \tilde{\gamma} \Big(-\mathcal{H}^2 + \frac{\mathcal{H} \mathcal{R}_0 '}{ \mathcal{R}_0}  + \frac{\mathcal{H}\tilde{\gamma}'}{\tilde{\gamma}}\nonumber\\
&&+2 \frac{\mathcal{R}_0 ' \tilde{\gamma}'}{ \mathcal{R}_0 \tilde{\gamma}}+2\frac{a''}{a}+\frac{\mathcal{R}_0 ''}{ \mathcal{R}_0}+\frac{\tilde{\gamma}''}{\tilde{\gamma}}\Big) \bigg]-\frac{8\tilde{\gamma} \mathcal{R}_0}{3 a^2}\partial^2 \delta \mathcal{R}+ 4\frac{\tilde{\gamma} \mathcal{R}_0 }{a^2}\delta \mathcal{R}'' +\frac{4}{a^2} \tilde{\gamma} \mathcal{R}_0 \delta \mathcal{R} ' \bigg[ \mathcal{H} + 2\frac{\mathcal{R}_0 '}{\mathcal{R}_0}+2 \frac{\tilde{\gamma}'}{\tilde{\gamma}}\bigg]\nonumber\\
&& + \mathcal{R}_0 ^2 \Phi \bigg[\frac{4}{a^2} \tilde{\gamma} \Big(\mathcal{H}^2-2 \frac{\mathcal{H}\mathcal{R}_0'}{\mathcal{R}_0}-2 \frac{\mathcal{R}_0 ^{\prime 2}}{\mathcal{R}_0 ^2}-\frac{\mathcal{H} \tilde{\gamma}'}{\tilde{\gamma}}-4\frac{\mathcal{R}_0 ' \tilde{\gamma}'}{\mathcal{R}_0 \tilde{\gamma}}-2\frac{a''}{a}-2\frac{\mathcal{R}_0 ''}{\mathcal{R}_0 }- \frac{\tilde{\gamma}''}{\tilde{\gamma}}\Big) + 2 \mu^2 \alpha_2 - \mu \alpha_6 \Big(\frac{2\mathcal{R}_0'}{a \mathcal{R}_0} \nonumber\\
&& + \frac{\alpha_6 '}{a\alpha_6}\Big) - \alpha_7\Big(19\frac{\mathcal{H}\mathcal{R}_0'}{a^2 \mathcal{R}_0} + 9 \frac{a''}{a^3} + 2 \frac{\mathcal{R}_0 ^{\prime 2}}{a^2 \mathcal{R}_0 ^2}-\frac{\mathcal{R}_0 ''}{a^2 \mathcal{R}_0} + \Big\{\frac{8 \mathcal{H}}{a^2} + 4 \frac{\mathcal{R}_0 '}{a^2 \mathcal{R}_0} \Big\} \frac{\alpha_7 '}{\alpha_7}+\frac{\alpha_7 ''}{a^2 \alpha_7}\Big) + 3 \alpha_8 \Big( - \frac{2\mathcal{H}^2}{a^2}\nonumber\\
&&  + 8 \frac{\mathcal{H}\mathcal{R}_0 '}{a^2 \mathcal{R}_0} + 4 \frac{a''}{a^3}+4\frac{\mathcal{H} \alpha_8 '}{a^2 \alpha_8}\Big) + 2 \alpha_9 \Big( - \frac{\mathcal{H}^2}{a^2}+ 4 \frac{\mathcal{H} \mathcal{R}_0 '}{a^2 \mathcal{R}_0} + 2 \frac{a''}{a^3} + 2 \frac{\mathcal{H}\alpha_9 '}{a^2 \alpha_9} \Big)
+2 \alpha_{10} \Big(2 \frac{\mathcal{H} \mathcal{R}_0 '}{a^2 \mathcal{R}_0}+\frac{\mathcal{H} \alpha_{10} '}{a^2\alpha_{10}}\Big)\bigg] \nonumber\\
&&
-2 \mathcal{R}_0 ^2 \Psi \bigg[\mu \alpha_6 \Big(3 \frac{\mathcal{H}}{a}+2 \frac{\mathcal{R}_0 '}{a \mathcal{R}_0} + \frac{\alpha_6 '}{a\alpha_6}\Big)+\alpha_7 \Big( 3 \frac{\mathcal{H}^2}{a^2}+10 \frac{\mathcal{H}\mathcal{R}_0 '}{a^2 \mathcal{R}_0} + 3 \frac{a''}{a^3}+2 \frac{\mathcal{R}_0 ^{\prime 2}}{a^2 \mathcal{R}_0 ^2} + 2 \frac{\mathcal{R}_0 ''}{a^2 \mathcal{R}_0} + 5 \frac{\mathcal{H} \alpha_7 '}{a^2 \alpha_7}  \nonumber\\
&&+ 4 \frac{\mathcal{R}_0' \alpha_7 '}{a^2 \mathcal{R}_0 \alpha_7} + \frac{\alpha_7 ''}{a^2 \alpha_7}\Big)-6 \alpha_8 \Big( \frac{\mathcal{H}^2}{a^2} + 2 \frac{\mathcal{H}\mathcal{R}_0 '}{a^2 \mathcal{R}_0}+ \frac{a''}{a^3}+\frac{\mathcal{H} \alpha_8 '}{a^2 \alpha_8}\Big)
- 2 \alpha_9 \Big(\frac{\mathcal{H}^2}{a^2} + 2 \frac{\mathcal{H} \mathcal{R}_0 '}{a^2 \mathcal{R}_0}+\frac{a''}{a^3}+\frac{\mathcal{H} \alpha_9 '}{a^2 \alpha_9}\Big) \nonumber\\
&&-\alpha_{10} \Big( \frac{\mathcal{H}^2}{a^2}+8 \frac{\mathcal{H} \mathcal{R}_0 '}{a^2 \mathcal{R}_0}+3 \frac{a''}{a^3}+4\frac{\mathcal{H} \alpha_{10} '}{a^2 \alpha_{10}}\Big)\bigg]+\frac{\mathcal{R}_0 ^2}{a} \tau \bigg[\mp m^2 \beta_2'-\lambda \beta_3'\mathcal{R}_0 ^2+2 \tilde{\gamma}' \Big( -2 \frac{\mathcal{H} \mathcal{R}_0 ' }{a^2 \mathcal{R}_0}+2\frac{\mathcal{R}_0 ^{\prime 2}}{a^2 \mathcal{R}_0 ^2} \nonumber\\
&&+ \frac{a''}{a^3}+2\frac{\mathcal{R}_0 ''}{a^2 \mathcal{R}_0}
- \frac{\mathcal{H}\tilde{\gamma} ''}{a^2 \tilde{\gamma} '}+4 \frac{\mathcal{R}_0 ' \tilde{\gamma} ''}{a^2 \mathcal{R}_0 \tilde{\gamma}'}+ \frac{\tilde{\gamma} '''}{a^2 \tilde{\gamma} '} \Big) \bigg]+\frac{ \mathcal{R}_0 ^2}{3 a^3}\partial^2 \tau \bigg[-4 \tilde{\gamma}' -6 \alpha_7 \Big(2 \frac{\mathcal{R}_0 '}{ \mathcal{R}_0} + \mathcal{H} + \frac{\alpha_7 '}{\alpha_7}\Big)+6 \alpha_8  \nonumber\\
&&\times \Big(2 \frac{\mathcal{R}_0'}{\mathcal{R}_0} +\mathcal{H}+\frac{\alpha_8 '}{\alpha_8}\Big)+2 \alpha_9 \Big(2 \frac{\mathcal{R}_0'}{ \mathcal{R}_0} +\mathcal{H}+\frac{\alpha_9 '}{\alpha_9}\Big)\bigg]-\frac{4  \mathcal{R}_0 ^2}{3 a^2}(\tilde{\gamma} + \alpha_{10} )   \partial^2 [\Phi-\Psi] +2  \frac{\mathcal{R}_0 ^2}{a} \tau' \bigg[\frac{\tilde{\gamma}'}{a^2} \Big(-\mathcal{H} \nonumber\\
&&+4 \frac{\mathcal{R}_0 '}{\mathcal{R}_0}+2 \frac{\tilde{\gamma} ''}{ \tilde{\gamma}'}\Big)     - \mu^2 \alpha_2 + \alpha_4 \Big(\frac{2 \mathcal{H}\mathcal{R}_0'}{a^2 \mathcal{R}_0}+\frac{\mathcal{H}\alpha_4 '}{a^2 \alpha_4}\Big)+\alpha_5 \Big( -3 \frac{\mathcal{H}^2}{a^2} +2 \frac{\mathcal{H}\mathcal{R}_0 '}{a^2 \mathcal{R}_0}+\frac{\mathcal{H} \alpha_5 ' }{a^2 \alpha_5}+\frac{a''}{a^3}\Big) \bigg] +2  \frac{\mathcal{R}_0 ^2}{a^2} \Phi' \nonumber\\
&& \times \bigg[-\tilde{\gamma} \Big( 2\mathcal{H}+2 \frac{\mathcal{R}_0 '}{\mathcal{R}_0}+\frac{\tilde{\gamma} '}{ \tilde{\gamma}}\Big) + \alpha_4\Big(2 \frac{\mathcal{R}_0 '}{ \mathcal{R}_0} + \frac{\alpha_4 '}{\alpha_4}\Big)+\alpha_5 \Big(2 \frac{\mathcal{R}_0 '}{\mathcal{R}_0}-\mathcal{H} +\frac{\alpha_5 '}{\alpha_5}\Big)-3 \mathcal{H}\alpha_7 +3 \mathcal{H}\alpha_8 +\mathcal{H}\alpha_9 \bigg] \nonumber\\
&&+ \frac{\mathcal{R}_0 ^{2}}{a^2} \Psi' \bigg[4 \tilde{\gamma} \Big( 2\mathcal{H}+2 \frac{\mathcal{R}_0 '}{\mathcal{R}_0}+\frac{\tilde{\gamma} '}{\tilde{\gamma}}\Big) -6 \alpha_7 \Big(2 \frac{\mathcal{R}_0 '}{\mathcal{R}_0}  +2 \mathcal{H}+\frac{\alpha_7 '}{\alpha_7}\Big) +6 \alpha_8 \Big(2 \frac{\mathcal{R}_0 '}{\mathcal{R}_0}  +2 \mathcal{H}+\frac{\alpha_8 '}{\alpha_8}\Big) + 2 \alpha_9 \Big(2 \frac{\mathcal{R}_0 '}{\mathcal{R}_0} \nonumber\\
&& +2 \mathcal{H}+\frac{\alpha_9 '}{\alpha_9}\Big) + \alpha_{10} \Big(10 \mathcal{H}+ 6 \frac{\mathcal{R}_0 '}{\mathcal{R}_0} + 3 \frac{\alpha_{10} '}{\alpha_{10}}\Big) \bigg]- 2 \frac{ \mathcal{R}_0 ^2}{a^3}\tau'' \bigg[ -\tilde{\gamma}\alpha_4   \Big(2 \frac{\mathcal{R}_0 '}{\mathcal{R}_0} + \frac{\alpha_4 '}{\alpha_4}\Big) + \alpha_5 \Big(2 \frac{\mathcal{R}_0 '}{\mathcal{R}_0} -3 \mathcal{H} + \frac{\alpha_5 '}{\alpha_5}\Big)\bigg] \nonumber\\
&&+2 \frac{\mathcal{R}_0 ^2 \alpha_5}{a^2}\Phi''+ 2\frac{ \mathcal{R}_0 ^2}{a^2} \Psi''(-2 \gamma \beta_4-3 \alpha_7 +3 \alpha_8 + \alpha_9+\alpha_{10} )  -2 \frac{\mathcal{R}_0 ^2}{3 a^3}\partial^2 \tau'(3 \alpha_7 - 3 \alpha_8 - \alpha_9) -2 \frac{\mathcal{R}_0 ^2 \alpha_5}{a^3} \tau''' ,\nonumber\\
\eea
\bea
&&{}^{\mathcal{R}}\mathfrak{C}_i ^{\text{int}}=  \mathcal{R}_0 ^2 \bigg[\frac{2 \tilde{\gamma}}{a \mathcal{R}_0 '}\partial_i \delta \mathcal{R}\Big(-4 \mathcal{H}^2 -2\mathcal{H} \frac{\mathcal{R}_0 '}{\mathcal{R}_0}-2 \mathcal{H} \frac{\tilde{\gamma} '}{\tilde{\gamma}}+2 \frac{\mathcal{R}_0 ' \tilde{\gamma} '}{\mathcal{R}_0 \tilde{\gamma}}+2 \frac{a''}{a}+2 \frac{\mathcal{R}_0 ''}{\mathcal{R}_0}+ \frac{\tilde{\gamma} ''}{\tilde{\gamma}}\Big)-\frac{4\tilde{\gamma} }{a\mathcal{R}_0}\partial_i \delta \mathcal{R}'\nonumber\\
&&+\partial_i \tau \bigg(\frac{2 \tilde{\gamma}}{a^2} \Big\{-2 \frac{\mathcal{R}_0 ^{\prime}\tilde{\gamma} '}{\mathcal{R}_0 \tilde{\gamma}}+2\frac{\tilde{\gamma}'}{\tilde{\gamma}} \mathcal{H} -\frac{\tilde{\gamma} ''}{\tilde{\gamma}}  \Big\}+2 \mu^2 \alpha_2 +\frac{2}{a} \mu \alpha_3 \Big\{3 \mathcal{H}+2\frac{\mathcal{R}_0 '}{\mathcal{R}_0}+\frac{\alpha_3 '}{\alpha_3}\Big\}+2\alpha_4 \Big\{-\frac{4\mathcal{H}\mathcal{R}_0'}{a^2 \mathcal{R}_0} \nonumber\\
&&-\frac{2\mathcal{R}_0^{\prime 2}}{a^2 \mathcal{R}_0 ^2}- \frac{2\mathcal{R}_0 ''}{a^2 \mathcal{R}_0}-\frac{2 \mathcal{H} \alpha_4 '}{a^2 \alpha_4}-\frac{4 \mathcal{R}_0 ' \alpha_4 '}{a^2 \mathcal{R}_0 \alpha_4} - \frac{\alpha_4 ''}{a^2 \alpha_4}\Big\}+2\alpha_5 \Big\{3\frac{\mathcal{H}^2}{a^2}+3 \frac{a''}{a^3}+6\frac{\mathcal{H}\mathcal{R}_0 '}{a^2 \mathcal{R}_0}+3 \frac{\mathcal{H} \alpha_5 '}{a^2 \alpha_5}\Big\}   \bigg)\nonumber\\
&&+\frac{2}{a} \partial_i  \tau' \bigg(- \frac{\tilde{\gamma}'}{a} -\mu \alpha_3 +\alpha_4 \Big\{2\frac{\mathcal{R}_0 '}{a \mathcal{R}_0}+\frac{\alpha_4'}{a\alpha_4}\Big\}-4\frac{\mathcal{H}}{a}\alpha_5  \bigg)+\frac{1}{a^2}\partial_i \partial^2 \tau  (\alpha_7-2\alpha_8 -2 \alpha_9)+\frac{2}{a^2}\alpha_5 \partial_i \tau'' \nonumber\\
&& +\partial_i \Phi \bigg(\frac{2 \tilde{\gamma}}{a} \Big\{2 \frac{\mathcal{R}_0 ^{\prime}}{\mathcal{R}_0}+2\mathcal{H}+\frac{\tilde{\gamma}'}{\tilde{\gamma}}\Big\}
+2\mu \alpha_3-2\alpha_4 \Big\{2\frac{\mathcal{R}_0 '}{a \mathcal{R}_0} +\frac{\alpha_4 '}{a\alpha_4}\Big\}+\frac{\mathcal{H}}{a}(6 \alpha_5 + 3 \alpha_7 - 6 \alpha_8 -2 \alpha_9) \bigg)\nonumber\\
&&-\frac{4\mathcal{H} \mathcal{R}_0 ^2 \alpha_{10}}{a}\partial_i \Psi-2\frac{\alpha_5}{a}\partial_i \Phi'+ \frac{1}{a }(4 \tilde{\gamma}+3 \alpha_7 -6 \alpha_8 -2 \alpha_9)  \partial_i \Psi' \bigg], \nonumber\\[20pt]
&&{}^{\mathcal{R}}\Sigma_{ij,( i \neq j)} ^{\text{int}}  = 4 \tilde{\gamma} \mathcal{R}_0 \partial_{i}\partial_{j} \delta \mathcal{R}+\frac{2}{a} \mathcal{R}_0 ^2\Big(\tilde{\gamma}'+\alpha_9 \Big[\mathcal{H}+2 \frac{\mathcal{R}'_0}{ \mathcal{R}_0}+ \frac{\alpha_9 '}{\alpha_9}\Big]\Big)  \partial_i\partial_j \tau +2  \mathcal{R}_0^2 \Big(\tilde{\gamma}+\alpha_{10}\Big) \partial_{i}\partial_{j} [\Phi-\Psi]\nonumber\\
&&+2 \alpha_9 \frac{\mathcal{R}_0 ^2}{a} \partial_i \partial_j \tau',\nonumber\\[20pt]
&&{}^{\mathcal{R}}\Sigma_{ii} ^{\text{int}}=-\frac{2}{3a} \mathcal{R}_0 ^2 \Big(\tilde{\gamma} '+\alpha_9 \Big[\mathcal{H}+2 \frac{\mathcal{R}'_0}{\mathcal{R}_0}+\frac{\alpha_9 '}{\alpha_9}\Big]\Big)\Big[\partial^2 \tau- 3\partial_i ^2 \tau \Big]-\frac{4}{3}\tilde{\gamma} \mathcal{R}_0  \Big[\partial^2 \delta \mathcal{R} - 3 \partial_i ^2 \delta \mathcal{R} \Big]\nonumber\\
&&-\frac{2}{3} \mathcal{R}_0 ^2 \Big(\tilde{\gamma}+\alpha_{10}\Big) \Big[\partial^2(\Phi-\Psi) -3 \partial_i ^2 (\Phi-\Psi) \Big]-\frac{2\mathcal{R}_0 ^2}{3a} \alpha_9 \Big[\partial^2 \tau'- 3\partial_i ^2 \tau' \Big],
\eea
where $\partial^2 \equiv \delta^{ij} \partial_i \partial_j$.

\section{The effective field theory dark matter operators} \label{sec.operator}

The effective action for the dark matter sector expressed in Eq. \eqref{eft.act.dm} consists of all the relevant and marginally relevant dark matter operators that independently contribute to the equations of motion
at linear order in scalar perturbations. Some of the operators
are expressed in their perturbed forms since their contributions to the
background equations of motion are degenerate with those of the unperturbed
operators. Our intention in this section is to show that any
effective dark matter operator that is quadratic in perturbations
can be expressed in terms of the operators listed in Table \ref{table.operator}.
To this end, we first analyse the operators that
are quadratic in the dark matter perturbations and then
move on to the ones that are linear in the dark matter perturbations.

\subsection{Second order operators in the dark matter sector}

We list all such operators up to dimension four in Table \ref{table.dm-dm.operators}. The indices are contracted using the background $n^a$, $h^{ab}$,
and $g^{ab}$.

\begin{table}[h]
\centering
\begin{tabular}{ccc|c|c}
\cline{4-4}
& & & dimension  \\ \cline{2-4}
& \multicolumn{1}{|c}{1}& \multicolumn{1}{|c|}{$ \delta \Pi^{\dagger} \delta \Pi$} & 2 &  \\ \cline{2-4}
& \multicolumn{1}{|c}{2}& \multicolumn{1}{|c|}{$\nabla_a \delta \Pi^{\dagger} \delta \Pi, \ \text{c.c}$} & 3 &  \\ \cline{2-4}
& \multicolumn{1}{|c}{3}& \multicolumn{1}{|c|}{$ \nabla_a  \delta \Pi^{\dagger} \nabla_b  \delta \Pi$} & 4 &  \\ \cline{2-4}
\end{tabular}
\caption{Operators that are quadratic in dark matter perturbations.}
\label{table.dm-dm.operators}
\end{table}

All operators listed in the table above, except the second one, directly result from perturbing the operators that are listed in Table \ref{table.operator}. The
second operator can enter the effective Lagrangian density \eqref{eft.act.dm}
in the following form
\be
\bar{\mu} n^a \big( \kappa \nabla_a  \delta \Pi^{\dagger}  \delta \Pi
+ \kappa ^*  \delta \Pi^{\dagger} \nabla_a  \delta \Pi   \big),
\ee
for some complex function $\kappa$ and some constant $\bar{\mu}$ of dimension one.
It is not difficult to see that a rescaling of the dark matter fields $\Pi$ and
$\Pi^{\dagger}$ by some complex function, together with a redefinition
of the functions $\alpha$, $\alpha_1$, $\tilde{m}^2$, and $\tilde{\lambda}$
removes this extra term.

\subsection{First order operators in the dark matter sector}

Up to integration by parts, these operators enter the effective action
in the form
\be
\mathcal{O} \times \ (\delta \Pi^{\dagger} \Pi_0 + \text{c.c})
\footnote{One might consider the more general linearly perturbed dark matter operator
$$
 \kappa \delta \Pi^{\dagger} \Pi_0 + \text{c.c},
$$
for some complex function $\kappa$. This term is consistent with the global $U(1)$ symmetry. However, such a term would have to result from perturbing an operator in the form of $(\Pi^{\dagger} \Pi)^m$, which then implies that $\kappa$ has to be a real function.},
\ee
where $\mathcal{O}$ is first order in the gravitational / dark energy perturbations and up to dimension two. All such operators must be perturbations to the operators that transform covariantly under spacetime diffeomorphisms, and
when expressed in the dark energy uniform density gauge they must transform
covariantly under the foliation preserving diffeomorphisms.
They are listed in Table \ref{table.non.operators}. The indices on the operators listed above are contracted using the background values of $g^{ab}$ and $h^{ab}$ (or alternatively $g^{ab}$ and $n^a n^b$). Also note that the covariant derivative is evaluated using the background FRW metric.

\begin{table}[h]
\centering
\begin{tabular}{ccc|c|c}
\cline{4-4}
& & & dimension  \\ \cline{2-4}
& \multicolumn{1}{|c}{1}& \multicolumn{1}{|c|}{$ \delta g_{ab}$, $\delta n_a$, $\delta f$} & 0 &  \\ \cline{2-4}
& \multicolumn{1}{|c}{2}& \multicolumn{1}{|c|}{$ \nabla_c \delta g_{ab}$, $\nabla_c \delta n_a$, $\nabla_c \delta f$, $\delta K_{ab}$} & 1 &  \\ \cline{2-4}
& \multicolumn{1}{|c}{3}& \multicolumn{1}{|c|}{$ \nabla_d \nabla_c \delta g_{ab}$, $ \nabla_d \nabla_c \delta n_a$, $\nabla_d \nabla_c \delta f$, $\nabla_d \delta K_{ab}$, $\delta R^{(3)}_{abcd} $, $\delta R_{abcd} $  } & 2 &  \\ \cline{2-4}
\end{tabular}
\caption{Operators that are linear in gravitational / dark energy perturbations.
Recall that $f \equiv g^{ab} \bar{\phi}_{,a} \bar{\phi}_{,b}$. $R^{(3)}_{abcd} $ is the Riemann curvature tensor on surfaces of constant $\bar{\phi}$.}
\label{table.non.operators}
\end{table}

We begin by noting that the operators associated with $\delta g_{ab}$
are generated in the first line of action \eqref{eft.act.dm}. For instance, the operator
\be
g^{ab} \delta g_{ab} \ (\delta \Pi^{\dagger} \Pi_0+ \text{c.c})
\ee
comes from the term $\sqrt{-g} \Pi^{\dagger} \Pi$.
Through simple algebraic operations, such as integration by parts, the
same can be seen to be true for $\nabla_c \delta g_{ab}$ and $\nabla_d \nabla_c \delta g_{ab}$.

The operator $\delta n_{a}$, which should appear in the
effective action as $n^a \delta n_a$, and $\delta f$ provide a similar
set of information about the spacetime foliation function $\bar{\phi}$, with both being proportional to the lapse function $\Phi$ defined in Eq. \eqref{frw-newt} in the dark energy uniform density gauge.

As for $\nabla_b \delta n_a$ and $\nabla_c \nabla_b \delta n_a$,
they encode information about the changes to the unit normal vector field to the
surfaces of $\bar{\phi}$. This information is provided by the extrinsic
curvature and changes to the extrinsic curvature along the unit normal
to surfaces of $\bar{\phi}$, as well as the operators $\nabla_a \delta f$
and $\nabla_a \nabla_b \delta f$. Therefore we regard the operators
associated with $\delta n_a$ as being redundant.

The operators associated with $\delta K_{ab}$ appear in the effective
action in one of the two following forms
\be
n^a n^b \delta K_{ab}, \hspace{2cm} h^{ab} \delta K_{ab}.
\ee
Note that $h^{ab} \delta K_{ab}$ is generated by $\delta K$,
and $ n^a n^b \delta K_{ab} $ vanishes.
This latter results from
\be \label{nnK}
n^a n^b K_{ab} =0 \rightarrow \delta \big(n^a n^b K_{ab} \big)
= \big( \delta n^{a} n^b + \delta n^b n^a \big) K_{ab} + n^a n^b \delta K_{ab}=0,
\ee
together with $\vec{n} = -\vec{\partial}_\eta /a $ and the fact that the background extrinsic curvature is purely spatial. As for the ones
corresponding to $\nabla_a \delta K_{bc}$, they should enter the action
in the following forms
\be \label{nnnK}
n^a h^{bc} \nabla_{a} \delta K_{bc}, \hspace{1cm} n^a n^b n^c \nabla_{a} \delta K_{bc}, \hspace{1cm} n^b h^{ac} \nabla_a \delta K_{bc}.
\ee
The middle term in Eq. \eqref{nnnK} vanishes by simple algebraic operations
and using Eq. \eqref{nnK}. The first term can be written as
\bea \label{nhk}
&&n^a h^{bc} \nabla_a \delta K_{bc} = n^a \nabla_a [ h^{bc} \delta K_{bc}]
-n^a \delta K_{bc} \nabla_a h^{bc} = n^a \nabla_a [\delta K - \delta h^{bc} K_{bc} ] - n^a \delta K_{bc} \nabla_a h^{bc} \nonumber\\
&& = n^a \nabla_a \delta K - \delta K_{bc} n^a \nabla_a [n^b n^c] - n^a \nabla_a [ \delta h^{bc} K_{bc}] = n^a \nabla_a \delta K - n^a \nabla_a [ \delta g^{bc} K_{bc}],
\eea
where we used $n^a \nabla_a n^b=0$ and $n^a K_{ab}=0$. It is evident
from the last expression in Eq. \eqref{nhk} that this operator is already accounted for in the effective action. Similarly, we write $n^b h^{ac} \nabla_a \delta K_{bc}$ as
\bea
&&n^b h^{ac} \nabla_a \delta K_{bc} = h^{ac}\nabla_a [n^b \delta K_{bc}]-h^{ac} \delta K_{bc} \nabla_a n^b = h^{ac}\nabla_a [n^b \delta K_{bc}] - \delta K_{bc} K^{bc} \nonumber\\
&& = - h^{ac} \nabla_a [ \delta n^b K_{bc}] - \delta K_{bc} K^{bc}.
\eea
This term can be generated from $\delta[K_{ab} K^{ab}]$ along with
operators associated with $\delta f$ in the effective action.

Finally, the remaining two geometric objects are the Riemann curvature tensors
for the leaves of foliation and for the spacetime.
It follows from
\be \label{spcurv}
R_{abcd} ^{(3)} =h_{a} \ ^{e} h_{b} \ ^{f} h_{c} \ ^{g} h_{d} \ ^{h} R_{efgh} -K_{ac} K_{bd} + K_{bc} K_{ad}
\ee
that the perturbations to the Riemann curvature tensor of the
leaves of foliation can be expressed in terms of
$\delta n_a$ (or $\delta f$), $\delta K_{ab}$, and
$\delta R_{abcd}$. As for the spacetime curvature tensor, the following terms
\be
\delta R_{abcd} \  n^a n^c g^{bd}, \hspace{1cm} \delta R_{abcd} \ g^{ac} g^{bd}
\ee
along with their various indices-permuted counterparts should
be included in the effective action. We instead choose
the perturbations to the Ricci tensor $\delta R_{ab}$ for
the spacetime. The difference is absorbed in the terms
involving the operators $\delta f$ and $\delta g^{ab}$.

\bibliographystyle{JHEP}
\bibliography{Reference}

\end{document}